\UseRawInputEncoding
\documentclass[11pt]{article}
\pdfoutput=1 
\usepackage{amssymb, amsmath,amsfonts}
\usepackage{graphicx}
\usepackage{xcolor}
\usepackage{color}
\usepackage[pdftex, bookmarks=true,colorlinks,linkcolor=red,urlcolor=blue,citecolor=blue]{hyperref}
\usepackage[normalem]{ulem}
\usepackage{cite}
\usepackage{subcaption}
\usepackage{slashed}

\def\arXiv#1{\href{http://arxiv.org/abs/#1}{arXiv:#1}}
\def\arXiv#1#2{\href{http://arxiv.org/abs/#1}{arXiv:#1}}

\def\doi#1#2{\href{http://doi.org/#1}{#2}}

\allowdisplaybreaks
\usepackage{enumerate}
\usepackage[normalem]{ulem}

\makeatletter\@addtoreset{equation}{section}\makeatother

\newcommand{\preprint}[1]{\begin{table}[t]  
             \begin{flushright}               
             {#1}                             
             \end{flushright}                 
             \end{table}}                     
\renewcommand{\title}[1]{\vbox{\center\LARGE{#1}}\vspace{5mm}}
\renewcommand{\author}[1]{\vbox{\center#1}\vspace{5mm}}

\renewcommand{\Im}{\operatorname{Im}}
\newcommand{\address}[1]{\vbox{\center\em#1}}

\usepackage{bm}
\def\mI{{\bf I}}

\def\be{\begin{eqnarray}}
\def\ee{\end{eqnarray}}
\def\bea{\begin{eqnarray}}
\def\eea{\end{eqnarray}}

\def\Dslash{\,\,{\raise.15ex\hbox{/}\mkern-12mu D}}
\def\Dbarslash{\,\,{\raise.15ex\hbox{/}\mkern-12mu {\bar D}}}
\def\delslash{\,\,{\raise.15ex\hbox{/}\mkern-9mu \partial}}
\def\delbarslash{\,\,{\raise.15ex\hbox{/}\mkern-9mu {\bar\partial}}}
\def\pslash{\,\,{\raise.15ex\hbox{/}\mkern-9mu p}}
\def\calDslash{\,\,{\raise.15ex\hbox{/}\mkern-12mu {\cal D}}}

\def\lae{\mathrel{\mathop{\smash{\lower .5 ex \hbox{$\stackrel<\sim$}}}}}
\def\lae{\mathrel{\mathop{\smash{\lower .5 ex \hbox{$\stackrel>\sim$}}}}}

\usepackage{ulem}

\begin{document}

\unitlength = .8mm

\begin{titlepage}
\vspace{.5cm}
\preprint{}
\begin{center}
\hfill \\
\hfill \\
\vskip 1cm

\title{\bf 
A Weyl-$\boldmath{Z}_2$
semimetal from holography}
\vskip 0.5cm

{Xuanting Ji$^{a,b}$}\footnote{Email: {\tt jixuanting@cau.edu.cn
}},
{Yan Liu$^{c}$}\footnote{Email: {\tt yanliu@buaa.edu.cn}},
{Ya-Wen Sun$^{b,d}$}\footnote{Email: {\tt yawen.sun@ucas.ac.cn}},
{Yun-Long Zhang$^{e,f,g}$}\footnote{Email: {\tt zhangyunlong@nao.cas.cn}}

\address{${}^a$Department of Applied Physics, College of Science, \\
China Agricultural University, Beijing 100083, China}
\vspace{-10pt}
\address{${}^b$School of Physical Sciences, and CAS Center for Excellence in Topological Quantum Computation, University of Chinese Academy of Sciences, 
Beijing 100049, China}
\vspace{-10pt}
\address{${}^c$Center for Gravitational Physics, Department of Space Science, \\ and International Research Institute of Multidisciplinary Science,
\\ Beihang University,  Beijing 100191, China}
\vspace{-10pt}
\address{${}^d$Kavli Institute for Theoretical Sciences, \\
University of Chinese Academy of Sciences, 
Beijing 100049, China }
\vspace{-10pt}
\address{${}^e$National Astronomy Observatories, Chinese Academy of Science, Beijing, 100101, China}
\vspace{-10pt}
\address{${}^f$School of Fundamental Physics and Mathematical Sciences, Hangzhou Institute for Advanced Study,
University of Chinese Academy of Sciences, Hangzhou 310024, China}
\vspace{-10pt}
\address{${}^g$International Centre for Theoretical Physics Asia-Pacific, Beijing/Hangzhou, China}

\end{center}
\vspace{-10pt}
\abstract{
We present effective field theories for the weakly coupled  Weyl-$\boldmath{Z}_2$ semimetal, as well as the holographic realization for the strongly coupled case. In both cases, the anomalous systems have both the chiral anomaly and the $\boldmath{Z}_2$ anomaly and possess topological quantum phase transitions from the Weyl-$\boldmath{Z}_2$ semimetal phases to partly or fully topological trivial phases. We find that the topological phase transition is characterized by the anomalous transport parameters, i.e. the anomalous Hall conductivity and the $\boldmath{Z}_2$ anomalous Hall conductivity. These two parameters are nonzero at the Weyl-$\boldmath{Z}_2$ semimetal phase and vanish at the topologically trivial phases. In the holographic case, the different behavior between the two anomalous transport coefficients is discussed. Our work reveals the novel phase structure of strongly interacting Weyl-$\boldmath{Z}_2$ semimetal with two pairs of nodes.}

\vfill

\end{titlepage}

\begingroup
\hypersetup{linkcolor=black}
\tableofcontents
\endgroup

\section{Introduction}
\label{introduction}

As a new kind of topological gapless system, Weyl semimetal was proposed about ten years ago \cite{Wan}, which has drawn much attention in recent yesrs \cite{rmb}. Up to now, Weyl semimetals have been identified in the laboratory in the TaAs family \cite{weng,B.Q.Lv,shuang,SXu}, YbMnBi$_{2}$ \cite{Borisenko} (Type-I Weyl semimetal) and Mo$_{x}$W$_{1-x}$Te$_{2}$ \cite{Soluyanov,Deng,HaoZheng}  (Type-II Weyl semimetal) and so on.

Weyl semimetal, equipped with the so-called Weyl nodes, which can be presented as the band crossings and behave like monopoles due to the Berry curvature in the three-dimensional momentum space, can be obtained by breaking time-reversal symmetry or spatial inversion symmetry. One Dirac node in the Dirac semimetal separates into two Weyl nodes.
Weyl nodes appear in pairs with opposite chirality. With chiral anomaly \cite{Ninomiya} in the system, anomalous transport coefficients caused by the anomaly are  important to physics in condensed matter physics as well as in high energy physics and astrophysics. This may manifest itself by a negative longitudinal magneto-resistance (or positive magneto-conductance).
Furthermore, the mixed axial gravitational anomaly in NbP has also been observed recently \cite{Gooth}, which may open a new window in theoretical aspects.

Although the low energy excitation of some condensed matter systems can be described by fundamental ``relativistic" equations of motion \cite{Grushin:2012mt}, the condensed matter systems with nontrivial topological energy band structures show richer physical behaviors than these equations themselves. As an example, a nontrivial $\boldmath{Z}_2$ topological invariant has been demonstrated in the Dirac semimetal\cite{yang1,yang2,sato,Gorbar,fangc}, namely $\boldmath{Z}_2$ Dirac semimetal, which can not be understood straightforwardly from the relativistic Dirac equation.\footnote{Multi-Weyl semimetal \cite{Dantas:2019rgp} is another example.}
In the $\boldmath{Z}_2$ Dirac semimetal with two Dirac nodes, similar to the chiral charge the $\boldmath{Z}_2$ topological invariant can be defined as
$C_{\small Z_{2}}=\left(C_{\uparrow}-C_{\downarrow}\right)/2$, where $C_{\uparrow,\, \downarrow}$ are the chiral charges of the spin-up and spin-down Weyl fermions in the system. The $Z_2$ anomaly\footnote{Note that with the name $\boldmath{Z}_2$ anomaly, it does not mean that the system possesses a $Z_2$ symmetry. It refers to the fact that there exists a new kind of $Z_2$ topological charge. The extra new $Z_2$ anomaly is associated with another $U(1)$ symmetry which is neither a vector gauge symmetry nor the axial gauge symmetry. In the original paper \cite{kimb}, this extra anomaly has been named as $Z_2$ anomaly, which indicates this anomaly is associated with a symmetry that produces a $Z_2$ topological charge and the symmetry is not a $Z_2$ symmetry.}  is a spin analogy of the chiral anomaly, and the presence of this extra anomaly brings interesting transport behavior.
The interplay between $\boldmath{Z}_2$ and  chiral anomalies will lead to observable effects in magnetotransport \cite{kimb}. Since Weyl semimetal can be obtained from Dirac semimetal, it is fairly interesting to investigate whether these far-reaching results still exist in Weyl semimetal.

The goal of this paper is to study the Weyl semimetals with both chiral anomaly and $\boldmath{Z}_2$  anomaly which will be named as Weyl-$\boldmath{Z}_2$ (or equivalently $\boldmath{Z}_2$ Weyl) semimetal in the following. One can start from a $\boldmath{Z}_2$ Dirac semimetal with two Dirac points and then split each of these two Dirac points into Weyl points along the transverse direction. In this way we obtain the  $\boldmath{Z}_2$ Weyl semimetal, in which there exist two pairs of Weyl points and each pair carry a nontrivial $\boldmath{Z}_2$ topological charge, thus each node carries both the chiral and the $\boldmath{Z}_2$ charge. {We can mark the chiral and $\boldmath{Z}_2$ charge of the four Weyl points as ($\pm,\uparrow$) or ($\pm,\downarrow$) to distinguish them, where $\pm$ means the positive and negative of the chirality, while the up (down) arrow means spin up (down). There is also another possibility for the Weyl-$\boldmath{Z}_2$ semimetal, which has four nodes and carries pairs of topological charges as $(\pm, 0)$ and $(0, \uparrow), (0, \downarrow)$, i.e. each pair of charges are completely independent.

By extending the Weyl semimetal model with one pair of Weyl nodes in \cite{Grushin:2012mt, Landsteiner:2015pdh},  two different $\boldmath{Z}_2$ Weyl semimetal models with two pairs of Weyl/$\boldmath{Z}_2$ nodes will be studied
in Sec. \ref{sec:Z2WSMft}. In this model, each node carries both the chiral charge and $\boldmath{Z}_2$ charge. In contrast to the three phases realized in the model with only one pair of Weyl nodes in \cite{Grushin:2012mt, Landsteiner:2015pdh}, more phases exist in the current model when tuning the parameters in the system.
More precisely, we have: {(a) $\boldmath{Z}_2$ Weyl semimetal phase, where four Weyl nodes exist in the spectrum (referred to as the Weyl-$\boldmath{Z}_2$ phase in Fig.~\ref{fig:ophase} and \ref{fig:phase}); (b) the critical point with all of the nodes annihilated to one single Dirac point (critical-critical in Fig.~\ref{fig:ophase} and \ref{fig:phase}), which is different from the critical point of a Weyl semimetal \cite{Grushin:2012mt, Landsteiner:2015pdh}; (c) topologically trivial phase, in which there is no band touching point in the spectrum (trivial-trivial). There are also four critical phases which are the phase transition lines between different phases, including the following ones: (d) only one pair of the nodes annihilate to a critical point while the other pair of nodes corresponds still exist (Weyl/$\boldmath{Z}_2$-critical). Depending on which two nodes with opposite charges (both the chiral charge and the $\boldmath{Z}_2$ charge) annihilate, there are two types of phases in this category. If the other pair of nodes vanish and form a gap, we obtain another two critical phases, i.e. (e) critical-trivial. Depending on which two nodes with opposite charges form a gap, there are also two kinds of phases in this category. We also have two more interesting phases (f) one pair of the nodes vanish to form a gap while the other pair of nodes still exist  (Weyl/$\boldmath{Z}_2$-trivial). Again depending on which two nodes with opposite charges form a gap, there are two kinds of phases in this category. We will show them in subsections \ref{owpb} and \ref{wpb}.

The physical picture above in the weak coupling regime is clear from
the energy spectrum. What about an inherently strongly coupled system where the concept of quasiparticle is not
applicable? AdS/CFT makes it possible to build a holographic model which can describe the physical properties in the strong coupling regime
for the topological phase transition that occurred in the weakly coupling regime.
Recently, a holographic Weyl semimetal
with one-pair of Weyl nodes was built in \cite{Landsteiner:2015lsa,Landsteiner:2015pdh}. Several transport parameters including negative magnetoconductivity \cite{Sun:2016gpy}, odd viscosity \cite{Landsteiner:2016stv}, axial Hall conductivity \cite{Copetti:2016ewq}, AC conductivity \cite{Grignani:2016wyz}, chiral vortical effect \cite{Ji:2019pxx} have been studied. People also investigated the exotic surface state  \cite{Ammon:2016mwa}, topological invariants \cite{Liu:2018djq, Song:2019asj}, non-local physical quantities \cite{Baggioli:2018afg, Baggioli:2020cld}, disorder and momentum dissipation effects \cite{Ammon:2018wzb, Zhao:2021pih}, phase transitions to insulators \cite{Liu:2018spp, fadafan}, topological nodal line semimetals \cite{Liu:2018bye, Liu:2020ymx} and so on. These works are quite suggestive to the physics of strongly coupled gapless topological states of matter and more details can be found in a recent review \cite{Landsteiner:2019kxb}.

Based on these developments, we will construct a holographic model to study the strongly coupled Weyl-$\boldmath{Z}_2$ semimetal with two pairs of Weyl/$\boldmath{Z}_2$ nodes. We will solve the model at zero temperature and study all its solutions which give rise to all different phases which are also presented in the weak coupling regime.
We will also calculate the transport parameters by the Kubo formula at both finite and zero temperature, i.e. anomalous Hall conductivity ($\sigma_{\text{AHE}}$) and $\boldmath{Z}_2$ anomalous Hall conductivity ($\sigma_{\boldmath{Z_{2}}\text{AHE}}$), which can be regarded as the order parameters of the topological phase transitions. We shall show that the holographic model shares both features of the two weakly coupled Weyl-$\boldmath{Z_{2}}$ semimetals.
This model can also be reduced to the model with only one pair of Weyl nodes in \cite{Landsteiner:2015pdh} in the limiting case.

This paper is organized as follows. In Sec. \ref{sec:Z2WSMft} we construct two effective field theory models of the Weyl-$\boldmath{Z}_2$ semimetals with two pairs of Weyl/$\boldmath{Z}_2$ nodes which encodes the chiral anomaly and $Z_2$ anomaly. In Sec. \ref{sec:holo}, we study a holographic model of the Weyl-$\boldmath{Z}_2$ semimetal where all different phases are uncovered and the analogy to the weakly coupled theory is discussed. Order parameters of the topological phase transition are calculated at both finite and zero temperatures. Sec. \ref{sec:cd} is devoted to conclusions and discussions.

\section{Weyl-$\boldmath{Z}_2$ semimetal from effective field theories
}
\label{sec:Z2WSMft}

Weyl semimetals are gapless topological states of matter that possess nontrivial chiral charge. Usually Weyl nodes appear in pairs with opposite chiral charge $\pm 1$. The topological charge of a node in a Weyl semimetal can be calculated by the integral of the Berry curvature in the three dimensional momentum space. Usually the topological charge should be $\pm 1$, however, there exist the so-called multi-Weyl semimetal states where the topological charge for each node could be $\pm n$ with an integer $n$ greater than 1.

Recently, a novel kind of gapless topological states of matter with a non-trivial $Z_2$ topological charge has been considered in \cite{kimb} and \cite{Morimoto}. $\boldmath{Z}_2$ charge is a spin analog of the chiral charge, which is defined by $C_{\small Z_{2}}=\left(C_{\uparrow}-C_{\downarrow}\right)/2$, where $C_{\uparrow,\, \downarrow}$ are the chiral charges of the spin-up and spin-down Weyl fermions in the system. Quite similar to the chiral charge, this $\boldmath{Z}_2$ charge measures the difference between the spin up and spin down fermions with same chiral charges. This definition is quite similar with the definition of the spin-Chern numbers in a two dimensional quantum spin Hall insulator\cite{kimb}. The Dirac semimetal with a topological $Z_2$ charge has been constructed in \cite{kimb} where there are two Dirac nodes separated along a direction in the momentum space.\footnote{A holographic model in the probe limit for the $Z_2$ Dirac semimetal was studied recently in \cite{Kiczek:2020qsw}.} They have also demonstrated that in Dirac semimetals with two Dirac nodes, there exists a corresponding $\boldmath{Z}_2$  anomaly, which is closely analogous to the chiral anomaly. $\boldmath{Z}_2$ anomaly related transport properties have also been studied in \cite{kimb}, e.g. the $\boldmath{Z}_2$ anomalous Hall conductivity. As mentioned in the introduction, there is also another possibility for the Weyl-$\boldmath{Z}_2$ semimetal, with four nodes carrying pairs of charges as $(\pm, 0)$ and $(0, \uparrow), (0, \downarrow)$.

In this paper we focus on a generalization of the Weyl semimetal state, which is a topological semimetal system possessing both the chiral and $\boldmath{Z}_2$ topological charges. In this case, in the topological non-trivial phase, the nodes need to possess two non-trivial topological charges, the chiral one and the $\boldmath{Z}_2$ one. This means we need to have at least four nodes in the topological non-trivial phase. Thus we name this system the Weyl-$\boldmath{Z}_2$ semimetal. In this section we will first present two Lorentz invariant Lagrangians for the system with both topological charges. We then analyze its phase structure in a very detailed way and we will finally check the anomaly by calculating the Ward identities.

\subsection{Effective model for the Weyl-$\boldmath{Z}_2$ semimetal}
In this subsection, we will build an effective field theory model for Weyl semimetals with four nodes, and each node carries both the chiral and the $\boldmath{Z}_2$ topological charges. In next subsection, we will construct another effective field theory model for Weyl semimetals with four nodes with each node carrying either chiral topological charge or the $\boldmath{Z}_2$ topological charge. The comparisons between these two models will be discussed.

We first review that for a Weyl semimetal with only one pair of Weyl nodes,
the weakly coupled quantum field theory Lagrangian is
\begin{equation}
\label{eq1}
\mathcal{L} = \bar\psi \left( i \slashed\partial -  e \slashed{A} - \gamma^{\mu}\gamma^5  b_{\mu} + M \right)\psi\,,
\end{equation} which has been studied in \cite{burkov-tns, Colladay:1998fq, Grushin:2012mt} in which a single Dirac spinor and a time-reversal odd axial gauge field were introduced. The term $\slashed X = \gamma^\mu X_\mu$, where $\gamma^\mu$ are the Dirac matrices, and $\gamma_5 = i \gamma_0\gamma_1\gamma_2\gamma_3$
allows us to define left- or right-handed spinors via $(1\pm\gamma_5)\psi = \psi_{L,R}$. $\slashed{A}\equiv\gamma^\mu A_\mu$, and $A_\mu$ is the electromagnetic gauge potential. Tuning the ratio between the mass parameter $M$ and the time-reversal symmetry breaking parameter $b$, there exists a topological phase transition from a Weyl semimetal to a trivial semimetal across a critical Dirac semimetal. Without loss of generality, we could choose $\vec{b}$ to be in the $z$ direction so that the two Weyl nodes in the Weyl semimetal phase are separated in the $k_z$ direction of the momentum space.

Now our goal is to generalize this set-up to the Weyl-$\boldmath{Z}_2$ semimetal system. Note that in the set-up above the spinor is a four-component spinor whose degrees of freedom describe the chirality and particle-hole degrees, which could be seen from the calculation of the Berry curvature where we treated half of the bands as occupied hole states. To have also the $\boldmath{Z}_2$ topological charge, we need to introduce the spin degrees of freedom to the system. In order to do so, we need to expand the four-component spinor $\psi$ into an eight-component spinor $\Psi$.

Following \cite{kimb,Jang}, we consider the following eight-component spinor $\Psi$
\begin{eqnarray}
\label{psi}
\Psi=\left(\Psi_{p,+,\uparrow},\Psi_{p,+,\downarrow},\Psi_{h,+,\uparrow},\Psi_{h,+,\downarrow},\Psi_{p,-,\uparrow},\Psi_{p,-,\downarrow},\Psi_{h,-,\uparrow},\Psi_{h,-,\downarrow}\right)^{T},
\end{eqnarray}
where $p,h$ in the index refers to particle-hole, $\pm$ refers to the chirality and the arrow means spin up and down, which will be confirmed later by the definition of the operators of the system.

For this eight-component spinor system, we need to also generalize the $4\times 4$ Gamma matrices into $8\times8$  matrices. Thus, we define the following new `Gamma' matrices
\be
\label{eq:88marix}
\Gamma^{\mu}\equiv\gamma^{\mu}\otimes \mathbb{I}_2\,,~~\hat{\Gamma}^{\mu}\equiv\gamma^{\mu}\otimes \mathbb{Z}_2\,,~~\Gamma^{5}\equiv\gamma^{5}\otimes \mathbb{I}_2\,,~~\hat{\Gamma}^{5}\equiv\gamma^{5}\otimes \mathbb{Z}_2\,,\ee
where $\mu=0,1,2,3$, $\gamma^\mu$ is the $4\times 4$ Dirac Gamma matrix and
\begin{align}
\label{eq:matrix}
\mathbb{I}_2 &=\left(\begin{array}{cc}1 & 0 \\0 & 1\end{array}\right) ,\qquad
\mathbb{Z}_2=\left(\begin{array}{cc}1 & 0 \\0 & -1\end{array}\right)\,.
\end{align}
Note that in four space-time dimensions, for eight-component spinors we should have eight Gamma matrices, which are $\Gamma^{\mu}$ and $\hat{\Gamma}^{\mu}$ here. From the definition above, we could see that ${\Gamma}^{\mu}$ is a direct generalization of the $\gamma^{\mu}$ to the eight-component system, while $\hat{\Gamma}^{\mu}$ is more like a $\boldmath{Z}_2$ generalization of the $\gamma^{\mu}$ to the eight-component system. $\Gamma^{5}$  defined above is in fact $-i \Gamma^{0}\Gamma^{1}\Gamma^{2}\Gamma^{3}$ and 
$\hat\Gamma^5=-i \hat{\Gamma}^{0}\hat{\Gamma}^{1}\hat{\Gamma}^{2}\hat{\Gamma}^{3} \,{\mI}_{Z}$ with  ${\mI}_{Z}=\mathbb{I}_4\otimes \mathbb{Z}_2$. Here, $\pm1$ in $\mathbb{Z}_{2}$ means spin up and down, respectively.
We have
$\left\{ \Gamma^{\mu},\Gamma^{\nu}\right\} =\left\{ \hat{\Gamma}^{\mu},\hat{\Gamma}^{\nu}\right\} =2\eta^{\mu\nu}\mathbb{I}_{8\times8}$
and $\left\{ \Gamma^{\mu},\Gamma^{5}\right\} =\left\{ \hat{\Gamma}^{\mu},\hat{\Gamma}^{5}\right\} =0$\,. {Now we can show that the first four components in \eqref{psi} have a positive chirality, while the remaining four components have a negative chirality. This can be confirmed from the matrix $\Gamma^{5}$, which is the chiral charge operator. The $\boldmath{Z}_2$ charge operator is defined by a direct product of a $4\times 4$ unit matrix with the $\mathbb{Z}_{2}$ matrix above. With this definition, we can confirm the role of each component in the eight component spinor \eqref{psi} under the $\boldmath{Z}_2$ charge. Therefore, the information of $\boldmath{Z}_2$ charge has been taken into consideration in the eight-component spinor. }

Motivated by  \cite{kimb} and \cite{burkov-tns, Colladay:1998fq, Grushin:2012mt}, we consider the following Lagrangian as the field theory model for the $\boldmath{Z}_2$ Weyl semimetal,
\begin{align}
\label{eq:oLagrangian}
\mathcal{L}=\Psi^{\dagger}\left[\Gamma^{0}\left(i\Gamma^{\mu}\partial_{\mu}-e\Gamma^{\mu}A_{\mu}-\Gamma^{\mu}\Gamma^{5}b_{\mu}+M_1{\mI}_{1}+ M_2{\mI}_{2}
\right)+\hat{\Gamma}^{0}\left(e\hat{\Gamma}^{\mu}\hat{A}_{\mu}-\hat{\Gamma}^{\mu}\hat{\Gamma}^{5}c_{\mu}\right)\right]\Psi\,,
\end{align}
where the `Gamma' matrices are defined in (\ref{eq:88marix}).
$ {\mI}_{1}$ and $ {\mI}_{2}$ are two diagonal matrices with diagonal elements as $ {\mI}_1=\text{diag}\left(1,0,1,0,1,0,1,0\right)$ and $ {\mI}_2=\text{diag}\left(0,1,0,1,0,1,0,1\right)$.
We have introduced four gauge fields in (\ref{eq:oLagrangian}). $A_{\mu}$ and $b_{\mu}$ are the electromagnetic gauge field and chiral gauge field with field strengths $F_{\mu\nu}=\partial_{\mu}A_{\nu}-\partial_{\nu}A_{\mu}$,
and
$F_{\mu\nu}^{5}=\partial_{\mu}b_{\nu}-\partial_{\nu}b_{\mu}$, respectively. These two gauge fields contribute to the chiral anomaly.
While $\hat{A}_{\mu}$ is the fictitious spin gauge field with field strength $\hat{F}_{\rho\sigma}=\partial_{\rho}\hat{A}_{\sigma}-\partial_{\sigma}\hat{A}_{\rho}$ and $c_\mu$ is the so-called $\boldmath{Z}_{2}$ gauge field with field
strength $\hat{F}_{\rho\sigma}^{5}=\partial_{\rho}c_{\sigma}-\partial_{\sigma}c_{\rho}$\cite{kimb},
both of which contribute to the $\boldmath{Z}_2$ anomaly.
Here $b_0$ and $c_0$ play the role of axial chemical potential and $\boldmath{Z}_2$ chemical potential which breaks inversion symmetry, while $b_i$ and $c_i$ play the role of separation in the momentum space which breaks time reversal symmetry.

 The mass terms ${M}_{1} {\mI}_{1}$ and ${M}_2 {\mI}_{2}$ are introduced in (\ref{eq:oLagrangian}) to break the chiral and the analog spin symmetry separately.
The two masses $M_1$ and $M_2$ could be thought of the gap in the degrees of freedom for spin up and down sector respectively.
Comparing to the model \eqref{eq1},
$M_1$ and $b$ describe half of the degrees of freedom in this system,  while $M_2$ and $c$ describe the other half degrees of freedom in this system.
Also note that in this section we work in the convention of the metric with most minus sign while in the next section with most plus sign.

In the following, we consider the Weyl-$\boldmath{Z}_2$ semimetal system with the Lagrangian above. Similar to the Weyl semimetal case, in the Weyl-$\boldmath{Z}_2$ semimetal system the parameter $b_{\mu}$ denotes the separation of the Weyl nodes in the momentum space while $c_{\mu}$ denotes the separation of the nodes with opposite $\boldmath{Z}_2$ charges in the momentum space.  Here we choose $b_{\mu}=b\delta^{z}_{\mu}$ and $c_{\mu}=c\delta^{y}_{\mu}$, i.e. the two separations are in different spatial directions, in order to make the physics more clear. We could as well choose the two separations in the same spatial direction, which we leave for future consideration.

With the choice of $b_{\mu}$ and $c_{\mu}$ above, we could calculate the energy spectrum of (\ref{eq:oLagrangian}) from the Hamiltonian of the system. The eight eigenvalues of the Hamiltonian are
\eqref{eq:oLagrangian}  are
\begin{eqnarray}\label{ospectrum}
\begin{split}
E_{1}&=\pm\sqrt{b_{z}^{2}+c_{y}^{2}+k_{y}^{2}+k_{z}^{2}+M_1^{2}\pm2\sqrt{2b_{z}c_{y}k_{y}k_{z}+c_{y}^{2}\left(k_{y}^{2}+M_1^{2}\right)+b_{z}^{2}\left(k_{z}^{2}+M_1^{2}\right)}}\,,\\
E_{2}&=\pm\sqrt{b_{z}^{2}+c_{y}^{2}+k_{y}^{2}+k_{z}^{2}+M_{2}^{2}\pm2\sqrt{-2b_{z}c_{y}k_{y}k_{z}+c_{y}^{2}\left(k_{y}^{2}+M_{2}^{2}\right)+b_{z}^{2}\left(k_{z}^{2}+M_{2}^{2}\right)}}\,.
\end{split}
\end{eqnarray}
In the above form, the two `$\pm$' in the same $E$ should be understood as independent choice of either plus or minus.
From \eqref{ospectrum}, we note that both $E_1$ and $E_2$ depend on the gauge fields $b$ and $c$ which are crucial for the topological charges. The mixing of $b$ and $c$ in the spectrum lead to that the nodes carry both the chiral $U(1)$ and the analog spin $Z_2$ charges.

We will analyze the phase structure of spectrum \eqref{ospectrum} in the following in detail. Before that note that a different while closely related set-up for the multi-Weyl semimetal has been studied in \cite{Dantas:2019rgp}. In that set-up, there could still be only one pair of Weyl nodes, but the Weyl nodes could have integer monopole charges larger than 1 and the low energy dispersion around the Weyl nodes is anisotropic.\footnote{A holographic model for this type of mulitiple Weyl semimetal has been studied in \cite{Juricic:2020sgg}.} In the multi-Weyl semimetal system, a non-abelian anomaly exists \cite{Dantas:2019rgp}.
Here we have two pairs of Weyl nodes and each Weyl node has both the chiral charge $\boldmath{Z}_2$ charge to be $\pm 1$. Thus, in summary they are focusing on a single pair of Weyl nodes with additional monopole charge $n>1$, and the effective field theory displays a U(1)$_A\times$SU(2) non-Abelian anomaly, while in our case, we have two pairs of nodes with the U(1)$_A$ anomaly and an extra Abelian $Z_2$ anomaly.

\subsubsection{Phase diagram from the field theory model}
\label{owpb}
In this subsection, we analyze the phase behavior of the energy spectrum (\ref{ospectrum}) obtained in the last subsection.
 Recall that in the Weyl semimetal case, depending on the relative values of $b$ and $M$, there are two phases and one critical point. Here with a fixed ``unimportant" parameter $c/b$\footnote{The ``unimportance"  refers to the fact that $c/b$ only affects the phase structure quantitatively. Different from the effect of tuning $M_1/b$ and $M_2/c$, tuning $c/b$ has no qualitative effects in the phase structure of the system.}, the system would still depend on two dimensionless parameters $M_1/b$ and $M_2/c$, thus, the phase diagram would be two dimensional. Phase boundaries would be lines on the phase diagram plane and the phase transition lines would intersect at a particular critical point. We will show that depending on the values of $M_1/b$ and $M_2/c$, there will be four distinct phases and four phase transition lines and one critical point.

Apparently the four bands in $E_{1,2}$ of (\ref{ospectrum}) with positive signs within the square root are gapped bands and in the following we focus on the other four bands which may produce interesting structure, i.e. the four bands that pick the negative sign within the square root. For these four bands, depending on the values of $b$, $c$, $M_1$ and $M_2$, the minimum values of $E_{1,2}$ could be either zero or larger than zero, thus the system could have crossing nodes or be in a gapped state for different parameters.

The behavior of the spectrum as a function of $k_y$ and $k_z$ in the nine different phases is summarized in Fig.~\ref{fig:ophase}, where $k_x$ is fixed to be zero. Note that a nonzero $k_x$ would immediately gap the system thus the crossing nodes in the figure are still nodes in the three dimensional momentum space. We explain in detail the four different phases, four phase transition lines and one critical point in the following.

\vspace{.25cm}
\noindent {\bf  $\bullet$ \em The Weyl-$\boldmath{Z}_2$ phase}

When both $M_1$ and $M_2$ are smaller than $\sqrt{b^2+c^2}$, we obtain the spectrum in Fig. \ref{fig:ophase}(b)
with four nodes at  $(k_x, k_y, k_z)=\left(0\,,~\pm\frac{c\sqrt{b^{2}+c^{2}-M_1^{2}}}{\sqrt{b^2+c^2}}\,,~\pm\frac{b\sqrt{b^{2}+c^{2}-M_{1}^{2}}}{\sqrt{b^2+c^2}}\right)$ and $\bigg(0\,,~\pm\frac{ c\sqrt{b^{2}+c^{2}-M_{2}^{2}}}{\sqrt{b^2+c^2}}\,,\\
~\mp\frac{b\sqrt{b^{2}+c^{2}-M_{2}^{2}}}{\sqrt{b^2+c^2}}\bigg)$. For this phase, a small perturbation in the mass terms would not gap the system.} This belongs to a topologically nontrivial phase with four nontrivial nodes. Different from the Weyl semimetal case, here we define a set of topological charges with two components as $(n_1,n_2)$, where $n_1=\pm 1$ gives the chiral charge of the nodes and $n_2=\,\uparrow$ or $n_2=\,\downarrow$ gives the $\boldmath{Z}_2$ charge of the nodes.

Note that in order to calculate the topological charges using the Berry curvature in this case, we need to first project the eight component spinor into a four component spinor, e.g. to calculate the chiral charge, we project the negative energy eigenstate of the Hamiltonian to a four component spinor composed of the (1,3,5,7)-th component of the full spinor while to calculate the $\boldmath{Z}_2$ charge, we need to project the negative energy eigenstate to the four component spinor with the (1,2,3,4)-th component of the original spinor. With this definition, the four nodes in this phase would have topological charges $(+,\uparrow)$,$(-,\uparrow)$,$(+,\downarrow)$ and $(-,\downarrow)$, respectively. With this viewpoint, we can view the four nodes as two pairs of Weyl nodes with opposite $\boldmath{Z}_2$ charges between the two pairs of nodes or we can as well view the four nodes as two pairs of $\boldmath{Z}_2$ nodes with opposite chiral charges between the two pairs. Here we emphasize again that in this system we have multiple nodes, but it is different from the multi-Weyl semimetal state mentioned above in which there could be only two nodes but with each node possess a topological number larger than 1.

\vspace{.25cm}
\noindent {\bf $\bullet$ \em The two Weyl/$\boldmath{Z}_2$-critical phases}

Without loss of generality, we fix $b, c$ and tune $M_1, M_2$ to obtain different phases. When we increase $M_1$ (or $M_2$) from (a) in Fig.~\ref{fig:ophase}, two nodes with opposite topological charges (both the chiral charge and the $\boldmath{Z}_2$ charge) would annihilate and form a critical Dirac node with topological charge $(0,0)$ and the other pair of nodes stay the same. That is, the two nodes with topological charges $(1,1)$ and $(-1,-1)$ (or $(1,-1)$ and $(-1,+1)$)\footnote{Note that we can also use $\pm 1$ to represent spin up and down, respectively.} would annihilate to a trivial Dirac node. Note that it is not that each pair of the Weyl nodes (or $\boldmath{Z}_2$ nodes) annihilate into a critical Dirac point. This is because in this case the Dirac point would have a chiral charge of 2 which is not allowed in this system and only nodes with opposite topological charges could annihilate.

Depending on which pair of nodes annihilate, there are two different cases, each having one critical point and a pair of nodes with opposite topological charges. Thus this case corresponds to two phase transition lines in the phase diagram and is shown in (d) of Fig.~\ref{fig:ophase}. As shown in (d) we now have a pair of Weyl/$\boldmath{Z}_2$ nodes and a critical Dirac point.

\vspace{.25cm}
\noindent {\bf $\bullet$ \em The two Weyl/$\boldmath{Z}_2$-gap phases}

Continue to increase $M_1$ (or $M_2$) from the two Weyl/$\boldmath{Z}_2$-critical phases transition lines in (d) of Fig.~\ref{fig:ophase}, the critical Dirac node becomes a trivial gap. This corresponds to two Weyl/$\boldmath{Z}_2$-gap phases, as shown in the case (f) in Fig.~\ref{fig:ophase}.

\vspace{.25cm}
\noindent {\bf $\bullet$ \em The double critical point}

Again starting from the two Weyl/$\boldmath{Z}_2$-critical phases transition lines in (d) of Fig.~\ref{fig:phase} and this time we increase the other parameter $M_2$ (or $M_1$), the two Weyl/$\boldmath{Z}_2$ nodes will also reach a critical point at which two nodes merge into one Dirac node. The system at this special set of parameters $M_1/b$ and $M_2/c$ corresponds to a double critical point on the phase diagram, which is the case (b) in Fig.~\ref{fig:ophase}.

\vspace{.25cm}
\noindent {\bf $\bullet$ \em The two gap-critical phases}

Starting from the double critical point, we increase one of the mass parameters $M_1$ (or $M_2$), the fourfold-degenerate critical point would split into a pair of gapped bands and one twofold-degenerate critical point. These two phases correspond to two phase transition lines between the gap-gap phase and the Weyl/$\boldmath{Z}_2$-gap phase. This spectrum is shown in (e) of Fig.~\ref{fig:ophase}.

\vspace{.25cm}
\noindent {\bf $\bullet$ \em The gap-gap phase}

Starting from any of the two gap-critical phases transition lines above, and increase the other mass parameters, the system would become fully gapped and is in a gap-gap phase with all bands gapped. This corresponds to (c) in Fig.~\ref{fig:ophase}.
\vspace{.25cm}\\
The behavior of the energy spectrum with different parameters is summarized in Fig.~\ref{fig:ophase}. There are nine different phases (including critical points or phase transition lines) which could be summarized into six types of spectrums: including the Weyl-$\boldmath{Z}_2$,  Weyl/$\boldmath{Z}_2$-critical, Weyl/$\boldmath{Z}_2$-gap, critical-critical, critical-gap, and gap-gap phases.

\vspace{0cm}
\begin{figure}[h!]
  \centering
  \begin{subfigure}[b]{0.33\textwidth}
\includegraphics[width=\textwidth]{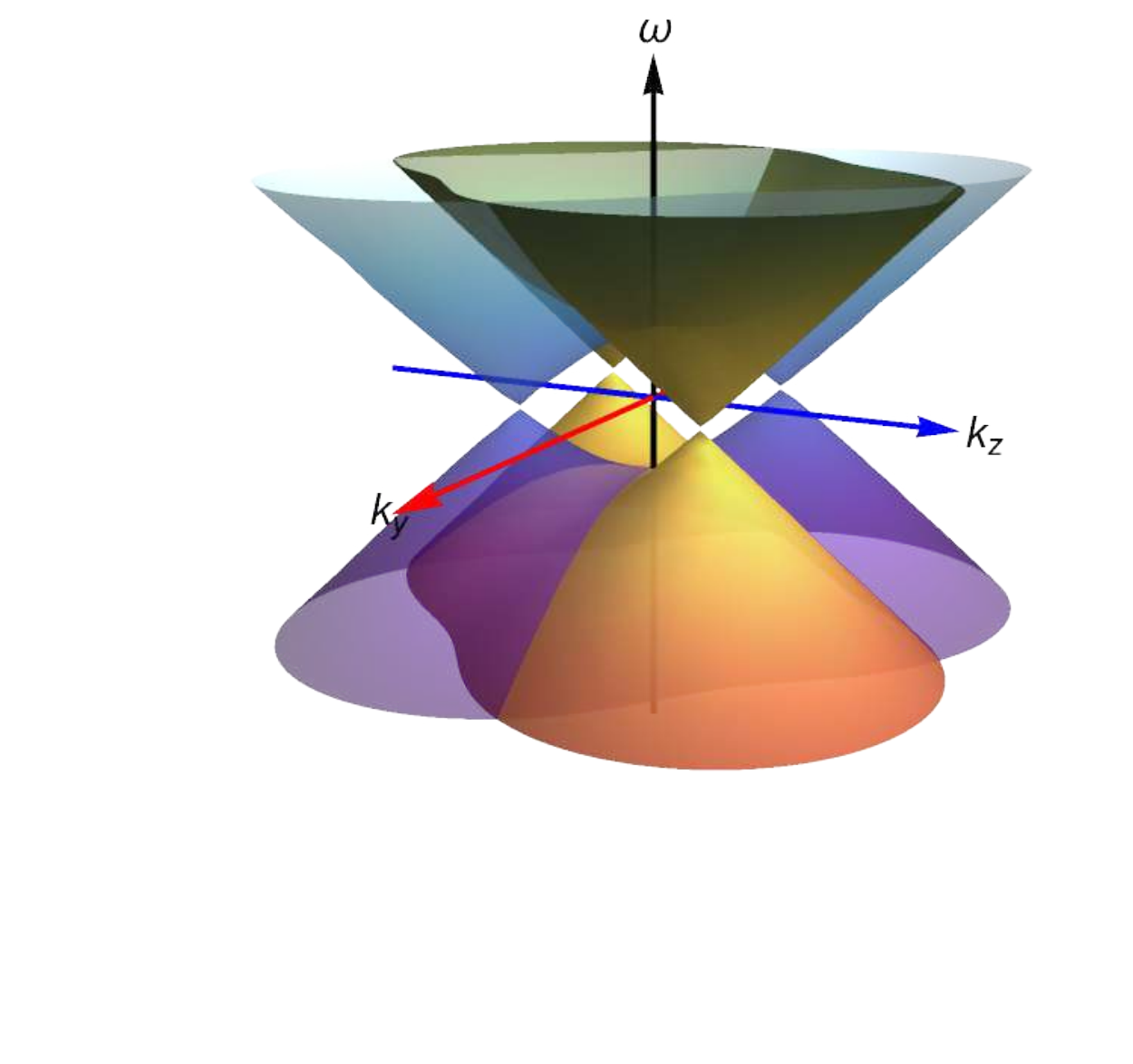}
\vspace{-1.7cm}
\caption{\small Weyl-$\boldmath{Z}_2$}
\end{subfigure}
\begin{subfigure}[b]{0.32\textwidth}
\includegraphics[width=\textwidth]{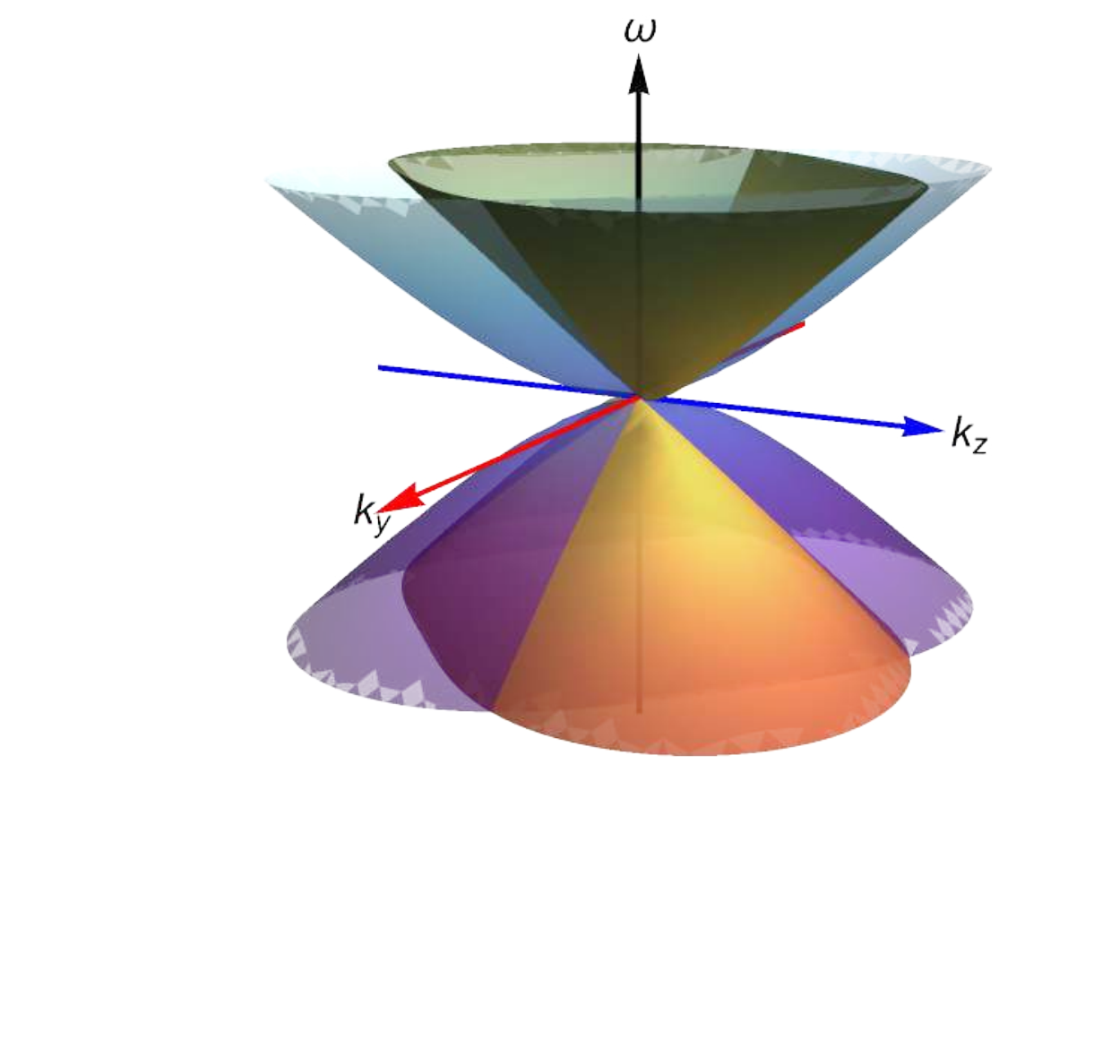}
\vspace{-1.7cm}
\caption{\small double critical}
\end{subfigure}
\begin{subfigure}[b]{0.33\textwidth}
\includegraphics[width=\textwidth]{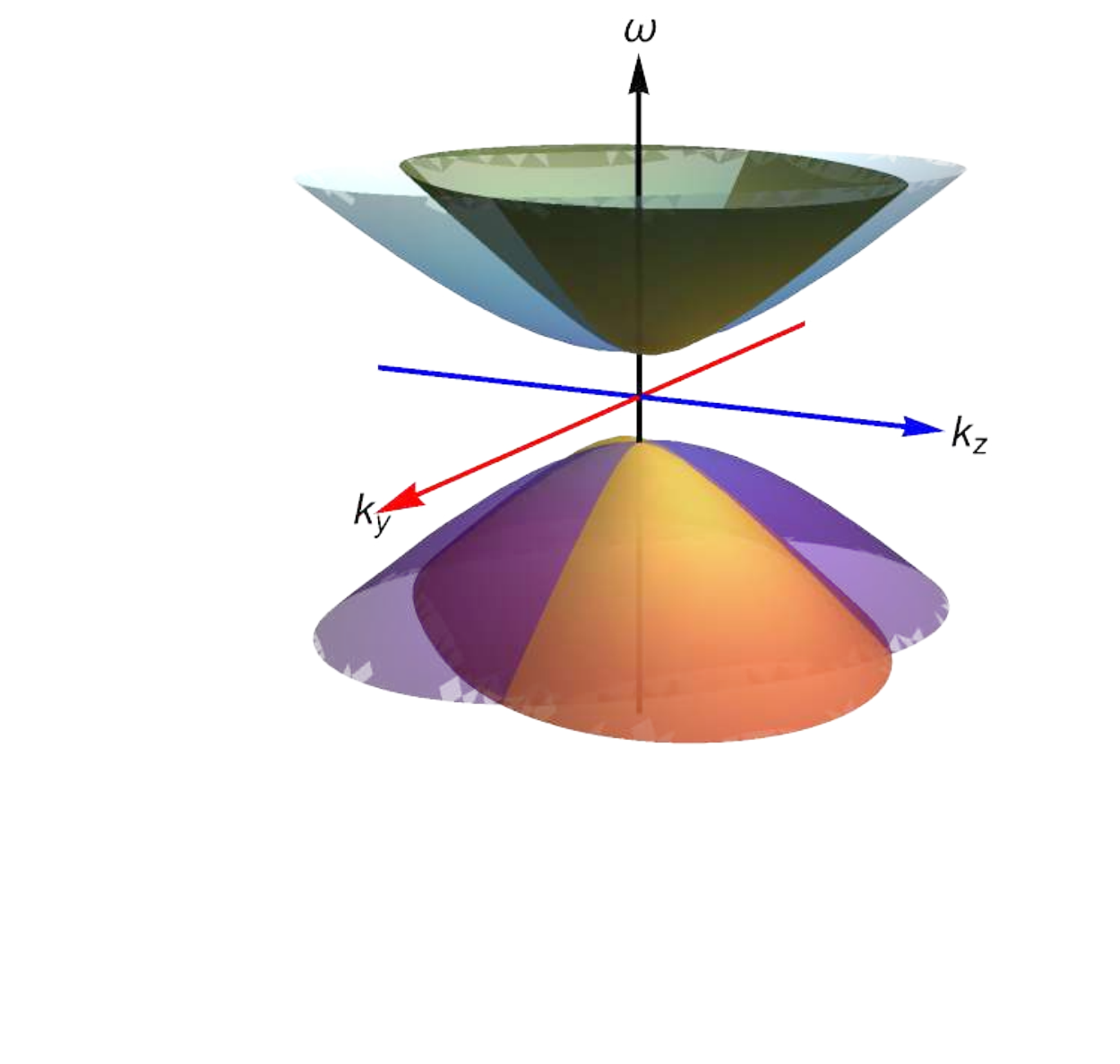}
\vspace{-1.7cm}
\caption{\small gap-gap}
\end{subfigure}
\begin{subfigure}[b]{0.32\textwidth}
\includegraphics[width=\textwidth]{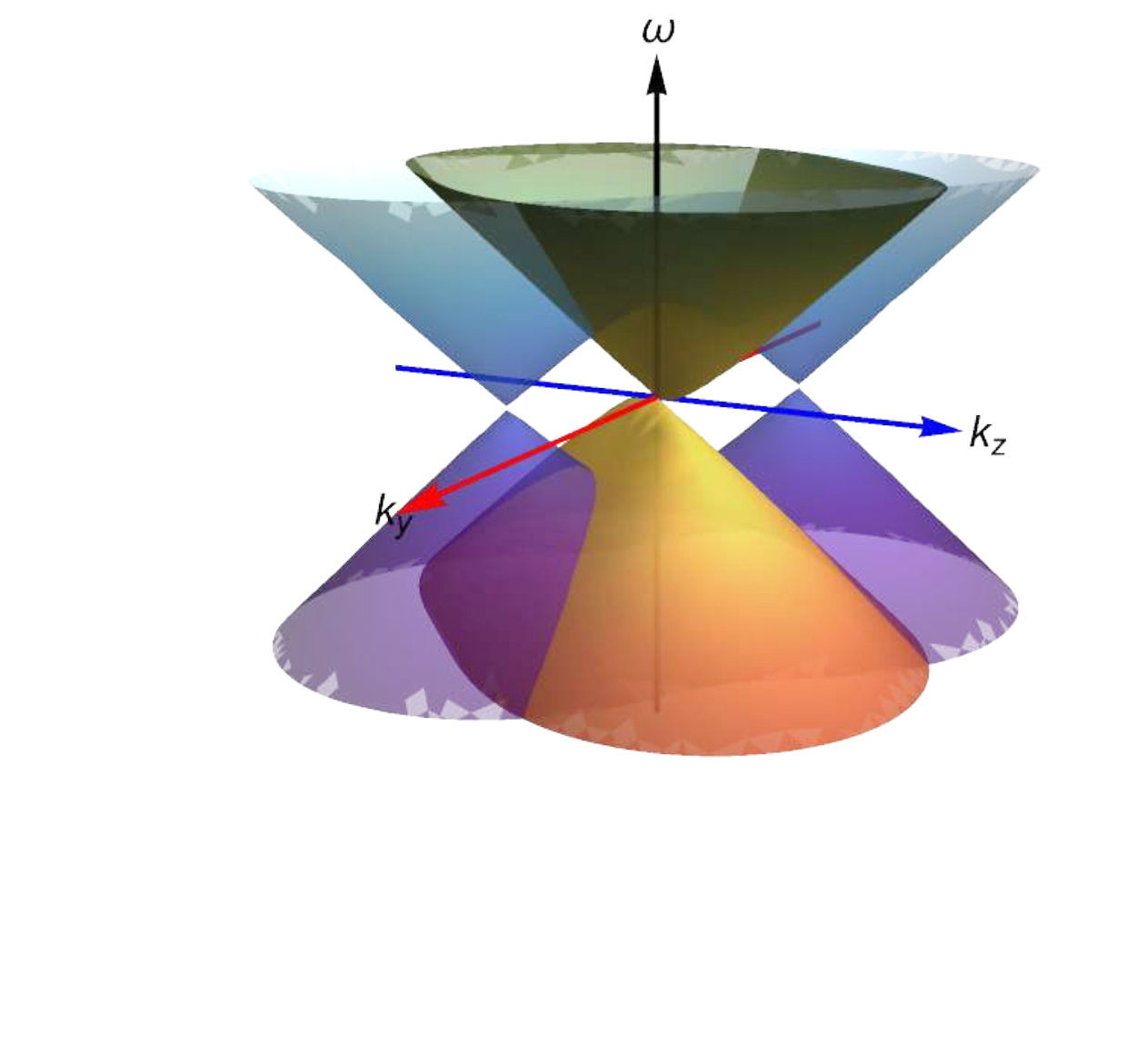}
\vspace{-1.7cm}
\caption{\small Weyl/$\boldmath{Z}_2$-critical}
 \end{subfigure}
\begin{subfigure}[b]{0.32\textwidth}
\includegraphics[width=\textwidth]{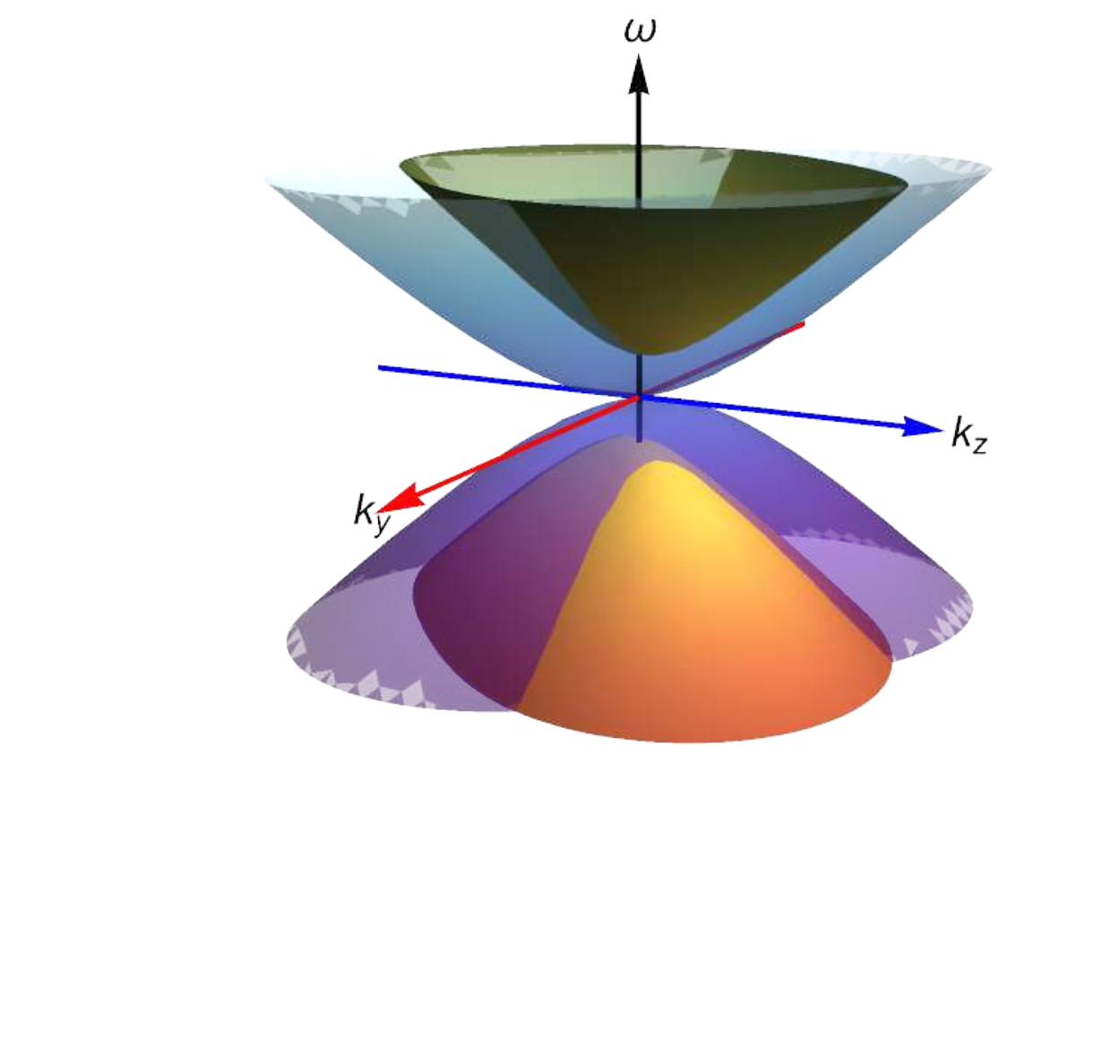}
\vspace{-1.7cm}
\caption{\small critical-gap or gap-critical}
\end{subfigure}
\begin{subfigure}[b]{0.32\textwidth}
\includegraphics[width=\textwidth]{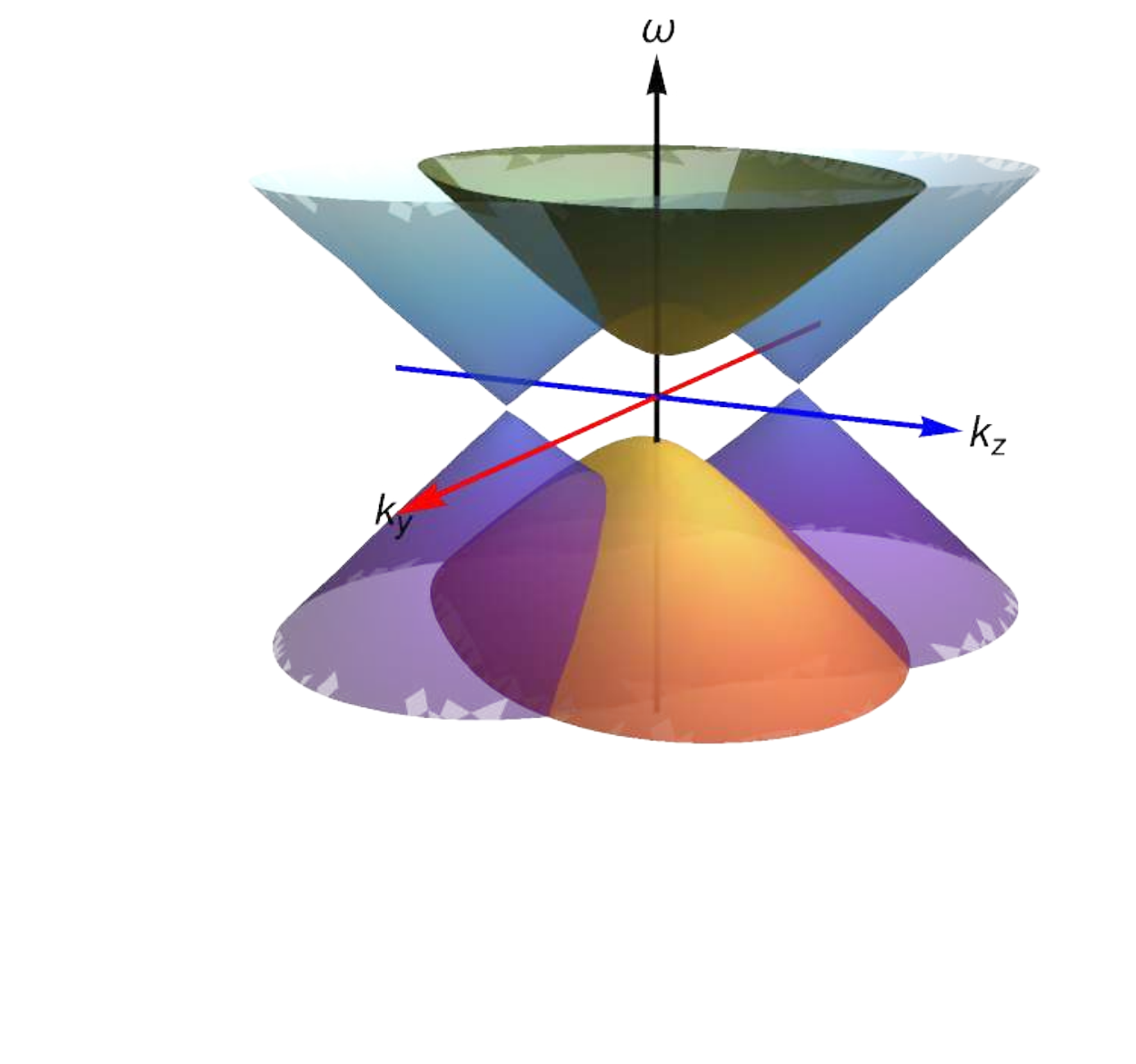}
\vspace{-1.7cm}
\caption{\small Weyl/$\boldmath{Z}_2$-gap
\label{fig:phasef}}
\end{subfigure}
  \caption{\small The energy spectrum of \eqref{eq:oLagrangian} as a function of $k_y$ and $k_z$ with $k_x=0$. From (a) to (f): the system has two pairs of Weyl/$\boldmath{Z}_2$ nodes (a), two critical Dirac nodes (b), fully gapped (c), one pair of Weyl/$\boldmath{Z}_2$ nodes and a critical Dirac node (d), a critical Dirac node and a gapped phase (e) and the case of one pair of Weyl/$\boldmath{Z}_2$ nodes with two gapped bands (f). Note that at nonzero $k_x$ the system is gapped, thus the crossing nodes are still nodes in the three dimensional momentum space.
  }
  \label{fig:ophase}
\end{figure}

The phase diagram can be plotted with the dimensionless parameters
\be
\hat{M}_1=M_1/b\,, ~~~\hat{M}_2=M_2/c
\ee and $c/b$. Now we have three dimensionless parameters, and the full phase diagram should be three dimensional in general. The phase diagrams with different $\hat{M}_1$ and $\hat{M}_2$, for $c/b=1$ and for generic $c/b$ (from $c/b=1/2$ to $c/b=2$) are shown in the left and right figures of Fig.~\ref{fig:phaseo2} respectively. In the left plot of Fig.~\ref{fig:phaseo2},
the red point is the critical point at which both pairs of Weyl/$\boldmath{Z}_2$ nodes become critical in Fig.~\ref{fig:ophase}(c). The blue dashed lines correspond to the phase transition lines where one pair of the Weyl/$\boldmath{Z}_2$ nodes annihilates into a critical Dirac node while the other pair of Weyl/$\boldmath{Z}_2$ nodes still exists in Fig.~\ref{fig:ophase}(d). The purple dot lines correspond to another type of phase transition lines where one pair of Weyl/$\boldmath{Z}_2$ nodes annihilates into a critical Dirac point while the other pair of Weyl/$\boldmath{Z}_2$ node becomes gapped in Fig.~\ref{fig:ophase}(e). The down-left portion of the phase diagram corresponds to the phase in Fig.~\ref{fig:ophase}(b). The up-left and down-right portions of the phase diagrams correspond to the phase in Fig.~\ref{fig:ophase}(f). The up-right portion of the phase diagram corresponds to the phase Fig.~\ref{fig:ophase}(a). In comparison to the location of the double critical point (red point in Fig.~\ref{fig:phaseo2}) for $c/b=1$, we also plot the location of the double critical point for $c/b=2$ by the blue point in Fig.~\ref{fig:phaseo2}.

\vspace{0cm}
\begin{figure}[h!]
  \centering
\includegraphics[width=0.4\textwidth]{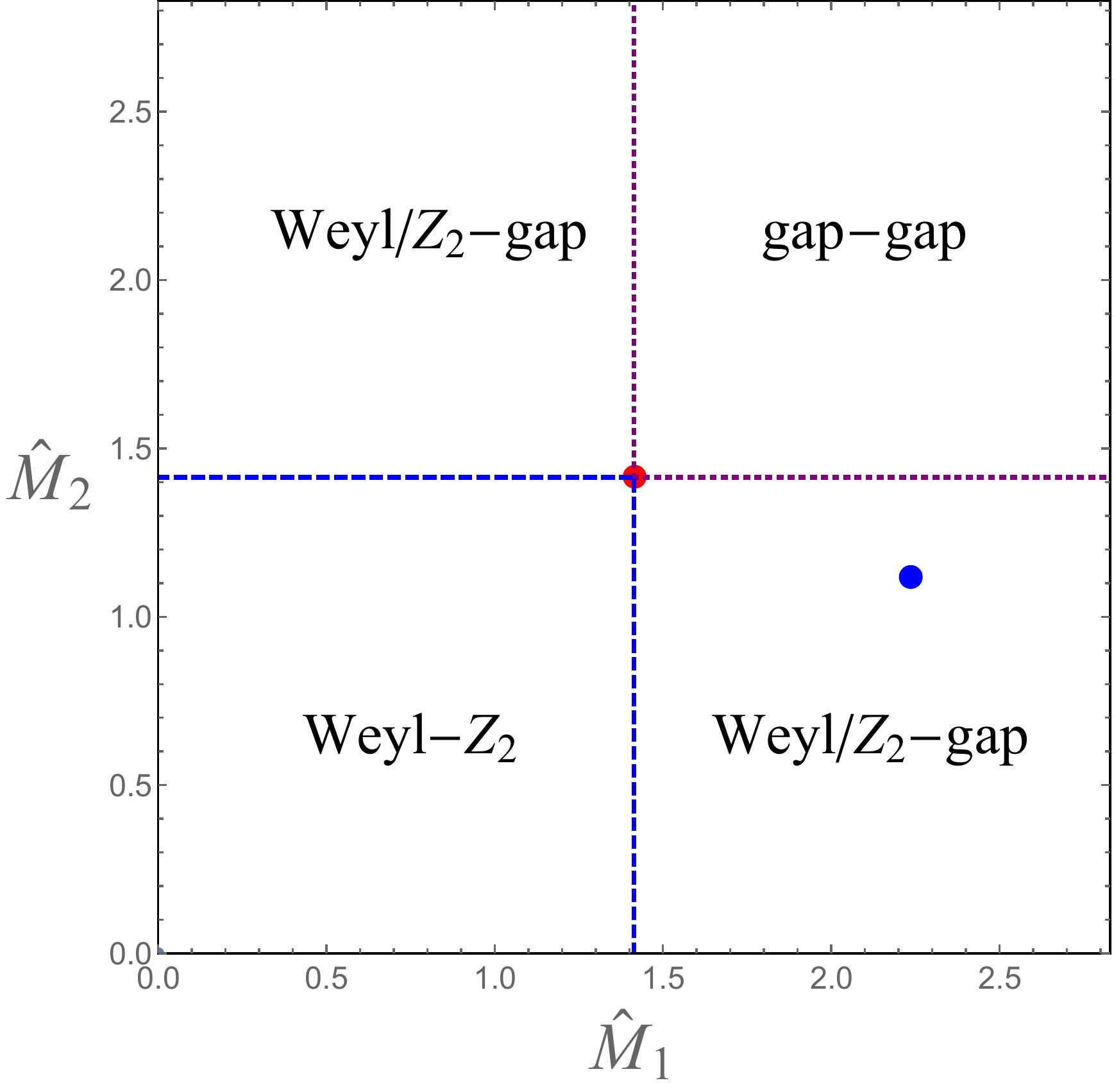}~~~~~
\includegraphics[width=0.4\textwidth]{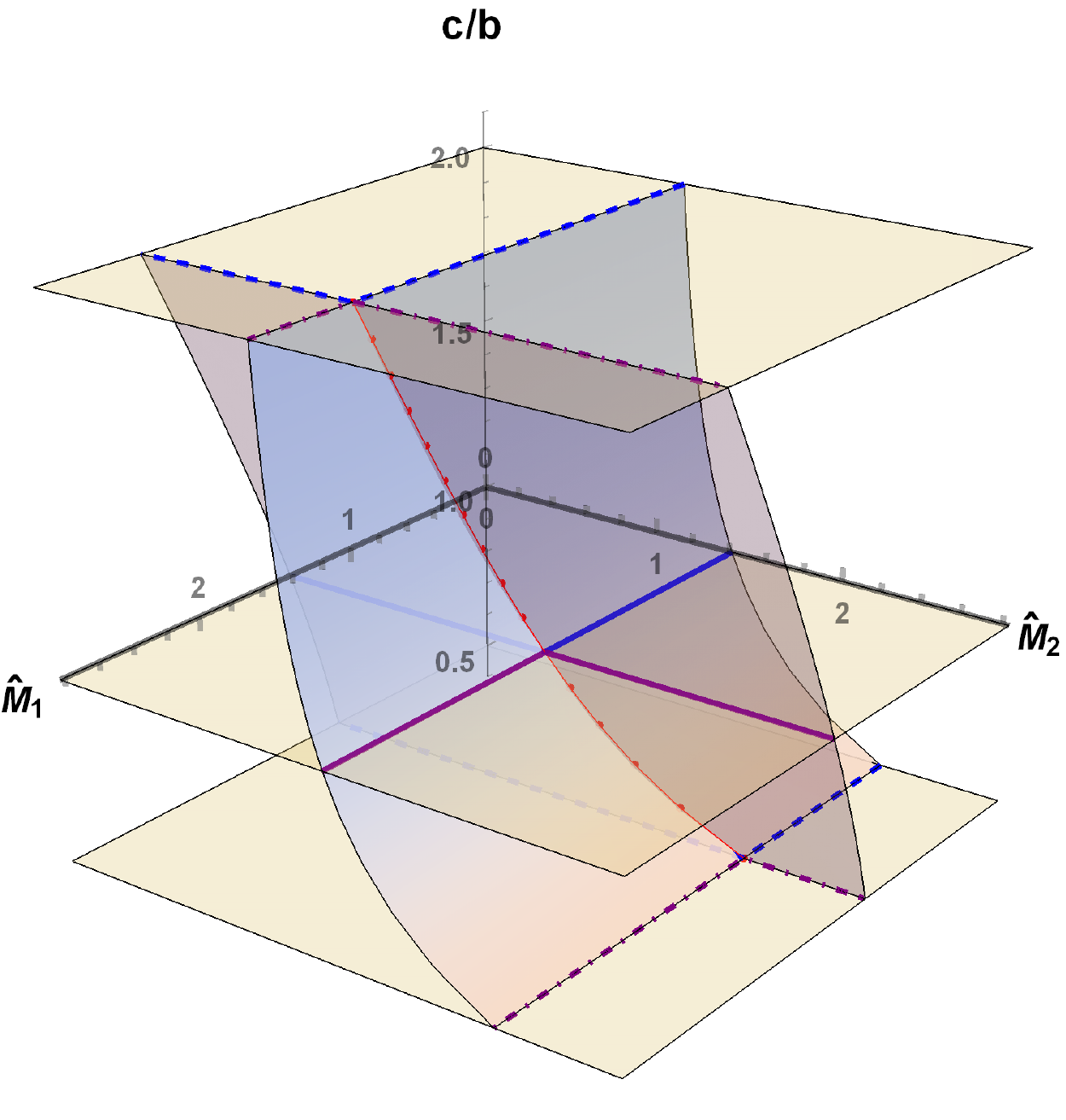}
\vspace{-0.3cm}
  \caption{\small The phase diagram of the system \eqref{eq:oLagrangian} with three dimensionless parameters $\hat{M}_1=M_1/b$, $\hat{M}_2=M_2/c$ and $c/b=1$ (left plot) and generic $c/b$ (right plot). The red point is the double critical point at which both two pairs of nodes become critical (Fig. \ref{fig:ophase}b). The vertical blue dashed lines correspond to the critical phase in which two of the Weyl/$\boldmath{Z}_2$ nodes with opposite topological charges annihilate into a critical Dirac node while the other two nodes still exist (Fig. \ref{fig:ophase}d, the Weyl/$\boldmath{Z}_2$-critical phases).  The vertical purple dotted lines correspond to the phase where one pair of Weyl/$\boldmath{Z}_2$ nodes annihilate into a critical Dirac point while the other pair becomes gapped (Fig. \ref{fig:ophase}e, the critical-gap phase).  The left-down portion of the phase corresponds to the phase in Fig. \ref{fig:ophase}a. The up left and down right portions of the phase diagram correspond to the phase in Fig. \ref{fig:ophase}f (i.e. the Weyl/$\boldmath{Z}_2$-gap phase). The up right portion of the phase diagram corresponds to the phase Fig. \ref{fig:ophase}c.
  The surfaces in the right figure are formed by moving the corresponding critical lines in the parameter space of $c/b$. The blue point is to show the location of the double critical point where $c/b=2$.
  }
 \label{fig:phaseo2}
\end{figure}

As could be seen from the spectrum,
there is an
 exchange  symmetry of $b$ and $c$ in the system  \eqref{eq:oLagrangian}. When we increase $c/b$, the location of the critical point $(\frac{M_1}{b})_c$ increases and $(\frac{M_2}{c})_c$ decreases. We could also see that the two critical values of $(\frac{M_1}{b})_c,(\frac{M_2}{c})_c$ exchange for $c/b=2$ and $c/b=1/2$.
The critical point in the left figure of Fig. \ref{fig:phaseo2} becomes the red critical line in the $c/b$ space in the right figure. The two critical lines become surfaces in the three-dimensional phase space, which are shown as separate surfaces in Fig. \ref{fig:phaseo2}.

\subsection{Another effective model for Weyl-$\boldmath{Z}_2$ semimetal with four nodes}
\label{subsec:model2}
The nodes in the model \eqref{eq:oLagrangian} carry topological charges $(+1,+1), (-1,-1)$ and $(+1,-1), (-1,+1)$. There is also another possibility that the nodes carry topological charges of $(+1,0), (-1,0)$ and $(0,+1), (0,-1)$ which will be constructed in this subsection.

Starting from the eight-component field \eqref{psi} and the Gamma matrix defined in \eqref{eq:88marix}, we consider the Lagrangian \begin{align}
\label{eq:Lagrangian}
\mathcal{L}=\Psi^{\dagger}\left[\Gamma^{0}\left(i\Gamma^{\mu}\partial_{\mu}-e\Gamma^{\mu}A_{\mu}-\Gamma^{\mu} \Gamma^{5}b_{\mu} {\mI}_{1}+ {M}_{1} {\mI}_{1}+ {M}_{2} {\mI}_{2} \right)+\hat{\Gamma}^{0}\left(e\hat{\Gamma}^{\mu}\hat{A}_{\mu}- \hat{\Gamma}^{\mu}\hat{\Gamma}^{5}c_{\mu} {\mI}_{2}\right)\right]\Psi\,,
\end{align}}
where the `Gamma' matrices are defined in (\ref{eq:88marix}).
Four gauge fields and two mass terms in (\ref{eq:oLagrangian}) have the same physical meaning with those in \eqref{eq:oLagrangian}. Different from \eqref{eq:oLagrangian}, the axial gauge fields here
only couple to one half of the eight-components of the field, which leads to different structure in the spectrum and different topological charges for each node, while qualitatively similar phase diagrams, as we shall show in the following.

Similar to the case in \eqref{eq:oLagrangian}, we choose $b_{\mu}=b\delta^{z}_{\mu}$ and $c_{\mu}=c\delta^{y}_{\mu}$.
Now the eight eigenvalues of the Hamiltonian of  \eqref{eq:Lagrangian} are
\begin{eqnarray}
\label{spectrum}
\begin{split}
E_{1}=\pm\sqrt{\left(b_{z}\pm\sqrt{k_{z}^{2}+M_1^{2}}\right)^{2}+k_{y}^{2}}\,,\quad~~~~
 E_{2}=\pm\sqrt{\left(c_{y}\pm\sqrt{k_{y}^{2}+M_2^{2}}\right)^{2}+k_{z}^{2}}\,.
\end{split}
\end{eqnarray}
Compared to \eqref{ospectrum}, now the spectrum \eqref{spectrum} is much simpler due to the decoupled effect of $b$ and $c$. More precisely, here $E_1$ only depends on $b$ and $M_1$ while $E_2$ only depends on $c$ and $M_2$. $E_1$ can be viewed as the energy spectrum of \eqref{eq1}, with $M$ replaced by $M_1$, and $E_2$ can be viewed as the energy spectrum of \eqref{eq1} with $\vec{b}$ and $M$ replaced by $\vec{c}$
and $M_2$.

Compared to the  Weyl-$Z_2$ semimetal described by \eqref{eq:oLagrangian}, here we have two independent pairs of nodes along $k_y$ and $k_z$ directions. Each pair of them carry topological charges $(1,0), (-1, 0)$ or $(0,1), (0, -1)$. One could identify the pair of nodes in one direction as carrying the charge of chiral $U(1)$, while the nodes in the other direction as carrying the charge of the analog spin $Z_2$ charge. For simplicity, one could also name this kind of Weyl semimetal with four nodes as a Weyl-$Z_2$  semimetal. In the following, we shall show that our holographic realization of Weyl-$Z_2$ semimetals with four nodes share both features of these two field theory models.

In the following we will analyze the phase behavior of this field theory model following the same strategy as in subsection \ref{owpb}.

\subsubsection{Phase diagram from the field theory model}
\label{wpb}
In this subsection, we analyze the phase behavior of the energy spectrum (\ref{spectrum}) obtained in the last subsection.
Similar to the case in \eqref{eq:oLagrangian}, the system depends on two dimensionless parameters $M_1/b$ and $M_2/c$. Compared to \eqref{eq:oLagrangian}, the crucial difference is that since two axial gauge fields couple to a four-component fermion separately, the value $c/b$ does not play any role in the phase diagram as we will show in the following.

\begin{figure}[h!]
  \centering
  \begin{subfigure}[b]{0.33\textwidth}
\includegraphics[width=\textwidth]{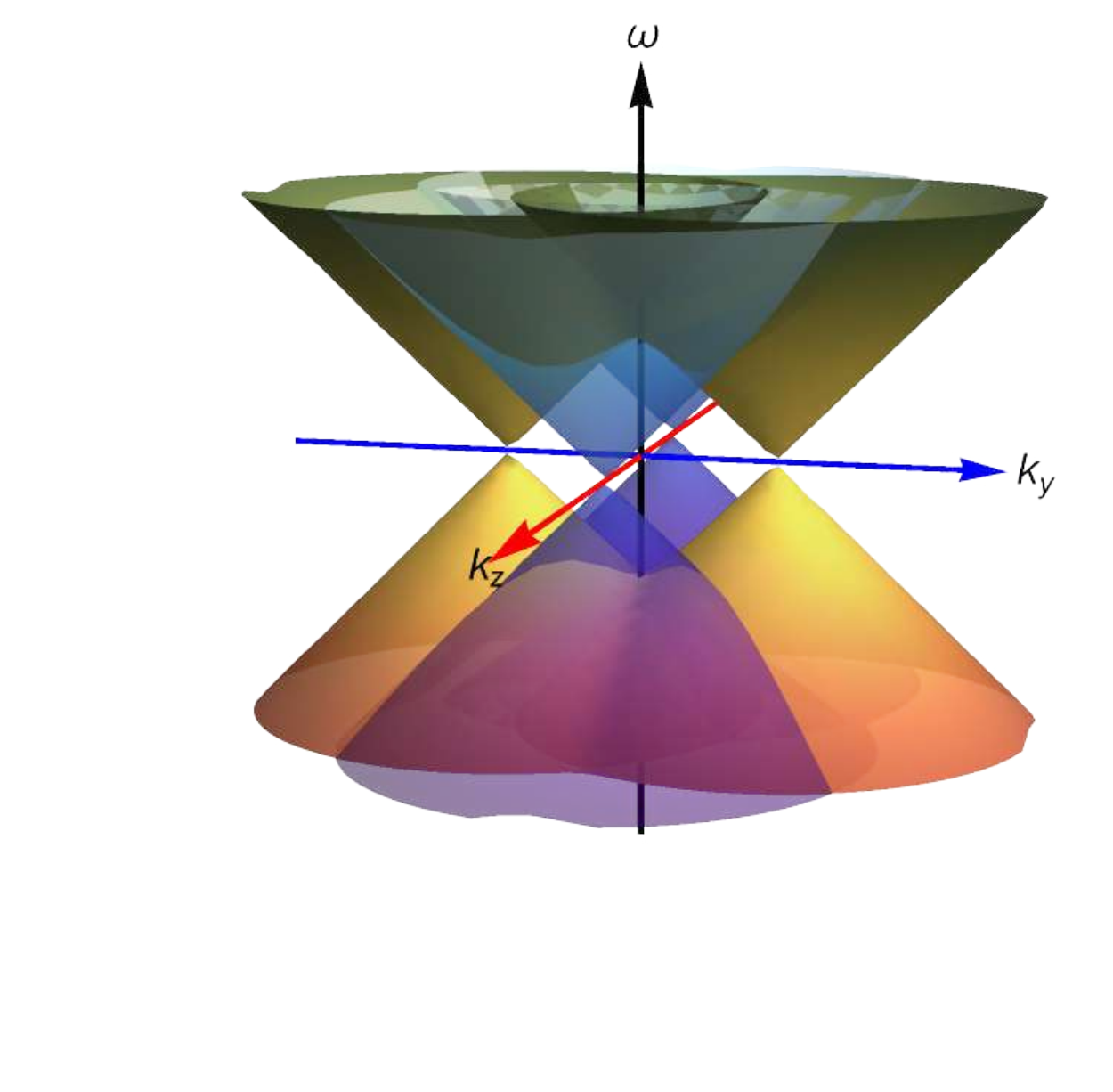}
\vspace{-1.7cm}
\caption{\small Weyl-$\boldmath{Z}_2$}
\end{subfigure}
\begin{subfigure}[b]{0.32\textwidth}
\includegraphics[width=\textwidth]{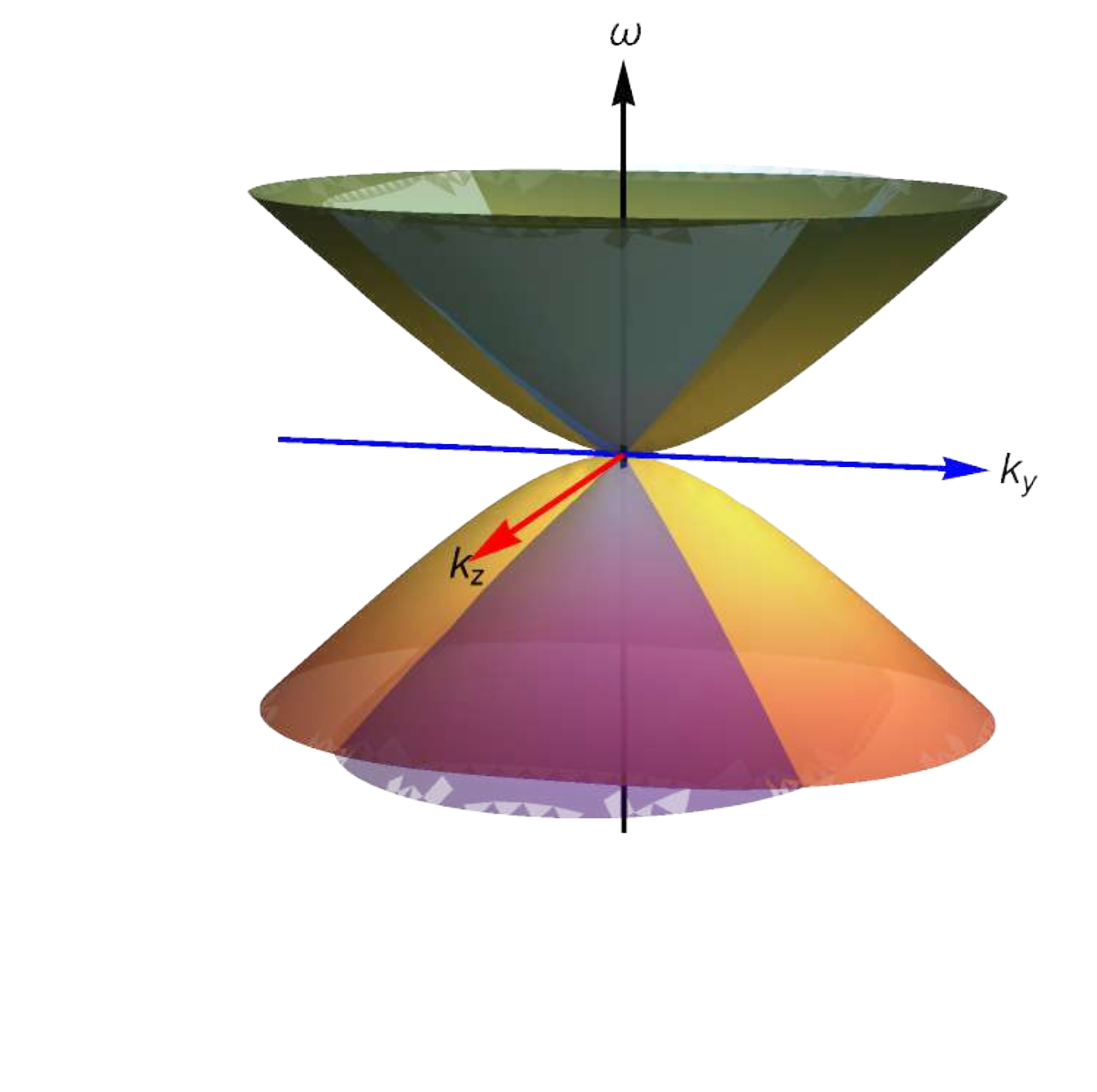}
\vspace{-1.7cm}
\caption{\small double critical}
\end{subfigure}
\begin{subfigure}[b]{0.33\textwidth}
\includegraphics[width=\textwidth]{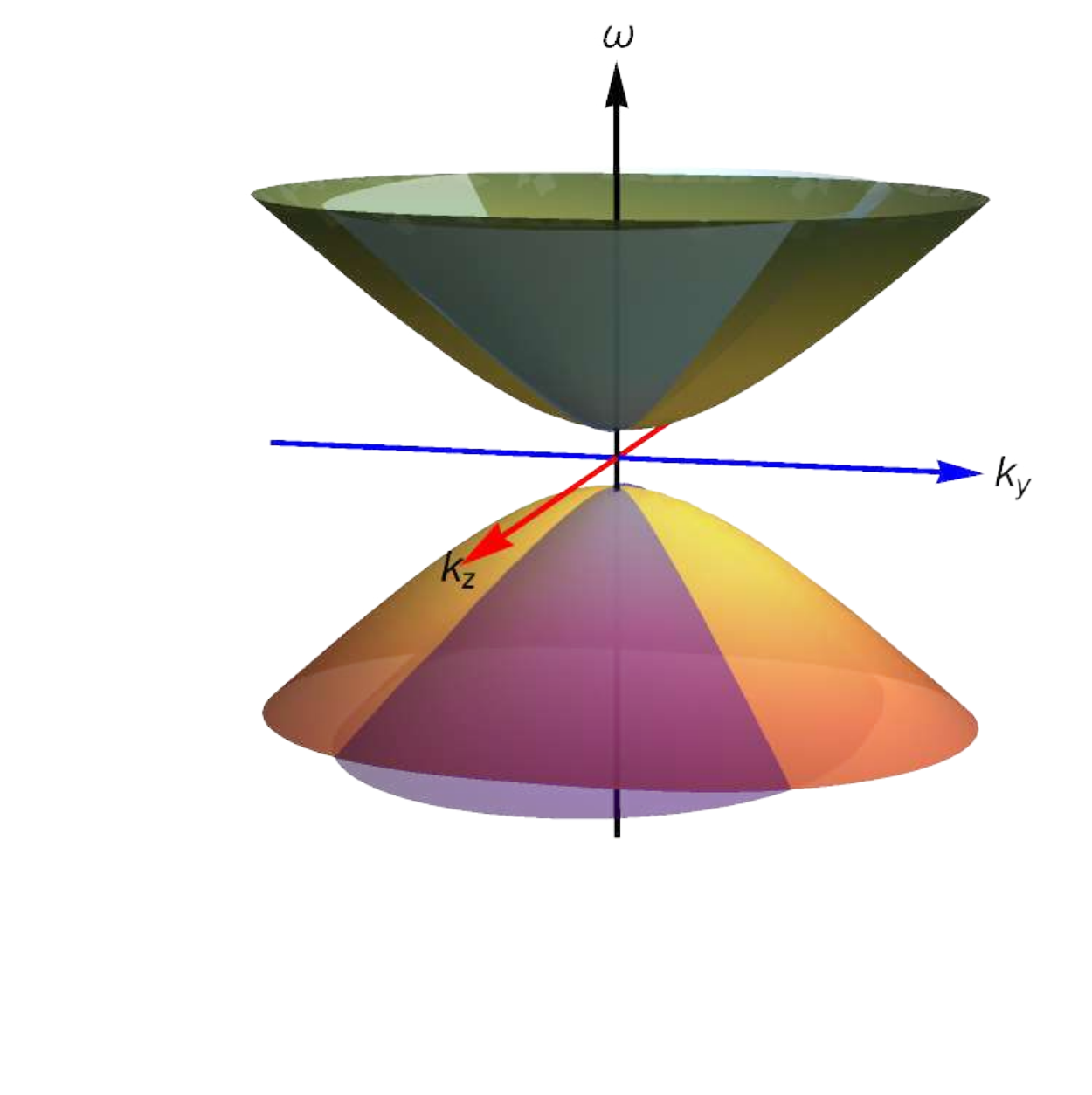}
\vspace{-1.7cm}
\caption{\small gap-gap}
\end{subfigure}
\begin{subfigure}[b]{0.32\textwidth}
\includegraphics[width=\textwidth]{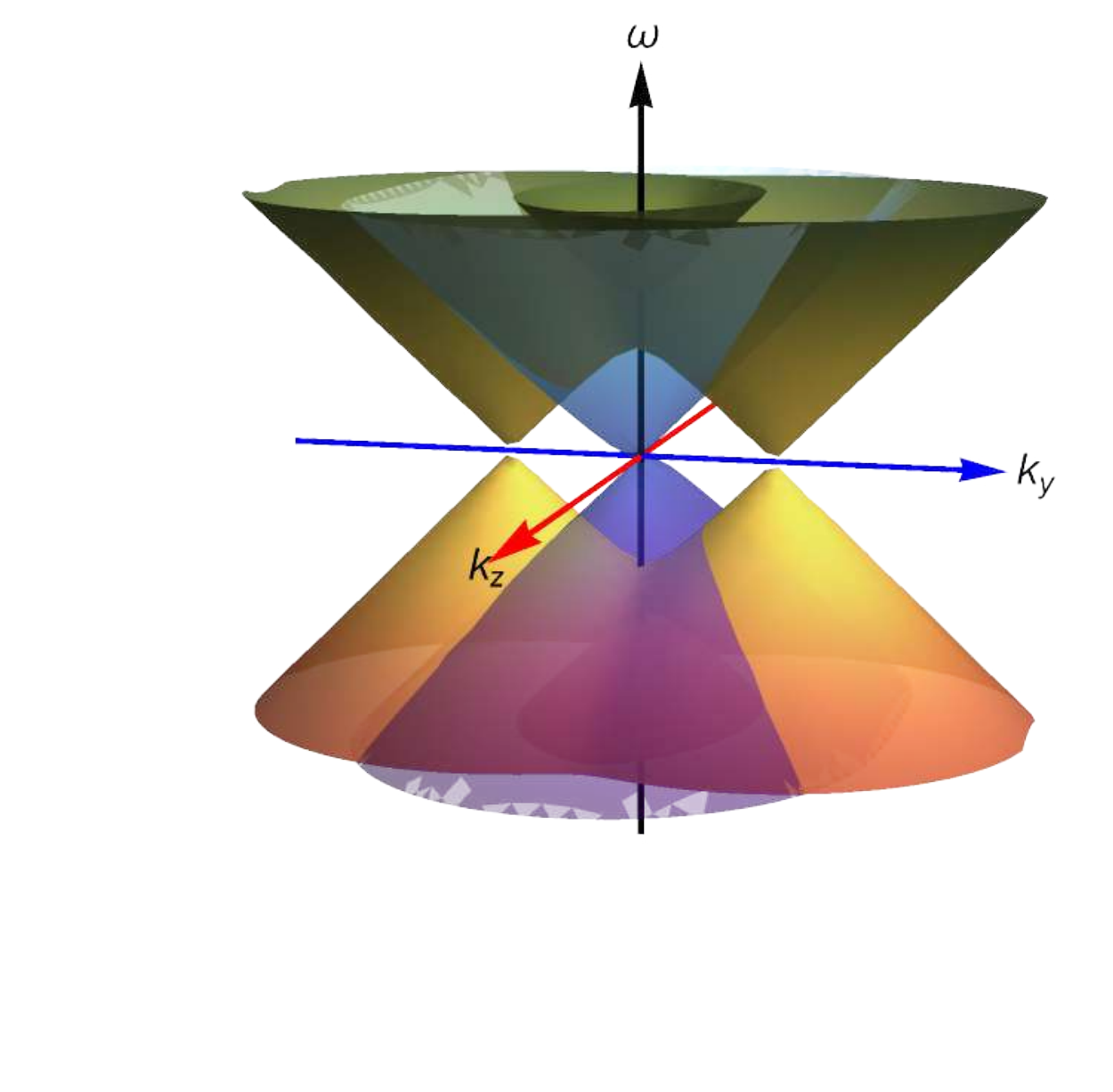}
\vspace{-1.7cm}
\caption{\small Weyl/$\boldmath{Z}_2$-critical}
 \end{subfigure}
\begin{subfigure}[b]{0.32\textwidth}
\includegraphics[width=\textwidth]{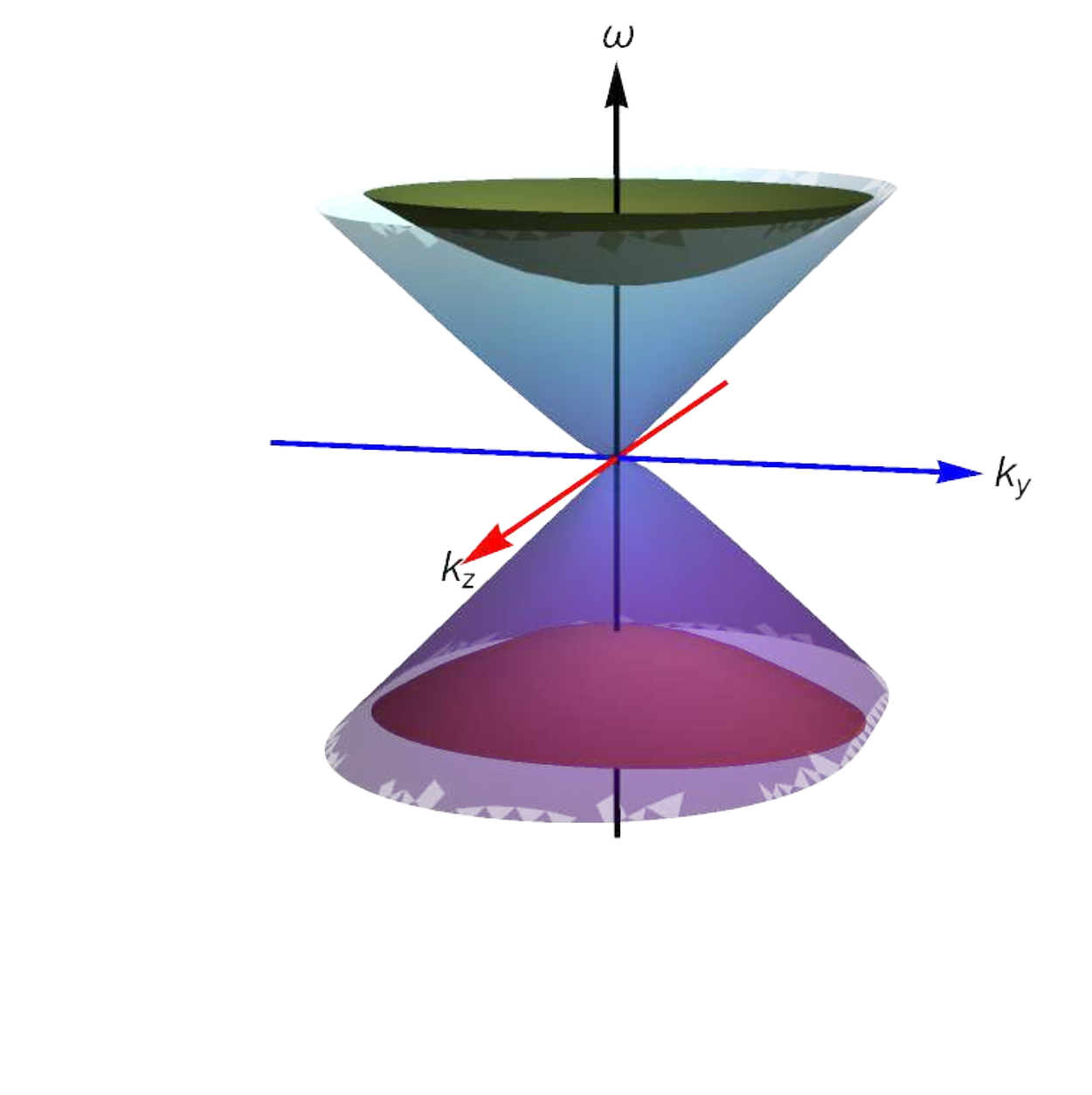}
\vspace{-1.7cm}
\caption{\small critical-gap or gap-critical}
\end{subfigure}
\begin{subfigure}[b]{0.32\textwidth}
\includegraphics[width=\textwidth]{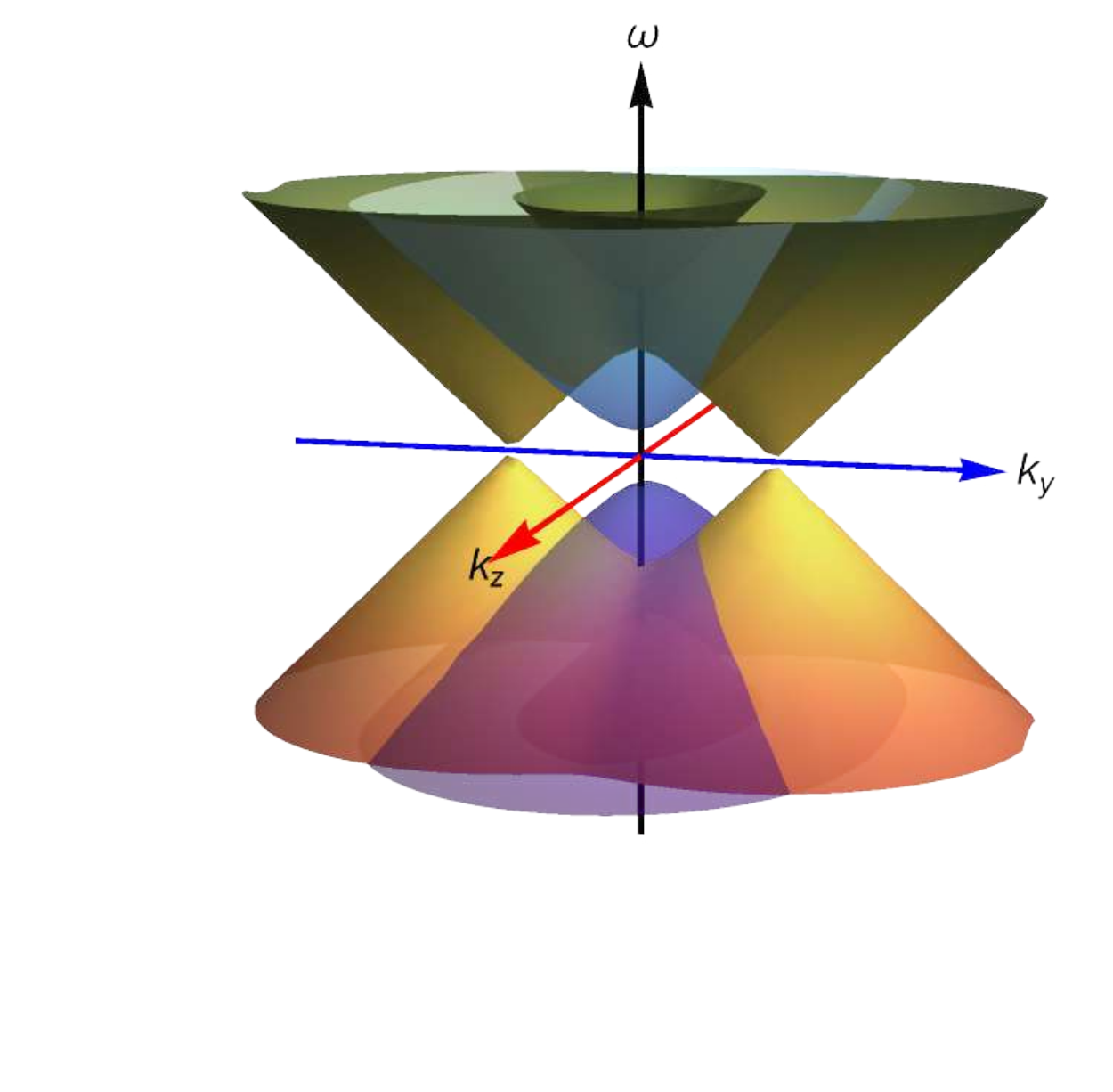}
\vspace{-1.7cm}
\caption{\small Weyl/$\boldmath{Z}_2$-gap
\label{fig:phasef}}
\end{subfigure}
  \caption{\small The energy spectrum \ref{ospectrum} as a function of $k_y$ and $k_z$ with $k_x=0$.  {From (a) to (f): the system has two pairs of Weyl/$\boldmath{Z}_2$ nodes (a), two critical Dirac nodes (b), fully gapped (c), one pair of Weyl/$\boldmath{Z}_2$ nodes and a critical Dirac node (d), a critical Dirac node and a gapped phase (e) and the case of one pair of Weyl/$\boldmath{Z}_2$ nodes with two gapped bands (f)}.
  }
  \label{fig:phase}
\end{figure}

We also fix $k_x$ to be zero and plot the spectrum as a function of $k_y$ and $k_z$. Now the topological charges for these two models are different, nevertheless we still describe the corresponding phases with the same name, which is still reasonable in this model.
The nine phases are summarized in Fig. \ref{fig:phase}, which could be summarized into six types of spectrum: including the Weyl-$\boldmath{Z}_2$,  Weyl/$\boldmath{Z}_2$-critical, Weyl/$\boldmath{Z}_2$-gap, critical-critical, critical-gap, and gap-gap phases.

\vspace{0cm}

In the following we follow the same order to explain in detail these phases.
\vspace{.25cm}\\
\noindent {\bf  $\bullet$ \em The Weyl-$\boldmath{Z}_2$ phase}

When both $M_1<b$ and $M_2<c$, we obtain the spectrum as shown in Fig.~\ref{fig:phase}(a).
From \eqref{spectrum}, the four nodes at the plane of $k_x=0$ are
\begin{eqnarray}
\begin{split}
\label{eq:pt}
(k_z, k_y)=\left(\pm\sqrt{b^{2}-M_1^{2}}\,,~ 0\right),~~~~~~~
(k_z, k_y)=\left(0\,,~\pm\sqrt{c^{2}-M_{2}^{2}}\right),
\end{split}
\end{eqnarray}
This is a topologically nontrivial phase with four nontrivial nodes with topological charges $(1,0), (-1, 0)$ and $(0,1), (0,-1)$.
This is a different model with four Weyl nodes from the one in the last subsection. Here not all the nodes carry two topological charges. Each pair of them only carries one type of nontrivial topological charge.

\vspace{.25cm}
\noindent {\bf $\bullet$ \em The two Weyl/$\boldmath{Z}_2$-critical phases}

When we fix $b,c$ and increase $M_1$ (or $M_2$) from Fig.~\ref{fig:phase}(a), two nodes would annihilate to form a critical Dirac node and the other pair of nodes stay the same. The difference is that now the location of the nodes will depend on the values of $M_1, M_2$. In this case, we now have a pair of Weyl/$Z_2$ nodes and a critical Dirac point, as shown in Fig. ~\ref{fig:phase}(d).

\vspace{.25cm}
\noindent {\bf $\bullet$ \em The two Weyl/$\boldmath{Z}_2$-gap phases}

When we further increase $M_1$ (or $M_2$) from the two Weyl/$Z_2$-critical phases transition lines in Fig. ~\ref{fig:phase}(d), the critical Dirac node becomes a trivial gap. This corresponds to two Weyl/$Z_2$- gap phases, as shown in Fig. ~\ref{fig:phase}(f).
\vspace{.25cm}\\
\noindent {\bf $\bullet$ \em The double critical point}

When we start from the two Weyl/$Z_2$-critical phases transition lines in Fig. \ref{fig:phase}(d) and increase the other parameter $M_2$ (or $M_1$), the two Weyl/$Z_2$ nodes will also reach a critical point at which two nodes merge into one Dirac node. The system at this special set of parameters $M_1/b$ and $M_2/c$ corresponds to a double critical point on the phase diagram, which is shown in Fig.\ref{fig:phase}(b).
\vspace{.25cm}\\
\noindent {\bf $\bullet$ \em The two gap-critical phases}

When we start from the double critical point in Fig.~\ref{fig:phase}(b) and increase one of the mass parameters, the fourfold-degenerate critical point would split into a pair of gapped bands and one twofold-degenerate critical point. These two phases correspond to two phase transition lines between the gap-gap phase and the Weyl/$Z_2$-gap phase as shown in Fig.~\ref{fig:phase}(e).
\vspace{.25cm}\\
\noindent {\bf $\bullet$ \em The gap-gap phase}

When we start from any of the two gap-critical phase transition lines above, and increase the other mass parameter, the system would become fully gapped and is in a gap-gap phase with all bands gapped. This corresponds to Fig.~\ref{fig:phase}(c).

\begin{figure}[h!]
  \centering
\includegraphics[width=0.4\textwidth]{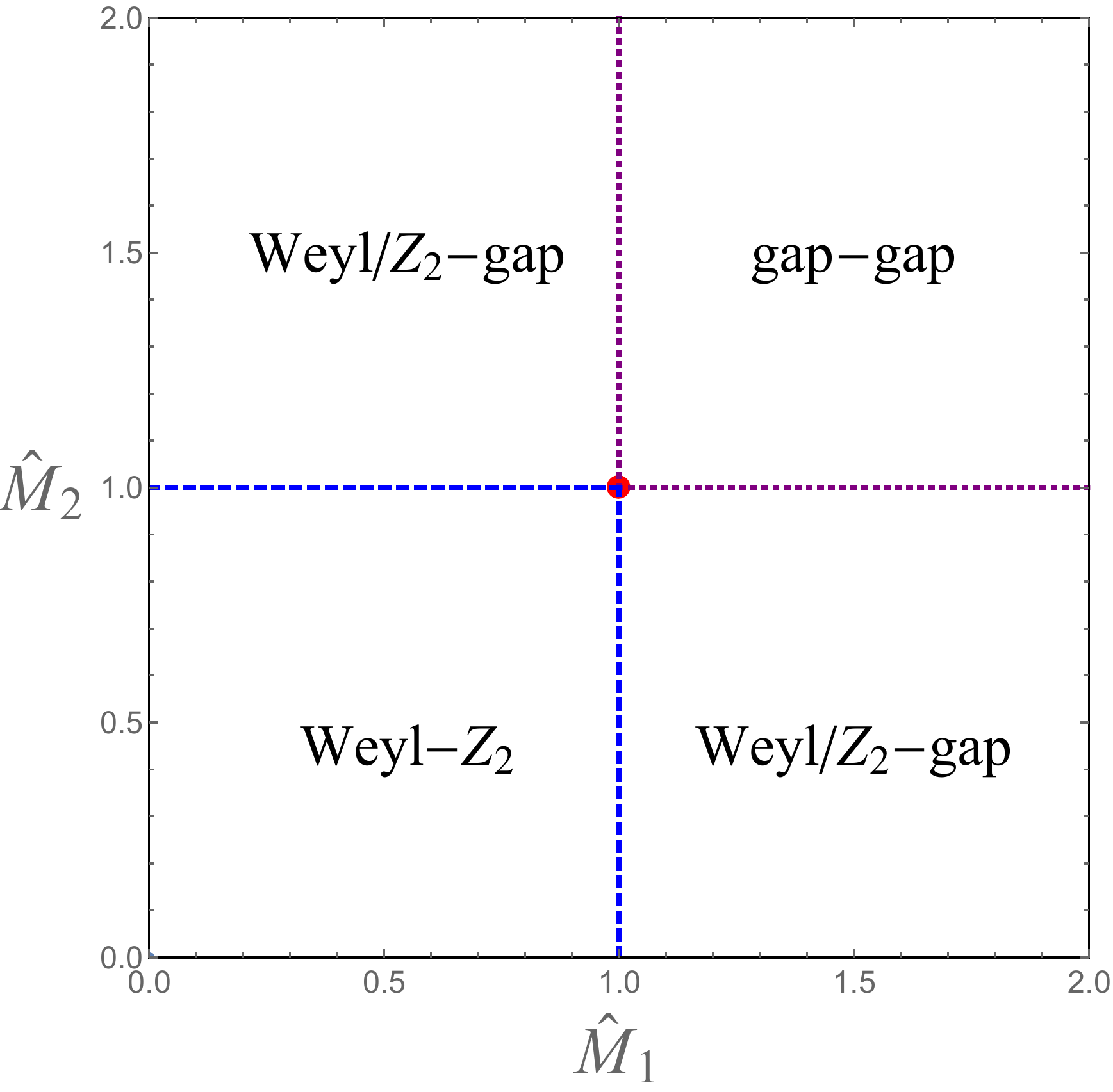}~~~~~
\includegraphics[width=0.4\textwidth]{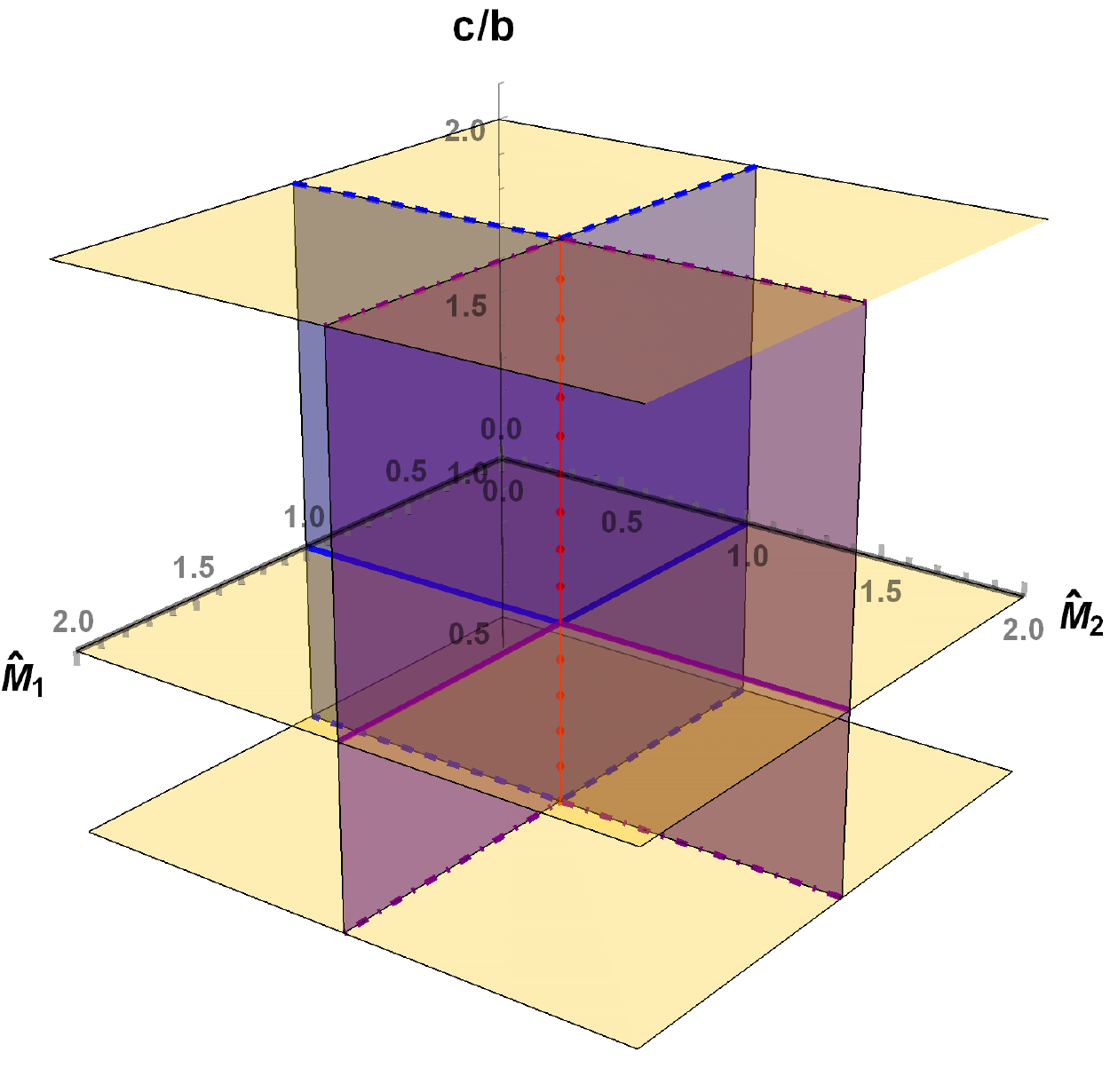}
\vspace{-0.3cm}
  \caption{\small {The phase diagram of the system \eqref{eq:Lagrangian} with three dimensionless parameters $\hat{M}_1=M_1/b$, $\hat{M}_2=M_2/c$ and $c/b=1$ ({\em left}) and generic $c/b$ ({\em right}). The red point is the double critical point at which both two pairs of nodes become critical (Fig.~\ref{fig:phase}(b)).
Obviously the phase diagram does not depend on the value of $c/b$. }}
 \label{fig:phase2}
\end{figure}

\vspace{.25cm}
Similar to the phase diagram shown in Fig. \ref{fig:phaseo2}, we could plot the full phase diagram of \eqref{eq:Lagrangian}.
Different from the one shown in Fig.~\ref{fig:phaseo2}, now $\hat{M}_{1c}$ and $\hat{M}_{2c}$ does not depend on $c/b$. This is due to the fact that the axial gauge fields are completely decoupled in \eqref{eq:Lagrangian}.
There is also an
 exchange  symmetry of $b$ and $c$ in the system.
The critical point in the left figure of Fig.~\ref{fig:phase2} becomes the red critical line in the $c/b$ space in the right figure. It is easy to see that the phase diagram does not depend on $c/b$, i.e. the critical line is exactly along the $z$-direction, which is different from the behavior in Fig.~\ref{fig:phaseo2}. The two critical lines become surfaces in the three-dimensional phase space, which are shown as separate surfaces in Fig.~\ref{fig:phase2}.

\vspace{0cm}

In summary, both of the two effective models \eqref{eq:oLagrangian} and \eqref{eq:Lagrangian} could describe the Weyl-$\boldmath{Z}_2$ semimetal with four nodes, although the underlying mechanisms are different. Four Weyl/$\boldmath{Z}_2$ nodes are obtained by 
splitting the eight-fold degeneracy of one 
node with different process. For the model \eqref{eq:oLagrangian}, each node carries both the chiral and $Z_2$ topological charges, while for the model
\eqref{eq:Lagrangian}, each node carries either the chiral or the $Z_2$ topological charge. In section \ref{sec:holo} we shall construct a holographic model for Weyl semimetals with four nodes which shares both features of these two different weakly coupled field theory models. Before that we will discuss the anomalous transports and Ward identities of the field theory models.

\subsection{Anomalous transports and Ward identities}

When tuning $M_1/b$ and $M_2/c$, we have phase transitions in the system \eqref{eq:oLagrangian} and \eqref{eq:Lagrangian}, as has already been discussed above. Note that there is no symmetry breaking during this process, and this phase transition is a topological quantum phase transition. Analogous to the Weyl semimetal case, the order parameters of this topological phase transition can be taken to be the anomalous Hall conductivity and $\boldmath{Z}_2$ anomalous Hall conductivity. To be specific, these two conductivities of a quantum many body systems can be computed via the Kubo formula
\begin{eqnarray}
\label{kubo}
\sigma_{ij}=\underset{\omega\to0}{\lim}\frac{1}{i\omega}\langle \mathcal{J}_{i}\mathcal{J}_{j}\rangle_{R}\,,~~~~
\hat{\sigma}_{ij}=\underset{\omega\to0}{\lim}\frac{1}{i\omega}\langle\mathcal{\hat{J}}_{i} \mathcal{J}_{j}\rangle_{R}\,,
\end{eqnarray}
where $\sigma_{ij}$ is the electric conductivity, $\mathcal{J}_{i}$ is the electric current, while $\hat{\sigma}_{ij}$ is the spin conductivity which characterizes the spin current $\mathcal{\hat{J}}_{i}$ generated under the electric field. To obtain these currents, we need to perform the functional derivative of the low energy effective action $S_\text{eff}$ of in the weak coupling regime with respect to the gauge fields.

We shall focus on the system \eqref{eq:Lagrangian} whose low energy effective theory could be easily constructed while the discussion for the other model \eqref{eq:oLagrangian} has been done in the four-component description in \cite{kimb} and physically the results should be equivalent to the eight-component description, which we will not discuss here and refer the readers to \cite{kimb}.
Starting from the low energy effective action $S_\text{eff}$ in (\ref{eq:massless}),
the currents are
\begin{eqnarray}\label{current111}
\begin{split}
\mathcal{J}^{\mu}=
\frac{e^{2}}{\pi^{2}}b_{\nu}\epsilon^{\mu\nu\alpha\beta}\partial_{\alpha}A_{\beta}+\frac{e^{2}}{\pi^{2}}c_{\nu}\epsilon^{\mu\nu\alpha\beta}\partial_{\alpha}\hat{A}_{\beta}\,,~~~~~
\mathcal{\hat{J}}^{\mu}=
\frac{e^{2}}{\pi^{2}}b_{\nu}\epsilon^{\mu\nu\alpha\beta}\partial_{\alpha}\hat{A}_{\beta}+\frac{e^{2}}{\pi^{2}}c_{\nu}\epsilon^{\mu\nu\alpha\beta}\partial_{\alpha}A_{\beta}\,.
\end{split}
\end{eqnarray}
More details on the derivation can be found in the appendix \ref{app:a}.

To calculate the anomalous transport, we shall focus on the low energy effective theory.
As in \cite{Landsteiner:2015pdh}, for the un-gapped phases, the Lagrangian for the low energy field theory of
\eqref{eq:Lagrangian} is
\begin{align}
\label{eq:Lagrangiana}
\mathcal{L}_{(a)}=\Psi^{\dagger}\left[\Gamma^{0}\left(i\Gamma^{\mu}\partial_{\mu}-e\Gamma^{\mu}A_{\mu}-\Gamma^{\mu} \Gamma^{5}  {b}_{\text{eff}} \hat{e}_{\mu} {\mI}_1\right)
+\hat{\Gamma}^{0}\left(e\hat{\Gamma}^{\mu}\hat{A}_{\mu}-\hat{\Gamma}^{\mu}\hat{\Gamma}^{5} {c}_{\text{eff}} \hat{e}_{\mu} {\mI}_2 \right)\right]\Psi\,,
\end{align}
where
$\vec{b}_\text{eff}=b_\text{eff}\hat{e}_{z}=\sqrt{b^{2}-M_{1}^{2}}\hat{e}_{z}$ and $\vec{c}_\text{eff}=c_\text{eff}\hat{e}_{y}=\sqrt{c^{2}-M_{2}^{2}}\hat{e}_{y}$.
The parameter regimes for this Lagrangian are $M_1\leq b$ and $M_2\leq c$.
The four nodes of the Weyl-$\boldmath{Z}_2$ phase in Fig.~\ref{fig:phase}(a) are simply given by $(k_z, k_y)=(\pm b_\text{eff},0)$ and $(k_z, k_y)=(0,\pm c_\text{eff})$.
The Weyl/$\boldmath{Z}_2$-critical phase in Fig.~\ref{fig:phase}(d) and double critical phase in Fig.~\ref{fig:phase}(b) can also be realized by varying the parameters $M_1$,$M_2$, $b$ and $c$.
We can calculate the anomalous transports as follows.
The currents in \eqref{current111} now become
\begin{eqnarray}
\vec{\mathcal{J}}=\frac{e^{2}}{\pi^{2}}\vec{b}_\text{eff}\times \vec{E}+\frac{e^{2}}{\pi^{2}}\vec{c}_\text{eff}\times\vec{\hat{E}}\,,~~~~~
\vec{\hat{\mathcal{J}}}=\frac{e^{2}}{\pi^{2}}\vec{b}_\text{eff}\times\vec{\hat{E}}+\frac{e^{2}}{\pi^{2}}\vec{c}_\text{eff}\times \vec{E}\,,
\end{eqnarray}
where $\vec{E}$ is the electric field while $\vec{\hat{E}}=-\text{\ensuremath{\nabla}}\hat{A}_{0}$ is the analogous generalization for the fictitious
spin gauge fields\cite{Landsteiner:2015lsa,kimb}. Thus, we can calculate the anomalous Hall conductivity $\sigma_{\text{AHE}}$ (from the expression of $\sigma_{ij}$) and the $\boldmath{Z}_2$ anomalous Hall conductivity $\sigma_{\boldmath{Z_{2}}\text{AHE}}$ (from the expression of $\hat{\sigma}_{ij}$) accordingly. In the Weyl-$\boldmath{Z}_2$ semimetal phase, we have
\begin{eqnarray}
\sigma_{\text{AHE}}\propto b_\text{eff}\,,~~~~~
\sigma_{\boldmath{Z_{2}}\text{AHE}}\propto c_\text{eff}\,.
\end{eqnarray}
From these two expressions, we note that if we choose $c=M_2=0$ (or $b=M_1=0$), there will be only one pair of Weyl/$\boldmath{Z_{2}}$ nodes left and we could see that  $\sigma_{\text{AHE}}\propto\sqrt{b^2-M_{1}^2}$ (or $\sigma_{\boldmath{Z_{2}}\text{AHE}}\propto\sqrt{c^2-M_{2}^2}$), which is consistent with \cite{Landsteiner:2015lsa}. Setting both these two anomalous transport coefficients to be zero, we obtain the critical value of $\left(\frac{M_1}{b}\right)_c=\left(\frac{M_2}{c}\right)_c=1$.

For the Weyl/$\boldmath{Z}_2$-gap phase in Fig.~\ref{fig:phase}(f),
we have the low energy effective Lagrangian
\begin{align}
\label{eq:Lagrangianb}
\mathcal{L}_{(b)}=\Psi^{\dagger}\left[\Gamma^{0}\left(i\Gamma^{\mu}\partial_{\mu}-e\Gamma^{\mu}A_{\mu}+ {M}_{1\text{eff}} {\mI}_1 \right)+\hat{\Gamma}^{0}\left(e\hat{\Gamma}^{\mu}\hat{A}_{\mu}-\hat{\Gamma}^{\mu}\hat{\Gamma}^{5} {c}_{\text{eff} }  \hat{e}_{\mu} {\mI}_2 \right)\right]\Psi\,,
\end{align}
where $M_{1\text{eff}}=\sqrt{M_{1}^{2}-b^{2}}$, and $\vec{c}_\text{eff}=c_\text{eff}\hat{e}_{y}=\sqrt{c^{2}-M_{2}^{2}}\hat{e}_{y}$.\footnote{Note that another effective theory can be obtained by keeping the
mass matrix ${M}_{2\text{eff}} {\mI}_{2}$ with $M_{2\text{eff}}=\sqrt{M_{2}^{2}-c^{2}}$ and $\vec{b}_\text{eff}=b_\text{eff}\hat{e}_{z}=\sqrt{b^{2}-M_{1}^{2}}\hat{e}_{z}$. }

The parameter region for model \eqref{eq:Lagrangianb} are $M_1\geq b$ and $M_2\leq c$. The critical-gap or gap-critical phase in Fig.~\ref{fig:phase}(e) and double critical phase in Fig.~\ref{fig:phase}(b) can also be realized by changing these parameters. In this phase, only the current which is constructed by the $\vec{c}_\text{eff}$ exist. Thus, we have $\sigma_{\boldmath{Z_{2}}\text{AHE}}\propto c_\text{eff}$, and $\sigma_{\text{AHE}}$ vanishes.

For the gap-gap phase in Fig.~\ref{fig:phase}(c), the effective Lagrangian (where the parameter region for model \eqref{eq:Lagrangianb} are $M_1\geq b$ and $M_2\geq c$) is
\begin{align}
\label{eq:Lagrangianc}
\mathcal{L}_{(c)}=\Psi^{\dagger}\left[\Gamma^{0}\left(i\Gamma^{\mu}\partial_{\mu}-e\Gamma^{\mu}A_{\mu}+ {M}_{1\text{eff}} {\mI}_1+ {M}_{2\text{eff}}{\mI}_2\right)\right]\Psi\,,
\end{align}
where the matrix element in the mass matrixes
are $M_{1\text{eff}}=\sqrt{M_{1}^{2}-b^{2}}$ and $M_{2\text{eff}}=\sqrt{M_{2}^{2}-c^{2}}$, respectively. The double critical phase in Fig.~\ref{fig:phase}(b) can also be realized with $M_1=b,M_2=c$. In this case  both of the anomalous transports vanish.

Finally, we calculate the Ward identities for the currents using Fujikawa's method
\cite{Fujikawa} and the corresponding Ward identities are
\begin{eqnarray}
\label{eq:qftwi}
\begin{split}
\partial_{\mu}\mathcal{J}^{\mu}&=0\,,\\
\partial_{\mu}\mathcal{\hat{J}}^{\mu}&=0\,,\\
\partial_{\mu}\mathcal{J}_{5}^{\mu}&=\frac{e^{2}}{32\pi^{2}}\epsilon^{\mu\nu\alpha\beta}\left(3F_{\mu\nu}F_{\alpha\beta}+F_{\mu\nu}^{5}F_{\alpha\beta}^{5}+3\hat{F}_{\mu\nu}\hat{F}_{\alpha\beta}+\hat{F}_{\mu\nu}^{5}\hat{F}_{\alpha\beta}^{5}\right)+2i\bar{\Psi}
{M}_{1} {\mI}_{1}\Gamma^5\Psi\,,\\
\partial_{\mu}\mathcal{\hat{J}}_{5}^{\mu}&=\frac{e^{2}}{32\pi^{2}}\epsilon^{\mu\nu\alpha\beta}\left(6F_{\mu\nu}\hat{F}_{\alpha\beta}+2F_{\mu\nu}^{5}\hat{F}_{\alpha\beta}^{5}\right)+2i\bar{\Psi}
{M}_{2} {\mI}_{2}\Gamma^5\Psi\,,
\end{split}
\end{eqnarray}
where $\mathcal{J}^{\mu}$ is the electric current,  ${\hat{\mathcal{J}}}^{\mu}$ is the spin current. $\mathcal{J}_{5}^{\mu}$ is the axial current which has a chiral anomaly, while $\hat{\mathcal{J}}_{5}^{\mu}$ is the so-called $\boldmath{Z_{2}}$ axial current which indicates the imbalance of the $\boldmath{Z_{2}}$ charge.The form of $\mathcal{J}_{5}^{\mu}$, $\mathcal{\hat{J}}_{5}^{\mu}$ and the details for the calculation can be found in appendix \ref{app:a}.
These Ward identities include the information of the anomalies of the system and they are crucial for the strongly coupled holographic model to be built in the next section, for which we need to check whether these identities are correctly realized.

\section{Holographic model for the Weyl-$Z_2$ semimetal}
\label{sec:holo}
In this section, we will consider the strong coupling regime of the Weyl-$Z_2$ semimetal using holography. We first construct the holographic model in section \ref{model} and then solve it at zero temperature in section \ref{ss:sol}. We find that there exist nine different kinds of solutions in holography, and they correspond to the strong coupling version of the nine phases discussed in the previous section. In section \ref{chfree}, we calculate the free energy of the system and show that the free energy of the system changes smoothly when crossing the phase transition point, which means that this phase transition is a continuous
one. We calculate the anomalous transport coefficients in section \ref{tp} according to the Kubo formula. Different behavior indicating phase transitions has been obtained. The difference between the two anomaly effects is also checked by varying the parameters in the system.

\subsection{Set-up}
\label{model}

In this section, we will extend the model
\cite{Landsteiner:2015lsa,Landsteiner:2015pdh,Landsteiner:2019kxb}
 for strongly coupled Weyl semimetal with a single pair of Weyl points to a holographic model for the Weyl-$Z_2$ semimetal and study its full phase diagram. In holography the symmetries of the boundary field theory can be mapped to the symmetries of gauge fields in the AdS space. Compared to the holographic Weyl semimetal mode, for the Weyl-$Z_2$ semimetal system \cite{kimb} we have an additional analogue spin current which could be broken by the new  mass term. Therefore, we need to introduce one more $U\left(1\right)$ gauge fields and one more scalar field to the holographic Weyl semimetal system in \cite{Landsteiner:2015lsa,Landsteiner:2015pdh}.

More precisely, the electromagnetic $U(1)$ symmetry and the axial $U(1)$ symmetry are represented by the AdS bulk gauge field $V_a$ with the field strength $F = dV$ and the gauge field $A_a$ with the field strength $F_5 = dA$ respectively. We shall also introduce two additional gauge fields: $\hat{V}_a$ with the field strength $\hat{F} = d\hat{V}$ and $\hat{A}_a$ with the field strength $ \hat{F}_5 = d\hat{A}$ to encode the fictitious spin $U(1)$ symmetry and the $\boldmath{Z}_2$ axial $U(1)$ symmetry.
The anomalies are represented by Chern-Simons terms in the action. Axial symmetry (including $\boldmath{Z}_2$ symmetry) can be broken by the mass terms. The deformation of the mass term in field theory model is introduced via the non-normalizable mode of the scalar field. With all the ingredients above, we consider the following holographic model
\begin{align}\
S&=\int d^5x \sqrt{-g}\bigg[\frac{1}{2\kappa^2}\Big(R+\frac{12}{L^2}\Big)-\frac{1}{4}F^2-\frac{1}{4}\hat{F}^2-\frac{1}{4}F_5^2-\frac{1}{4}\hat{F}_5^2\nonumber\\
&+\frac{\alpha}{3}\epsilon^{abcde}A_\mu^{5} \Big(F^5_{bc} F^5_{de}+3 F_{bc} F_{de}+3 \hat{F}_{bc} \hat{F}_{de}+\hat{F}^5_{bc}\hat{F}^5_{de}\Big)+\frac{2\beta}{3}\epsilon^{abcde}\hat{A}_{\mu}^{5}\Big(3\hat{F}_{bc}F_{de}+\hat{F}_{bc}^{5}F_{de}^{5}\Big)\nonumber\\
&-(D^{a}\Phi_{1})^{*}(D_{a}\Phi_{1})-(\hat{D}^{a}\Phi_{2})^{*}(\hat{D}_{a}\Phi_{2})-V(\Phi_1,\Phi_2)\bigg]\,, \label{eq:holomodel}
\end{align}
where $\kappa^2$ is the five dimensional gravitational  constant, $L$ is the AdS radius,  $\alpha$ and $\beta$ are the Chern-Simons coupling constants. Scalar field $\Phi_1$ (or $\Phi_2$) is  charged only under the
axial (or $\boldmath{Z}_2$ axial) gauge symmetries. The gauge covariant derivatives are
\be
D_{a}\Phi_{1}=\left(\partial_{a}-iq_1A_{a}\right)\Phi_{1} \,,~~~\hat{D}_{a}\Phi_{2}=\left(\partial_{a}-iq_2\hat{A}_{\mu}\right)\Phi_{2}\,,
\ee
where $q_1$ and $q_2$ are the axial charges of the scalar operators.
 The scalar field potential is
 \be
 V(\Phi_1,\Phi_2)=m^2\left(|\Phi_1|^2+|\Phi_2|^2\right)+\frac{\lambda_1}{2}|\Phi_1|^4+\frac{\lambda_2}{2}|\Phi_2|^4\,.
 \ee
The scalar bulk mass is set to be $m^2 L^2=-3$. With this choice, the dual operator has dimension three and its source has dimension one. In the following we set $2\kappa^2=L=1$.

{To obtain the corresponding currents and the equations of motion, we can expand the four gauge fields around the asymptotically AdS background. From the first order in fluctuations we can obtain the expressions for the currents from the boundary terms as
\begin{eqnarray}
\begin{split}
J^{\mu}&=\underset{r\rightarrow\infty}{\lim}\sqrt{-g}\Big(F^{\mu r}+4\alpha\epsilon^{r\mu\nu\rho\sigma}A_{\nu}F_{\rho\sigma}+4\beta\epsilon^{r\mu\nu\rho\sigma}\hat{A}_{\nu}\hat{F}_{\rho\sigma}\Big)\,,\\
\hat{J}^{\mu}&=\underset{r\rightarrow\infty}{\lim}\sqrt{-g}\Big(\hat{F}^{\mu r}+4\alpha\epsilon^{r\mu\nu\rho\sigma}A_{\nu}\hat{F}_{\rho\sigma}+4\beta\epsilon^{r\mu\nu\rho\sigma}\hat{A}_{\nu}F_{\rho\sigma}\Big)\,,\\
J_{5}^{\mu}&=\underset{r\rightarrow\infty}{\lim}\sqrt{-g}\Big(F_{5}^{\mu r}+\frac{4\alpha}{3}\epsilon^{r\mu\nu\rho\sigma}A_{\nu}F_{\rho\sigma}^{5}+\frac{4\beta}{3}\epsilon^{r\mu\nu\rho\sigma}\hat{A}_{\nu}\hat{F}_{\rho\sigma}^{5}\Big)\,,\\
\hat{J}_{5}^{\mu}&=\underset{r\rightarrow\infty}{\lim}\sqrt{-g}\Big(\hat{F}_{5}^{\mu r}+\frac{4\alpha}{3}\epsilon^{r\mu\nu\rho\sigma}A_{\nu}\hat{F}_{\rho\sigma}^{5}+\frac{4\beta}{3}\epsilon^{r\mu\nu\rho\sigma}\hat{A}_{\nu}F_{\rho\sigma}^{5}\Big)\,.
\end{split}
\end{eqnarray}
}

After using the on-shell condition, we would obtain the Ward identities as
\begin{align}
\begin{split}
\nabla_{\mu}J^{\mu}&=0\,,\\
\nabla_{\mu}\hat{J}^{\mu}&=0\,,\\
\nabla_{\mu}J_{5}^{\mu}&=\lim_{r\to\infty}\left[-\frac{\alpha}{3}\epsilon^{r\nu\rho\sigma\tau}\left(3F_{\nu\rho}F_{\sigma\tau}+F_{\nu\rho}^{5}F_{\sigma\tau}^{5}\right)-\frac{\beta}{3}\epsilon^{r\nu\rho\sigma\tau}\left(3\hat{F}_{\nu\rho}\hat{F}_{\sigma\tau}+\hat{F}_{\nu\rho}^{5}\hat{F}_{\sigma\tau}^{5}\right)\right. \\
&~~~~ -iq_1\left[\Phi_{1}^*(D^{r}\Phi_{1})-\Phi_{1}(D^{r}\Phi_{1})^{*}\right]\bigg]\,,\\
\nabla_{\mu}\hat{J}_{5}^{\mu}&=\lim_{r\to\infty}\left[-\frac{\alpha}{3}\epsilon^{r\nu\rho\sigma\tau}\left(3\hat{F}_{\nu\rho}F_{\sigma\tau}+F_{\nu\rho}^{5}\hat{F}_{\sigma\tau}^{5}\right)-\frac{\beta}{3}\epsilon^{r\nu\rho\sigma\tau}\left(3F_{\nu\rho}\hat{F}_{\sigma\tau}+\hat{F}_{\nu\rho}^{5}F_{\sigma\tau}^{5}\right)\right.\\
&~~~~ -iq_2\left[\Phi_{2}^{*}(\hat{D}^{r}\Phi_{2})-\Phi_{2}(\hat{D}^{r}\Phi_{2})^{*}\right]\bigg]\,.
\end{split}
\end{align}
The vector-like currents are conserved as expected.
If we set $\alpha=\beta=-\frac{3e^{2}}{32\pi^{2}}$, these identities are exactly the same as (\ref{eq:qftwi}) from the weakly coupled theory.
Note that the Ward identities do not depend on the coupling constants of the system, therefore it is expected that this holographic model describes a strong interacting Weyl-$\boldmath{Z}_{2}$ semimetal model with two pairs of Weyl-$\boldmath{Z}_{2}$ nodes. However, we could not conclude if our holographic model describes the strongly coupled model for \eqref{eq:oLagrangian} or \eqref{eq:Lagrangian}. In fact, the holographic model shares both essential features of \eqref{eq:oLagrangian} and \eqref{eq:Lagrangian}. To precisely calculate the topological charges of the nodes in the dual field theory, one has to study the dual Fermionic spectrum functions in the holographic model following \cite{Liu:2018djq} and we leave it for future work.

Note that we need to distinguish between semimetals with multiple nodes and the so-called multi-Weyl semimetal. The former has multiple nodes but each with a topological charge of $\pm 1$ while the latter might have only one pair of Weyl nodes, but each node possesses a topological charge of $\pm n$ with integer $n>1$. Multi-Weyl semimetals could be produced from a Lagrangian with an additional SU(2) non-Abelian field whose components need to be appropriately chosen \cite{Dantas:2019rgp}. A different special choice of the configuration of the SU(2) field could also give a Weyl-Z$_2$ semimetal. In holography, the multi-Weyl semimetal have been discussed also in \cite{Juricic:2020sgg}.

\subsection{Zero temperature solutions for different phases}
\label{ss:sol}
We shall focus on the zero temperature physics to obtain the gravitational solutions which correspond to different quantum phases. Our Ansatz for the zero temperature solution is
\begin{eqnarray}\label{eq:ansatz}
\begin{split}
ds^2&=u\left(- dt^2+ dx^2\right)+f dy^2+h dz^2+\frac{dr^2}{u}\,,\\
 A&=A_z dz\,,~~~ \Phi_1=\phi_1 \,,\\
 \hat{A}&=C_y dy \,,~~~\Phi_2=\phi_2\,,
 \end{split}
\end{eqnarray} where fields $u, f, h, A_z, C_y, \phi_1$ and $\phi_2$ are functions of the radial coordinate $r$. The corresponding equations of motion are shown in appendix \ref{app:b} and we have seven independent
ordinary differential equations for seven fields to solve. Note that here the axial gauge fields are along $y$ and $z$ directions, which is similar to the weakly coupled model \eqref{eq:Lagrangian}. However, as we shall show later, our holographic model also has features of the model \eqref{eq:oLagrangian} because the two gauge fields talk to each other through gravitons in the bulk.

If we consider the case $q_1=q_2$ and $\lambda_1=\lambda_2$, we have a permutation symmetry for the background solutions
\be
\label{eq:per-sym}
(f, A_z, \Phi_1) \leftrightarrow (h, C_y, \Phi_2)\,.
\ee Note that this permutation symmetry is due to the fact that the Chern-Simons factors do not enter the equations of motion. However, at the level of fluctuations we do not have such symmetry.

Close to the UV boundary we can introduce the holographic analogues of the mass terms and the time-reversal breaking parameters by imposing the following boundary condition
\begin{equation}\label{eq:bcs}
 \lim_{r\rightarrow \infty}\,r\Phi_1 = M_1~,~~~\lim_{r\rightarrow \infty}\,r\Phi_2 = M_2~,~~~\lim_{r\rightarrow \infty}A_z = b\,,
 ~~~\lim_{r\rightarrow \infty}  C_y = c\,.
\end{equation}
$M_1$ and $M_2$ correspond to the sources of the dual scalar operator $\Psi^{\dagger}\Gamma^{0}\mathbb{M}_1\Psi$ and $\Psi^{\dagger}\Gamma^{0}\mathbb{M}_2\Psi$, $b$ corresponds to the source of the chiral current $\Psi^{\dagger}\Gamma^{0}\Gamma^{\mu}\Gamma^{5}\Psi$ and $c$ corresponds to the source of the $\boldmath{Z_2}$ current $\Psi^{\dagger}\hat{\Gamma}^{0}\hat{\Gamma}^{\mu}\hat{\Gamma}^{5}\Psi$, respectively.
We shall study the phase diagrams of strong interacting  $Z_2$-Weyl semimetals by turning on these four sources.

Due to the symmetries at zero temperature,
we can choose $\hat{M}_1$,$\hat{M}_2$ and $c/b$ as the tunable parameters of the system where we define
\be
\hat{M}_1=\frac{M_1}{b}\,,~~~\hat{M}_2=\frac{M_2}{c}\,.
\ee
In the following we shall study the bulk geometry and its free
energy by tuning these parameters in the UV. For the case with $q_1=q_2$, $\lambda_1=\lambda_2$, from the permutation symmetry (\ref{eq:per-sym}) the geometry with boundary values $(\hat{M}_1, \hat{M}_2, c/b)$ should be the same as the case $(\hat{M}_2, \hat{M}_1, b/c)$.
Thus, at the background level, the solutions are symmetric to each other. Specifically when $c/b=1$, the geometries are symmetric under interchange of $M_1\leftrightarrow M_2$. We will comment more on these properties when we discuss the phase diagrams.
\vspace{.4cm}\\
\emph{\textbf{Critical solution}.}~~
In the weakly coupled theory, we have the critical points as shown in Fig.~\ref{fig:phase}(b). The corresponding geometry in the holographic system is as follows.

The near horizon solution of the critical point is the following Lifshitz-type solution, which is an exact solution of the system
\begin{eqnarray}
\begin{split}
\label{eq:nhcri}
&ds^{2}=u_{0}r^{2}\left(-dt^{2}+dx^{2}\right)+\frac{dr^{2}}{u_{0}r^{2}}+f_{0}r^{2\alpha_{1}}dy^{2}+h_{0}r^{2\alpha_{2}}dz^{2}\,,\\
&A_{z}=r^{\alpha_{1}}\,,~~\phi_{1}=\phi_{10}\,,\\
&C_{y}=r^{\alpha_{2}}\,,~~
\phi_{2}=\phi_{20}\,.
\end{split}
\end{eqnarray}
This solution has an anisotropic Lifshitz-type symmetry $\left(t,x\right)\rightarrow s\left(t,x\right),~y\to s^{\alpha_{1}}y,~z\to s^{\alpha_{2}}z$. The constants $\left\{ u_{0},f_{0},h_{0},\alpha_{1},\alpha_{2},\phi_{10},\phi_{20}\right\}$ in (\ref{eq:nhcri}) can be fully determined by the values of the parameters $m,q_1,q_2,\lambda_1,\lambda_2$.
In the following we will consider the simple case $q_1=q_2$ and $\lambda_1=\lambda_2$ where we have the permutation symmetry (\ref{eq:per-sym}). In this case we have
$\alpha_1=\alpha_2$, $f_0=h_0$, $\phi_{10}=\phi_{20}$, and
(\ref{eq:nhcri}) can be simplified as
\begin{eqnarray}
\label{socritical}
\begin{split}
&ds^{2}=u_{0}r^{2}\left(-dt^{2}+dx^{2}\right)+\frac{dr^{2}}{u_{0}r^{2}}+f_{0}r^{2\alpha_{1}}\left(dy^{2}+dz^{2}\right)\,,\\
&A_{z}=C_{y}=r^{\alpha_{1}}\,,~~\phi_{1}=\phi_2=\phi_{10}\,.
\end{split}
\end{eqnarray}

As the solution above is an exact solution of the system, we need to introduce irrelevant deformations to flow this geometry to asymptotic AdS in the UV. In the IR the leading order solutions with irrelevant perturbations are
\begin{eqnarray}
\label{eq:nh-criticalsol}
\begin{split}
u&=u_{0}r^{2}\left(1+\delta u_{1}\,r^{\beta_{1}}+\delta u_{2}\,r^{\beta_{2}}\right)\,,\\
f&=f_{0}r^{2\alpha_{1}}\left(1+\delta f_{1}\,r^{\beta_{1}}+\delta f_{2}\,r^{\beta_{2}}\right)\,,\\
h&=f_{0}r^{2\alpha_{1}}\left(1+\delta h_{1}\,r^{\beta_{1}}+\delta h_{2}\,r^{\beta_{2}}\right)\,,\\
C_{y}&=r^{\alpha_{1}}\left(1+\delta c_{1}\,r^{\beta_{1}}+\delta c_{2}\,r^{\beta_{2}}\right)\,,\\
\Phi_{2}&=\phi_{10}\left(1+\delta\phi_{21}\,r^{\beta_{1}}+\delta\phi_{22}\,r^{\beta_{2}}\right)\,,\\
A_{z}&=r^{\alpha_{1}}\left(1+\delta a_{1}\,r^{\beta_{1}}+\delta a_{2}\,r^{\beta_{2}}\right)\,,\\
\Phi_{1}&=\phi_{10}\left(1+\delta\phi_{11}\,r^{\beta_{1}}+\delta\phi_{12}\,r^{\beta_{2}}\right)\,.
\end{split}
\end{eqnarray}

Taking into account the scaling symmetries of the system, There are only two free parameters above: the sign of $\delta a_{1}$ and the value of $\delta a_{2}$. Only the choice of $\delta a_{1}=-1$ could flow the geometry to asymptotic AdS boundary. The value of $\delta a_{2}$ can be freely changed to tune the value of $c/b$ and we could obtain the corresponding critical values of $M_{1c}/b, M_{2c}/c$. Note that in the holographic Weyl semimetal system, there is no free parameter for the critical solution as the system has a fixed critical point. However, here we have a free tuning parameter of $c/b$ in the phase diagram and the critical point becomes a critical line in the three dimensional phase diagram. Thus here we have an extra tuning parameter to tune the value of $c/b$ and the critical values of $M_{1c}/b, M_{2c}/c$ will be determined accordingly.

We fix $q=q_2=1$ and $\lambda_1=\lambda_2=1/10$ without loss of generality. We have
$\{u_0, f_0, \phi_{10}, \alpha_1, \beta_1, \beta_2\}$\\$\to \{2.367,\,0.347,\,1.081,\,0.410,\,1.063, \,1.203\}$,
$\{\delta u_{1}, \delta f_{1}=\delta h_{1}, \delta c_{1}, \delta\phi_{11}=\delta\phi_{21}\}\to$ \{8.111, \,-27.212, \,1, \, 8.832\}$\delta a_1$ and $\{\delta u_{2}, \delta f_{2}=-\delta h_{2}, \delta c_{2}, \delta\phi_{12}=-\delta\phi_{22} \}\to\{1.624, ~262.692, ~-1,~68.031\}\delta a_2$. Shooting to $c/b=1$ we have the critical values $\hat{M}_1=\hat{M}_2=0.908$, which corresponds to the critical phase where four Weyl-$Z_2$ nodes gather together as a single Dirac point.

\begin{figure}[h!]
  \centering
\includegraphics[width=0.5\textwidth]{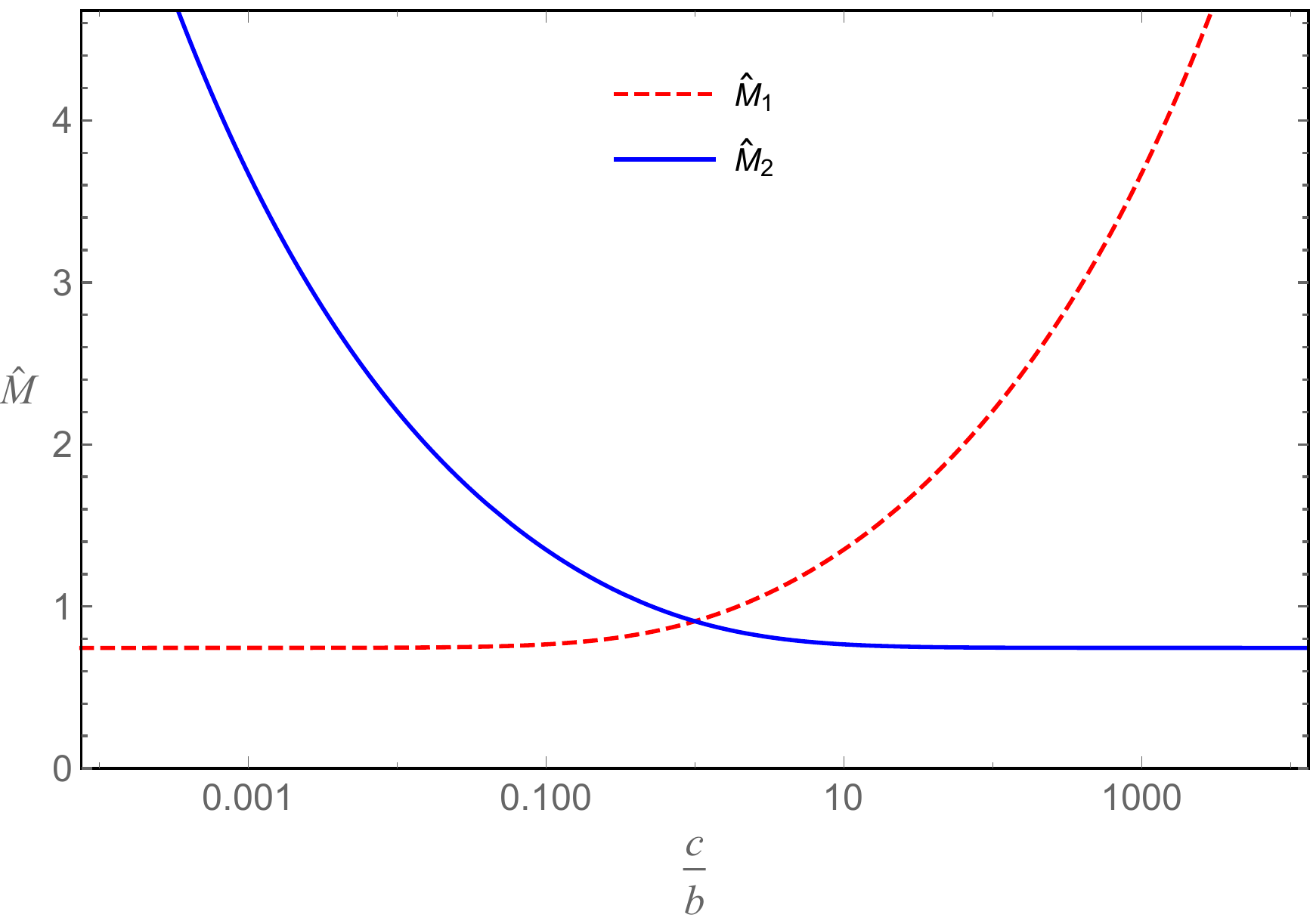}
\vspace{-0.3cm}
  \caption{\small The critical value $\hat{M}_1=M_1/b$ (red), $\hat{M}_2=M_2/c$ (blue) as a function of $c/b$.}
     \label{fig:critical}
\end{figure}

As shown in Fig.~\ref{fig:critical}, we numerically check that
at large $c/b$ the value of $\hat{M}_2$ approaches $0.744$ where only one pair of Weyl nodes exist as in \cite{Landsteiner:2015pdh}, while
$\hat{M}_1$ goes to the same value when $c/b$ tends to zero which is consistent with the permutation symmetry of the background. These two curves cross at $c/b=1$. This physical picture is qualitatively similar to the case in the weak coupling theory.

\vspace{0cm}

\vspace{.4cm}

Next we will give the two phase transition solutions which correspond to the case where only one pair of Weyl-$Z_2$ nodes annihilates to form a critical point. In the phase diagram these two solutions
correspond to critical lines with fixed value of $c/b$.
\vspace{.4cm}\\
\emph{\textbf{Weyl/$\boldmath{Z}_2$-Critical.}}~~
In the weak coupling case, we have the two phase transition surfaces where only one pair of Weyl/$\boldmath{Z}_2$ nodes annihilates to a critical Dirac point while the other pair of nodes still exists, as shown in Fig.~\ref{fig:ophase}(d) and ~\ref{fig:phase}(d). For this case it is critical along one of the $k_y$ and $k_z$ directions while the two Weyl/$\boldmath{Z}_2$ nodes are separated along the other direction.

\begin{figure}[h!]
  \centering
\includegraphics[width=0.45\textwidth]{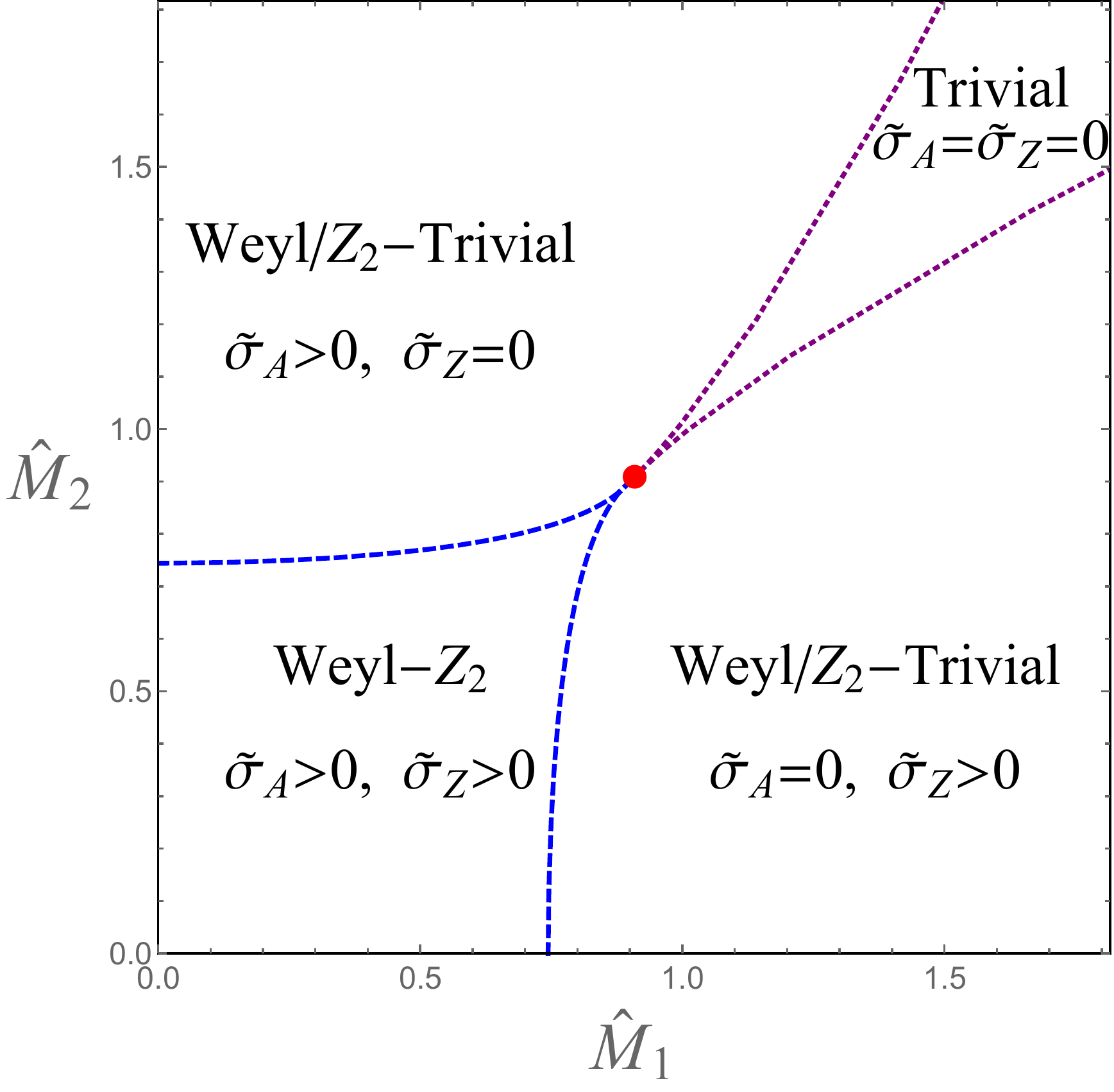}
\vspace{-0.2cm}
\caption{\small {The holographic phase diagram at $c/b=1$. $\tilde{\sigma}_{A}=\frac{\sigma_{\text{AHE}}}{8\alpha b}$ and
$\tilde{\sigma}_{Z}=\frac{\sigma_{\boldmath{Z}_2 \text{AHE}}}{8\beta c}$. The red point corresponds to the critical phase, at which both $\tilde{\sigma}_{A}$ and $\tilde{\sigma}_{Z}$ are zero. The blue dashed lines and the purple dotted lines are two critical phase transition lines with one critical Dirac node and one pair of nodes or gapped bands. The up-left and bottom-right parts of the phase diagram correspond to the Weyl/$\boldmath{Z}_2$-Trivial phase where $\tilde{\sigma}_{Z}=0$ while $\tilde{\sigma}_{A}\neq 0$ or $\tilde{\sigma}_{Z}\neq 0$ while $\tilde{\sigma}_{A}= 0$.
The right-up part in the phase diagram corresponds to the phase where both the Weyl-Z$_2$ nodes vanish and become trivial, i.e. the Trivial-Trivial phase. In this case, both of the two anomalous transport coefficients vanish. The left-down part in the phase diagram corresponds the phase where all of the four Weyl-$\boldmath{Z}_2$ nodes exist. In this phase,  $\tilde{\sigma}_{A}\neq 0$ and $\tilde{\sigma}_{Z}\neq 0$ (i.e. the Weyl-$\boldmath{Z}_2$ phase).}}
\label{fig:phase22}
\end{figure}

For the one of these two phase transition surfaces with the critical node in the $k_y$ direction, the holographic solution in the IR takes the following form
\begin{eqnarray}
\begin{split}
&u=h=u_{0}\,r^{2}\left(1+ \delta u_{1}\,r^{\beta_{1}}\right)\,,\\
&f=f_{0}\,r^{2\alpha_{1}}\left(1+ \delta f_{1}\,r^{\beta_{1}}\right)\,,\\
&C_{y}=r^{\alpha_{1}}\left(1+ \delta c_{1}\,r^{\beta_{1}}\right)\,,\\&
\Phi_{2}=\phi_{20}\left(1+ \delta\phi_{21}\,r^{\beta_{1}}\right)\,,\\
&A_{z}=a_{0}+\phi_{10}^{2}\frac{\sqrt{u_{0}}}{qa_{0}^{2}\pi r^{\alpha_{1}}}\exp\left(-\frac{2q_1a_{0}}{r\sqrt{u_{0}}}\right)\,,\\
&\Phi_{1}=\phi_{10}\left(\frac{1}{r}\right)^{1+\frac{\alpha_{1}}{2}}\exp\left(-\frac{q_1a_{0}}{r\sqrt{u_{0}}}\right)\,.
\label{dwc}
\end{split}
\end{eqnarray}
With the above choice of $q_1,q_2,\lambda_1,\lambda_2$, we have $\{u_0, f_0, \phi_{20},\alpha_1, \beta_1\}$
$\to \{1.468$,\, $0.344,$
$\, 0.947$,\, $0.407$,\, $1.315\}$ and $\{\delta u_{1}, \delta f_{1}, \delta\phi_{21}, \delta c_{1}\} \to \{0.021,\, -0.159,\,0.057, \,0.0078\}c_0$. Only the sign of $c_0$ is important for the flow and it has to be chosen to be $c_0=-1$. There are only two shooting parameters $a_0$ and $\phi_{10}$ corresponding to a surface in the three dimensional phase diagram. When we fix $c/b$ we will get a curve in the plane $M_1/b$-$M_2/c$. Shooting to $c/b=1$ we obtain the (dashed blue) curve as the critical line with two endpoints as (0.908,~0.908) and (0,~0.744) in the phase diagram of the $\hat{M}_1$-$\hat{M}_2$ plane (see
Fig.~\ref{fig:phase22}).

For the case where it is critical along the $k_z$ direction while the Weyl-$\boldmath{Z}_2$ nodes are in the $k_y$ direction,
we could just change the IR solution (\ref{dwc}) by $(f, A_z, \Phi_1)\leftrightarrow (h, C_y, \Phi_2)$ following the permutation symmetry and integrating the system to the boundary,
we obtain another (dashed blue) curve with two endpoints as (0.908,~0.908) and (0.744,~0) on the $\hat{M}_1$-$\hat{M}_2$ plane at $c/b=1$.
\vspace{.2cm}\\
\emph{\textbf{Critical-Trivial}}.~~
In the weak coupling system, when only one pair of Weyl/$\boldmath{Z}_2$ nodes annihilate to a critical Dirac point while the other pair of nodes becomes a trivial gap, we have two critical-trivial phases, as shown in Fig.~\ref{fig:phase}(e). In holography, the corresponding geometry for the
case where the $y$ direction is critical while the $z$ direction is trivial is as follows. In the IR, we have
\begin{eqnarray}
\begin{split}
&u=u_{0}\,r^{2}\left(1+\delta u_{1}\,r^{\beta_{1}}+a_0^2 \,\delta u_2\, r^{2 a_1-2}+\phi_{10}^2 \,\delta u_3 \,r^{2\phi_{11}}\right)\,,\\
&f=f_{0}\,r^{2\alpha_{1}}\left(1+\delta f_{1}\,r^{\beta_{1}}
+a_0^2\, \delta f_2\, r^{2 a_1 -2}+\phi_{10}^2\, \delta f_3\, r^{2\phi_{11}}
\right)\,,\\
&h=u_{0}\,r^{2}\left(1+\delta u_{1}\,r^{\beta_{1}}+a_0^2\, \delta h_2\, r^{2 a_1-2}+\phi_{10}^2\, \delta h_3\, r^{2\phi_{11}}\right)\,,\\
&C_{y}=r^{\alpha_{1}}\left(1+\delta c_{1}\,r^{\beta_{1}}
+a_0^2\, \delta c_2 \,r^{2 a_1-2}+\phi_{10}^2\, \delta c_3 \,r^{2\phi_{11}}\right)\,,\\
&\Phi_{2}=\phi_{20}\left(1+\delta\phi_{21}\,r^{\beta_{1}}
+a_0^2\, \delta \phi_2 \,r^{2 a_1-2}+\phi_{10}^2\, \delta \phi_3 \,r^{2\phi_{11}}\right)\,,\\
&A_{z}=a_{0}\,r^{a_1}\,, \\
&\Phi_{1}=\sqrt{\frac{3}{\lambda_{1}}}+\phi_{10}\,r^{\phi_{11}}\,,
\label{dgc}
\end{split}
\end{eqnarray}
where
\bea
\begin{split}
a_1&=\frac{-u_{0}\lambda_{1}\left(1+\alpha_{1}\right)+\sqrt{24q^{2}u_{0}\lambda_{1}+u_{0}^2\lambda_{1}^2\left(1+\alpha_{1}\right)^{2}}}{2u_{0}\lambda_{1}}\,,\\
\phi_{11}&=\frac{-u_{0}(3+\alpha_{1})+\sqrt{24u_0+u_{0}^2(
3+\alpha_1)^2}}{2u_{0}}\,.
\end{split}
\eea
Because the backreactions of $A_z$ and $\Phi_1$ to the other fields are of the same order, here we have considered higher order terms in the expansion to make the solution more accurate.
With the choice of $q_1=q_2=1$ and $\lambda_1=\lambda_2=1/10$,
we have $\{u_0, f_0, \phi_{20},\alpha_1, \beta_1\}\to$\{6.914, \, 0.393, \, 2.137,\, 0.44,\, 0.886\},
$\{\delta u_1, \delta f_1, \delta \phi_{21}\}$
$\to$\{-2.496,  28.77, -2.659\}$\delta c_1$,
$\{\delta u_2, \delta f_2, \delta h_2, \delta c_2, \delta \phi_2\}$
$\to$ \{0.032,  0.024, -0.3,  0.005, -0.0004\}, $\{a_1, \phi_{11}\}\to \{2.312, 0.236\}$
 and $\{\delta u_3,\delta f_3, \delta h_3, \delta c_3, \delta \phi_3\}$
 $\to$\{-2.241,  6.885,  -2.241,   0.988,   -1.053\}.
Note that $\delta c_1$ has to be set to $-1$. The two shooting parameters are $(a_0, \phi_{10})$. Therefore, for fixed $c/b$ we obtain the solutions as a curve in the plane $\hat{M}_1$-$\hat{M}_2$. We find that when shooting to $c/b=1$, this solution only exist at $\hat{M}_1 \geq 0.908$, which is the (dashed red) curve in the phase diagram of Fig.~\ref{fig:phase22}.
Similar to the case above,
the solution for the case that the $z$ direction is critical while $y$ direction is trivial can be obtained by replacing
$(f, A_z, \Phi_1)\leftrightarrow (h, C_y, \Phi_2)$. Shooting to $c/b=1$ we obtain the red dashed curve in Fig.~\ref{fig:phase22}.

\vspace{.4cm}
In the following we study the remaining three kinds of solutions (for four phases), each of the phase corresponds to a two dimensional regime in the phase diagram for fixing $c/b$ and to a three dimensional regime in the three dimensional phase diagram.
\vspace{.4cm}\\
\emph{\textbf{Weyl-$\boldmath Z_2$}.}~~ This kind of solution corresponds to the phase
with four Weyl nodes as shown in Fig.~\ref{fig:phase}(b).
The leading order geometry in the IR has the following form,
\begin{eqnarray}
\begin{split}
&u=f=h=r^{2}\,,\\
&
C_{y}=c_{0}+\frac{\pi c_{0}^{2}\,\phi_{20}^{2}}{16 r}\exp\left(-\frac{2q_2c_{0}}{r}\right)\,,\\
&\Phi_{2}=\phi_{20}\sqrt{\frac{\pi}{8}}\left(\frac{q_2c_{0}}{r}\right)^{\frac{3}{2}}\exp\left(-\frac{q_2c_{0}}{r}\right)\,,\\
&A_{z}=a_{0}+\frac{\pi a_{0}^{2}\,\phi_{10}^{2}}{16 r}\exp\left(-\frac{2q_1a_{0}}{r}\right)\,, \\
&\Phi_{1}=\phi_{10}\sqrt{\frac{\pi}{8}}\left(\frac{qa_{0}}{r}\right)^{\frac{3}{2}}\exp\left(-\frac{q_1a_{0}}{r}\right)\,.
\label{fwp}
\end{split}
\end{eqnarray}
We have shooting parameters $a_0$, $c_0,\phi_{10},\phi_{20}$. Either $a_0$ or $c_0$ can be fixed to be one and we have three independent parameters. Shooting to $c/b=1$ and with our choice of $q_1,q_2,\lambda_1,\lambda_2$ we find that this kind of solutions indeed only exists in the area of the left-down portion in Fig.~\ref{fig:phase22}. In particular, we can start
from (\ref{fwp}) with the choice $a_0=c_0,q_1=q_2,\lambda_1=\lambda_2$ and $\phi_{10}=\phi_{20}$, a straight line which connects the point (0,\,0) and the point (0.908,\,0.908) in the phase diagram of Fig.~\ref{fig:phase22} can be obtained. Varying the shooting parameters, we can obtain different $\hat{M}_1$ and $\hat{M}_2$ in the phase diagram.
\vspace{.4cm}\\
\emph{\textbf{Topological trivial solutions.}}~~
In the case without any Weyl/$Z_2$ nodes as shown in Fig.~\ref{fig:phase}(a), the dual theory is topologically trivial along all the spatial directions.

In the IR the leading order geometry of the holographic solution
takes the following form
\begin{eqnarray}
\begin{split}
&u=\left(1+\frac{3}{8\lambda_{1}}+\frac{3}{8\lambda_{2}}\right) r^2\,,\\
&f=h=r^{2}\,,\\
&
C_{y}=c_{0}r^{-1+\sqrt{1+\frac{48q_2^{2}\lambda_{1}}{3\lambda_{2}+\lambda_{1}\left(3+8\lambda_{2}\right)}}}\,,\\
&\Phi_{2}=\sqrt{\frac{3}{\lambda_{2}}}+\phi_{20}r^{-2+\sqrt{2+\frac{4\lambda_{1}\lambda_{2}}{3\lambda_{2}+\lambda_{1}\left(3+8\lambda_{2}\right)}}}\,,\\
&A_{z}=a_{0}r^{-1+\sqrt{
1+
\frac{48q_{1}^{2}\lambda_{2}}{3\lambda_{2}+\lambda_{1}\left(3+8\lambda_{2}\right)}}}\,,\\
&\Phi_{1}=\sqrt{\frac{3}{\lambda_{1}}}+\phi_{10}r^{-2+\sqrt{2+\frac{4\lambda_{1} \lambda_{2}}{3\lambda_{2}+\lambda_{1}\left(3+8\lambda_{2}\right)}}}\,.
\label{fgp}
\end{split}
\end{eqnarray}
We take $a_0$, $c_0,\phi_{10},\phi_{20}$ as the shooting parameters. Similar to the previous case, we have three independent parameters and we find this kind of solution only exists in right-up portion of the phase diagram of Fig.~\ref{fig:phase22}.
\vspace{.4cm}\\
\emph{\textbf{Weyl/$Z_2$-Trivial phases}}.~~ This kind of solution corresponds to two phases where only one pair of Weyl/$Z_2$ nodes becomes trivial while the other pair of nodes still exists as shown in Fig.~\ref{fig:phase}(f). For the case where the $y$-direction is topologically nontrivial while the $z$-direction is topologically trivial, we have the leading order holographic solution in the IR to be
\begin{eqnarray}
\begin{split}
&u=\left(1+\frac{3}{8\lambda_1}\right)r^{2}\,,\\
&f=h=r^{2}\,,\\
&
C_{y}=c_{0}+\frac{\pi \, c_{0}^{2}\, \phi_{20}^{2}}{16r}\exp\left(-\frac{2c_{0}}{r}\right)\,,\\
&\Phi_{2}=\frac{\phi_{20}}{c_{0}^{1/2}q_2^{1/2}u_0^{3/4}r^{3/2}}\sqrt{\frac{2}{\pi}}\left(1+\frac{3}{8\lambda_1}\right)
\text{\ensuremath{\exp}\ensuremath{\left(-\frac{c_{0}q_2}{r\sqrt{u_{0}}}\right)}}\,,\\
&A_{z}=a_{0}r^{-1+\sqrt{1+\frac{48 q_{1}^2}{3+8\lambda_1}}}
\,,\\
&\Phi_{1}=\sqrt{\frac{3}{\lambda_{1}}}+\phi_{10}\,r^{
2\left(-1+\sqrt{\frac{3+20\lambda_1}{3+8\lambda_1}}\right)}\,.
\label{inter}
\end{split}
\end{eqnarray}
With three independent shooting parameters since one of $\{a_0, c_0\}$ could be set to be 1 and with our choice of $q_1,q_2,\lambda_1,\lambda_2$, we find that this kind of solution only exists in the right-down part in the phase diagram of Fig.~\ref{fig:phase22}. The solution for the case that the $z$ direction is critical while the $y$ direction is trivial can be obtained by replacing $(f, A_z, \Phi_1)\leftrightarrow (h, C_y, \Phi_2)$. Thus we can obtain the left-up part in the phase diagram Fig.~\ref{fig:phase22} from the procedure above.

\subsubsection{Phase diagram}

We have shown the solutions for the nine different phases in the previous subsection. From the IR solution, we could tell if the solution has Weyl/$Z_2$ nodes or if the solution is gapped from the IR behavior of the fields. As mentioned before, the anomalous transports $\sigma_\text{\text{AHE}}$ and $\sigma_{\boldmath{Z}_2 \text{\text{AHE}}}$ calculated from the Kubo formula in \eqref{kubo} can be employed as the order parameters. To be specific, we have a non-zero $\sigma_\text{\text{AHE}}$ and $\sigma_{\boldmath{Z}_2 \text{\text{AHE}}}$ in the Weyl-$\boldmath{Z}_2$ phase, while the $\sigma_{\boldmath{Z}_2 \text{\text{AHE}}}$ (or $\sigma_\text{\text{AHE}}$) vanishes when we reach the Weyl/$\boldmath{Z}_2$-critical phase or the Weyl/$\boldmath{Z}_2$-gap phase. Both of the two anomalous transports become zero at the critical-critical point, the critical-gap phase and the gap-gap phase. In Sec. \ref{tp}, we will confirm the corresponding phase for each solution from the behavior of the two anomalous Hall transport coefficients $\sigma_\text{\text{AHE}}$ and $\sigma_{\boldmath{Z}_2 \text{\text{AHE}}}$ from the holographic calculation. In this subsection, we will first summarize the phase diagram of the nine different phases/critical lines/critical points above.

\begin{figure}[h!]
  \centering
\includegraphics[width=0.5\textwidth]{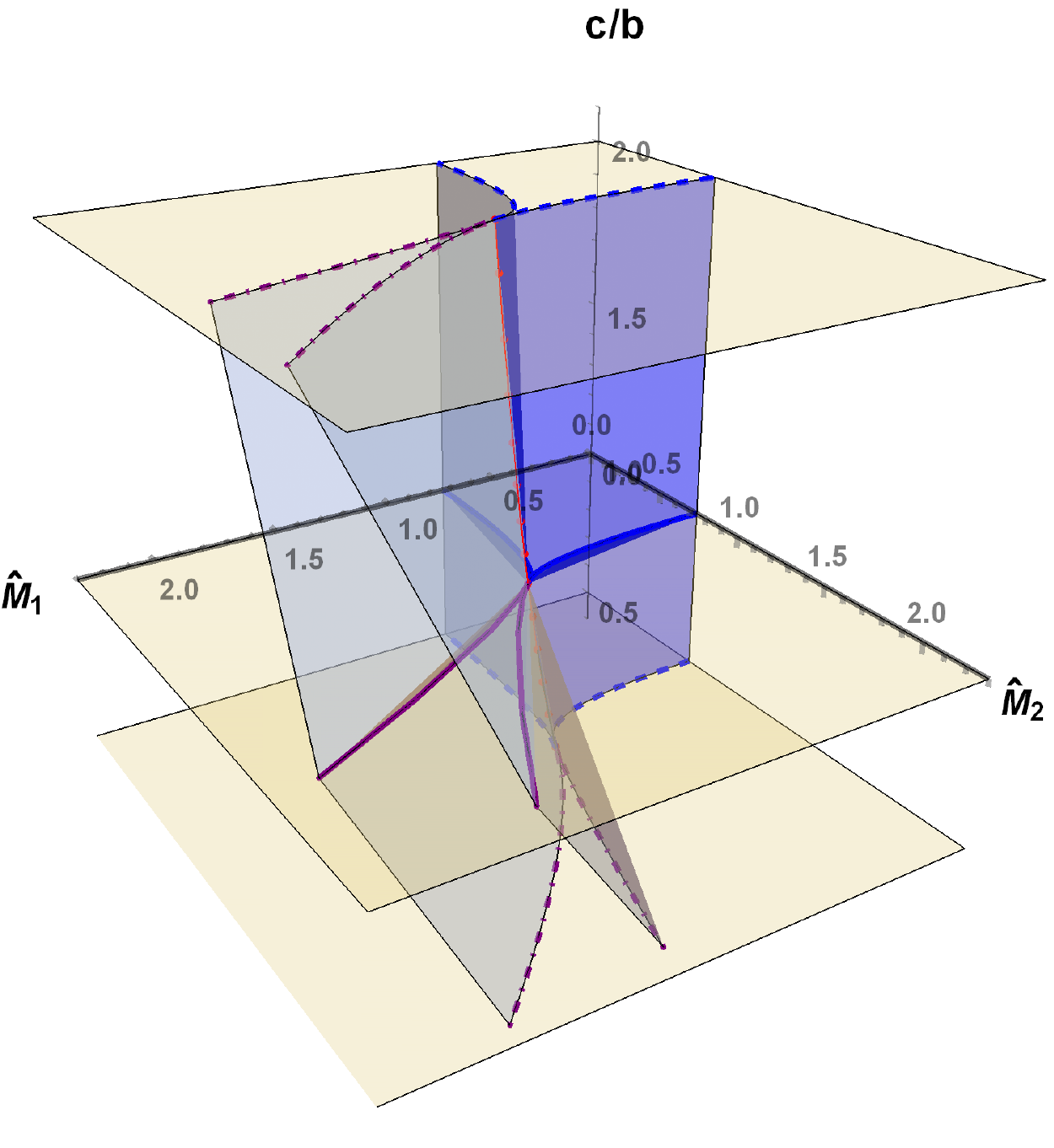}
\vspace{-0.3cm}
\caption{\small The phase diagram of the system with three dimensionless parameters $\hat{M}_1=M_1/b$, $\hat{M}_2=M_2/c$ and $c/b$. The red points are critical points forming a line, the blue dashed (blue solid for $c/b=1$) lines correspond to the critical phases where one pair of the nodes annihilate to a Dirac point while the other pair of nodes still exist (so is the light blue surface) and the purple dotted (purple solid at $c/b=1$) lines correspond to another critical phase where one pair of nodes vanish and become trivial while the other pair becomes a Dirac node (so is the light purple surface).}
\label{fig:phase21}
\end{figure}

The phase diagram for the holographic model with $c/b=1$ is shown in  Fig.~\ref{fig:phase22}, which is two dimensional in the parameter space. One can relax the ratio $c/b=1$ to arbitrary value and a three dimensional phase diagram can be obtained in the space of $c/b, \hat{M}_1, \hat{M}_2$, as shown in  Fig.~\ref{fig:phase21}. At each value of $c/b$, the qualitative behavior of the phase diagram is the same. The nontrivial dependence on the value of $c/b$ is similar to the behavior of weakly coupled result in the right plot of Fig.~\ref{fig:phaseo2}. This is due to the fact that in the holographic model the two axial gauge fields are no longer independent.

\vspace{0cm}

In the following we shall focus on the case of $c/b=1$ and point out the different behavior of the anomalous transport coefficients in different phases, whose detailed calculations will be presented in the next section. In section \ref{chfree} we will compute the free energy across the phase separations to show that the phase transition is a continues phase transition.

\begin{figure}[h!]
  \centering
\includegraphics[width=0.4\textwidth]{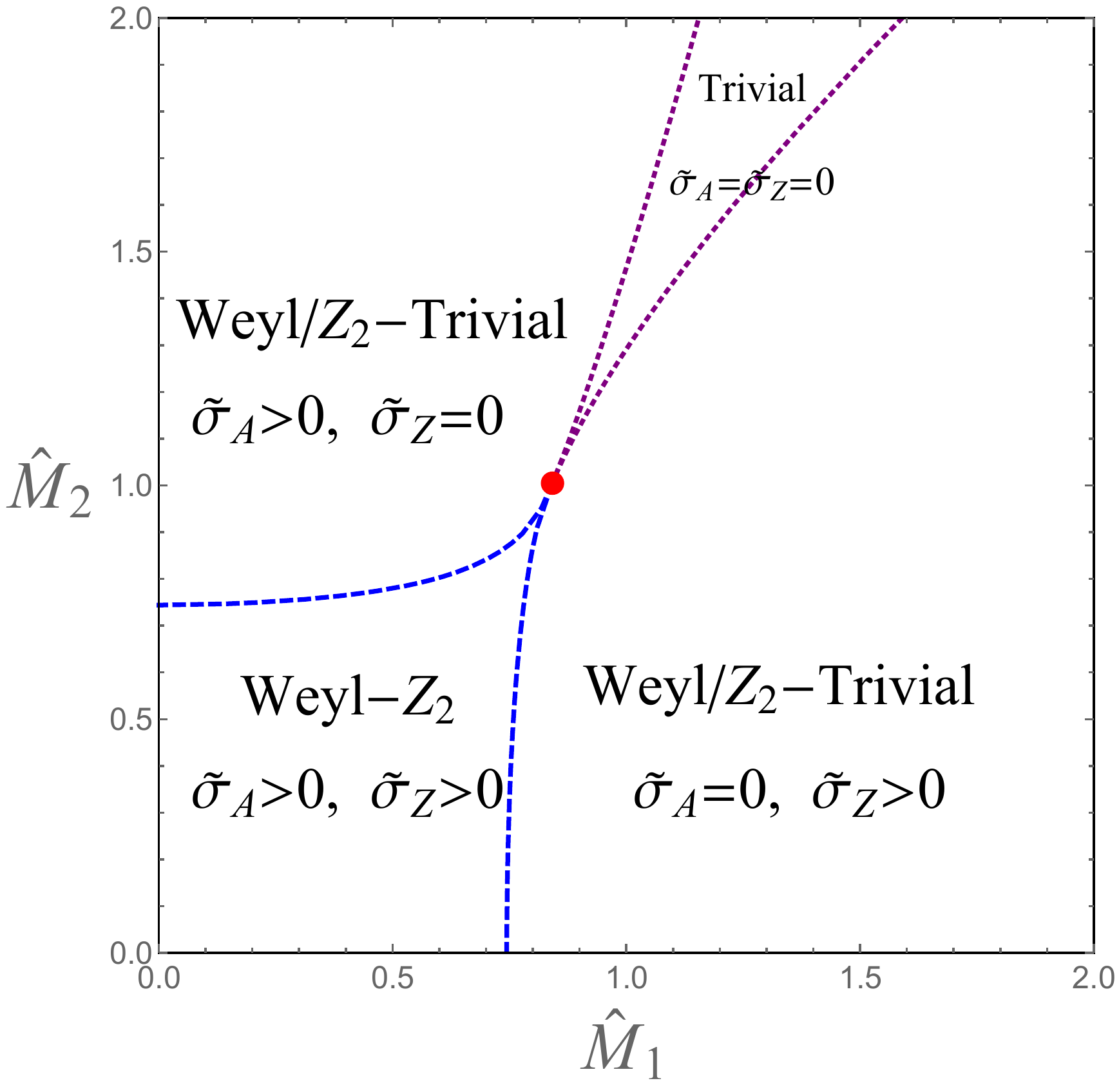}~~~
\includegraphics[width=0.4\textwidth]{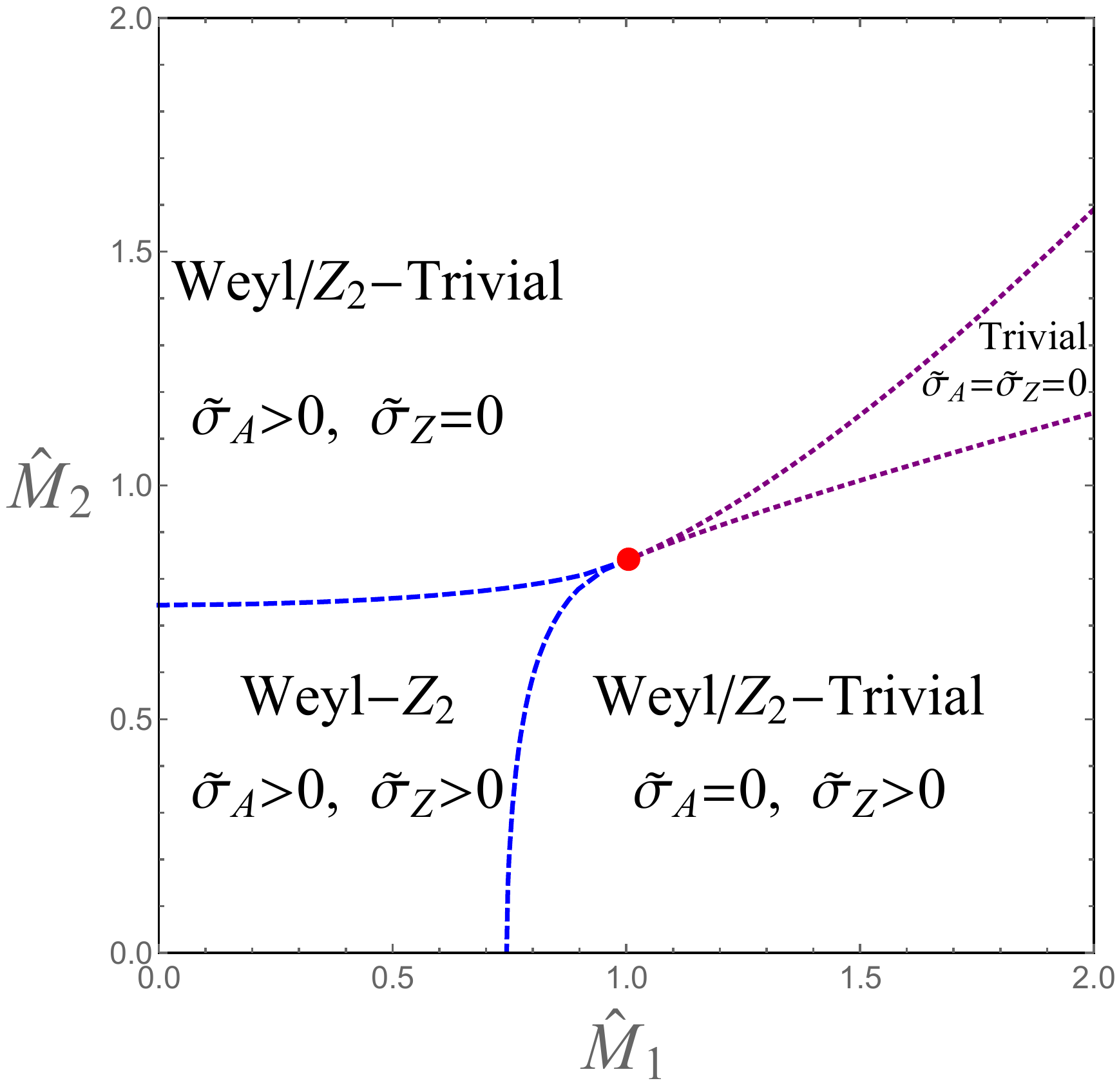}
\vspace{-0.2cm}
\caption{\small The phase diagram of the holographic system
at $c/b=1/2$ (left) and $c/b=2$ (right) respectively.}
\label{fig:phase23}
\end{figure}

{In Fig.~\ref{fig:phase22}, the red point corresponds to the solution with the near horizon solution (\ref{eq:nh-criticalsol}). This phase is a critical phase, both $\sigma_\text{\text{AHE}}$ and $\sigma_{\boldmath{Z}_2 \text{AHE}}$ are equal to zero. The blue dashed lines and the purple dotted lines are two phase transition lines with two nodes annihilating into a Dirac point and two nodes left. The left blue dashed lines correspond to the solution in the IR regime in (\ref{dwc}) with $\sigma_{\boldmath{Z}_2 \text{AHE}}=0$ while $\sigma_\text{AHE}\neq 0$ and vice verse.}\footnote{Note that we use dimensionless transports $\tilde{\sigma}_{A}$ and $\tilde{\sigma}_{Z}$ in Fig.~\ref{fig:phase22} and Fig.~\ref{fig:phase23}. The dimensionless transports are defined in  \eqref{sigmaAZ}.}

{Note that, the solution (\ref{inter})  corresponds to the phase where only one pair of Weyl/$\boldmath{Z}_2$ nodes exist while the other pair of nodes vanish. This phase is located at the up-left part in the phase diagram of Fig.~\ref{fig:phase22} (i.e. the Weyl/$\boldmath{Z}_2$-Trivial phase). Using the permutation symmetry of the system we can plot the right blue dashed line which corresponds to $\sigma_{\boldmath{Z}_2 \text{AHE}}\neq 0$ but $\sigma_\text{AHE}=0$, i.e. the Weyl/$\boldmath{Z}_2$-Critical phase. The right-down part can also be obtained through the permutation symmetry and we can only measure the anomalous Hall conductivity in this phase.}

{The right purple dotted line corresponds to the solution with the near horizon geometry (\ref{dgc}) with both $\sigma_{\boldmath{Z}_2 \text{AHE}}$ and $\sigma_{\text{AHE}}$ equal to zero (i.e. the Critical-Trivial phase). The left purple dotted line can be obtained by the permutation symmetry. The right-up part in the phase diagram corresponds to the phase where all the nodes vanish and become trivial, which has the near IR solution in (\ref{fgp}). In this case, both of the two anomalous transport coefficients vanish, i.e. the Trivial-Trivial phase.  The left-down part in the phase diagram has the near IR solution in (\ref{fwp}) where all of the Weyl/$\boldmath{Z}_2$ nodes exist. In this phase, both $\sigma_{\text{AHE}}$ and $\sigma_{\boldmath{Z}_2 \text{AHE}}$ are nonzero, i.e. the Weyl-$\boldmath{Z}_2$ phase. Finally we have a non-vanishing $\sigma_{\boldmath{Z}_2 \text{\text{AHE}}}$ and a zero $\sigma_\text{\text{AHE}}$ for the solution (\ref{inter}). It corresponds to the right-down part in the phase diagram of Fig.~\ref{fig:phase22}.}

We show the phase diagram at $c/b=1/2$ and $c/b=2$ in Fig.~\ref{fig:phase23}. As expected from the permutation symmetry, these two plots are symmetric under $\hat{M_1} \leftrightarrow \hat{M}_2$.  We can see that when $c/b$ changes, the qualitative behavior of the phase diagram does not change.

{Several different paths in the phase diagram in Fig.~\ref{fig:phase23}, say left panel without loss of generality, can lead to different phase transition processes. For the diagonal line, the system undergoes a phase transition from the Weyl-$\boldmath{Z}_2$ to the Weyl/$\boldmath{Z}_2$-trivial phase. While for the lines parallel to the vertical axis, the system may undergo different phase transition processes according to the location of the critical point. For example, when we study the line with fixed $\hat{M}_1=0.5$, the system goes from the Weyl-$\boldmath Z_{2}$ to the other Weyl/$\boldmath{Z}_2$-trivial phase. While a fixed $\hat{M}_1=1.3$ gives a Weyl/$\boldmath{Z}_2$-trivial to the trivial-trivial phase transition. When the fixed value approaches the critical point, say $\hat{M}_1=1$, we would have a more interesting process. The system first goes from the Weyl/$\boldmath{Z}_2$-trivial phase to the trivial-trivial phase then to the other Weyl/$\boldmath{Z}_2$-trivial phase. While at $\hat{M}_1=(\hat{M}_1)_c$, the system goes from the Weyl/$\boldmath{Z}_2$-trivial to the other Weyl/$\boldmath{Z}_2$-trivial phase with increasing $\hat{M}_2$. Similar processes happen when we consider the path parallel to the horizontal axis. In this paper, we only study the diagonal line without loss of generality. Note that the different Weyl/$\boldmath{Z}_2$-trivial/critical phases are characterised by different anomalous Hall conductivities being zero, i.e. $\sigma_\text{AHE}=0$ or $\sigma_{\boldmath{Z}_2 \text{AHE}}= 0$.}
\vspace{0cm}

\subsection{Free energy across the phase transition}
\label{chfree}

In this section, we compute the free energy of this system to study whether the phase transition is a continuous one. To compute the free energy, we need to be careful with the boundary
counter-terms. The renormalized action is
\begin{eqnarray}
S_\text{ren}=S_{\text{on-shell}} +S_\text{GH}+S_{c.t.},
\end{eqnarray}
where {$S_{\text{on-shell}}$ is the bulk on-shell action, which can be calculated from \eqref{eq:holomodel} in section \ref{model},} and $S_\text{GH}$ is the Gibbons-Hawking term
$S_\text{GH}=2\int_{r=r_{\infty}}d^{4}x\sqrt{-\gamma}K$. The counter-term $S_{c.t.}$ is
\begin{eqnarray}
\begin{split}
&S_{c.t.}=\int_{r=r_{\infty}}d^{4}x\sqrt{-\gamma}\left(-6-|\Phi_{1}|^{2}-|\Phi_{2}|^{2}+\frac{1}{2}\left(\log r^{2}\right)\left[\frac{1}{4}F^{2}+\frac{1}{4}F_{5}^{2}+\frac{1}{4}\hat{F}^{2}+\frac{1}{4}\hat{F}_{5}^{2}+\right.\right. \\
&~~~\left.\left.+|D_{1\mu}\Phi_{1}|^{2}+|D_{2\mu}\Phi_{2}|^{2}+\left(\frac{1}{3}+\frac{\lambda_{1}}{2}\right)|\Phi_{1}|^{4}+\left(\frac{1}{3}+\frac{\lambda_{2}}{2}\right)|\Phi_{2}|^{4}+\frac{2}{3}|\Phi_{1}|^{2}|\Phi_{2}|^{2}\right]\right),
\end{split}
\end{eqnarray}
where $\gamma_{\mu\nu}$ is the induced metric on the boundary $r=r_{\infty}$, $K$ is the trace of the extrinsic curvature with  $K=\gamma^{\mu\nu}\nabla_\mu n_\nu$ and $n_\nu$ is the outward unit
vector normal to the boundary. Compared to the counter-terms in \cite{Landsteiner:2015pdh}, an extra term $|\Phi_{1}|^{2}|\Phi_{2}|^{2}$ appears.

The bulk on-shell action {is calculated to be} a total derivative, which is
\begin{eqnarray}
S_{\text{on-shell}}=\int d^{4}xdr \sqrt{-g}\mathcal{L}=-\int d^{4}x\intop_{0}^{r_{\infty}}dr\left(u^{\prime}\sqrt{fhu}\right)^{\prime}.
\label{free}
\end{eqnarray}
With the field expansion {near the UV boundary $r\to \infty$, we obtain the free energy density $\frac{\Omega}{V}=- \frac{1}{V}S_{\text{ren}}=\frac{1}{8}(8M_1O_1+8M_2O_2+2b^2M_1^2q_1^2+2c^2M_2^2q_2^2+4b\eta_1+4c\eta_2+\lambda_1M_1^4+\lambda_2M_2^4)$, as shown in \eqref{eC3}.}
 The  detailed calculations are shown in appendix \ref{app:c}.
In the case that all the Z$_2$ parameters vanish, i.e. $q_2=\lambda_2=M_2=c=0$, this goes back to the same result as the holographic Weyl semimetal system \cite{Landsteiner:2015lsa}.

\vspace{0cm}
\begin{figure}[h!]
  \centering
\includegraphics[width=0.55\textwidth]{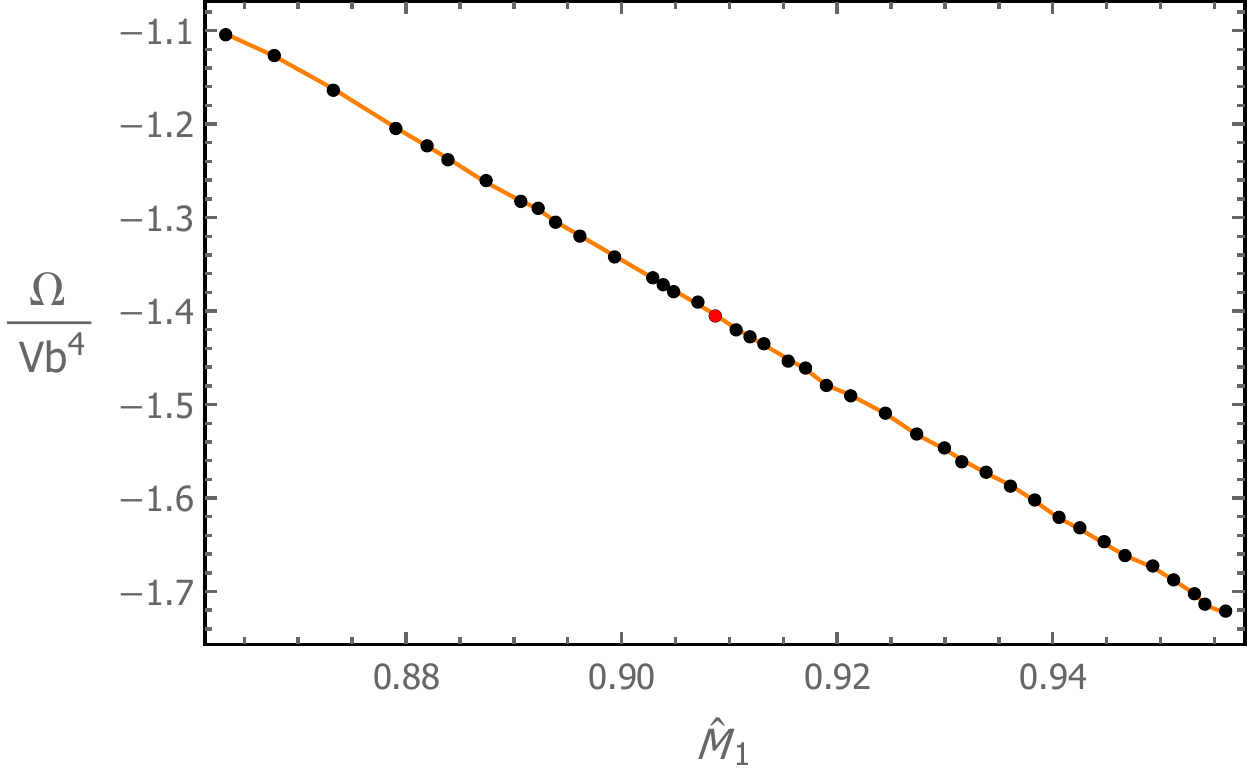}
\vspace{-0.1cm}
\caption{\small The free energy density as a function of $M_1/b=M_2/c$ at $c/b=1$ across the critical point. The red dot is the value of free energy density at the critical red dot in Fig.~\ref{fig:phase22}.
We found the phase transition is continuous and smooth (up to our numerical accuracy).}
\label{fig:freeenergy}
\end{figure}

As an example, in Fig.~\ref{fig:freeenergy}, we show the free energy as a function of $M_1/b=M_2/c$ at $c/b=1$ close to the phase transition point. We find that though the solutions in the IR are not continuous, however, the result of the free energy indicates that the quantum phase transition is a continuous one. Similar feature of phase transition also appears in \cite{Landsteiner:2015pdh, Liu:2018bye, Liu:2020ymx}. We could expect that in general the system is smooth when it crosses the phase transition.

\subsection{Order parameters: $\sigma_{\text{AHE}}$, $\sigma_{\boldmath{Z}_2 \text{AHE}}$ and finite temperature solutions}
\label{tp}
{In this subsection, we will give the detailed calculations of the anomalous transports $\sigma_{\text{AHE}}$ and $\sigma_{\boldmath{Z}_2 \text{AHE}}$. Since the system undergoes a topological phase transition, we employ the anomalous transports as the order parameters \cite{Landsteiner:2015pdh}. We then calculate the anomalous transports at both zero temperature and finite temperature. The different behavior of the anomalous transports confirms the phase diagrams in previous sections.
To be general, we start with the finite temperature solution with the Ansatz as follows}
\begin{eqnarray}
\begin{split}
ds^2&=-u dt^2+v dx^2+f dy^2+h dz^2+\frac{dr^2}{u}\,,\\
A&=A_z dz\,,~~~ \Phi_1=\phi_1 \,,\\
 \hat{A}&=C_y dy \,,~~~\Phi_2=\phi_2\,,
 \end{split}
\end{eqnarray}
where all the fields are functions of the radial coordinate $r$. Compared to the zero temperature case (\ref{eq:ansatz}), we have introduced another field $v$ in the Ansatz. At $T=0$ we have $u=v$. We will calculate the anomalous Hall conductivities which serve as the order parameters of the phase transition.

The conductivities of a quantum many body system can be computed via the Kubo formula which is shown in \eqref{kubo}. In holography, the current-current retarded correlators can be computed by studying the fluctuations of the gauge fields dual to the currents around the background with in-falling boundary conditions near the horizon.
We consider the following six fluctuations to calculate the correlators: $\text{\ensuremath{\delta}}V_{x}=e^{-i\omega t}v_{x}\left(r\right)$, $\text{\ensuremath{\delta}}V_{y}=e^{-i\omega t}v_{y}\left(r\right)$, $\text{\ensuremath{\delta}}V_{z}=e^{-i\omega t}v_{z}\left(r\right)$, $\delta\hat{V}_{x}=e^{-i\omega t}\hat{v}_{x}\left(r\right)$, $\delta\hat{V}_{y}=e^{-i\omega t}\hat{v}_{y}\left(r\right)$ and $\delta\hat{V}_{z}=e^{-i\omega t}\hat{v}_{z}\left(r\right)$. The detailed calculations for the anomalous Hall transport coefficients can be found from appendix \ref{app:e}, and it turns out only the fluctuations $\ensuremath{\delta}V_{x}, \ensuremath{\delta}V_{y},\ensuremath{\delta}\hat{V}_{z}$ are crucial for the Hall conductivities.  For both nonzero and zero temperatures we have
\be
\sigma_{\text{AHE}}=8\alpha A_z\left(r_0\right)\,,~~~~~~\sigma_{\boldmath{Z}_2 \text{AHE}}=8\beta C_y\left(r_0\right)\,.
\ee
In the following we shall focus on the dimensionless normalized transport coefficients
\be\label{sigmaAZ}
\tilde{\sigma}_{A}=\frac{\sigma_{\text{AHE}}}{8\alpha b}\,,~~~~~~
\tilde{\sigma}_{Z}=\frac{\sigma_{\boldmath{Z}_2 \text{AHE}}}{8\beta c}\,.
\ee

\begin{figure}[h!]
  \centering
\includegraphics[width=0.55\textwidth]{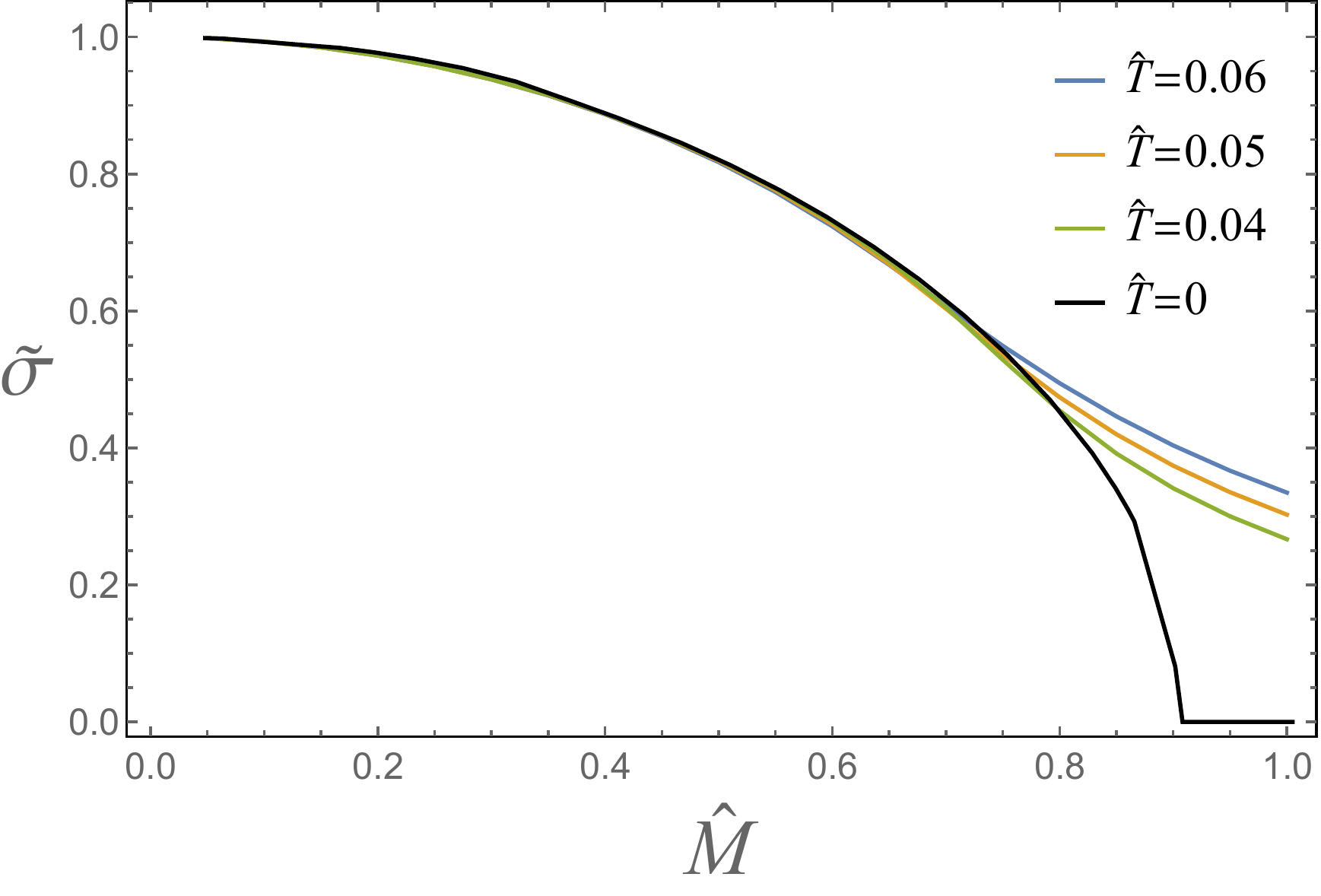}
\vspace{-0.1cm}
  \caption{\small For $\hat{M}_2=\hat{M}_1=\hat{M}$, the order parameter $\tilde{\sigma}$ (we use $\tilde{\sigma}$ as the axis label for simplification since $\tilde{\sigma}_{A}=\tilde{\sigma}_{Z}=\tilde{\sigma}$) as a function of $\hat{M}$ at $T/b=0.06,0.05,0.04$ and zero temperature with $c/b=1$. $q_1=q_2=1$ and $\lambda_1=\lambda_2=1/10$.}
  \label{fig:temperature}
\end{figure}

In Fig.~\ref{fig:temperature} we plot $\tilde{\sigma}_{A}$ as a function of $\hat{M}_1=\hat{M}_2$ with different temperatures when $c/b=1$. We have chosen $q_1=q_2=1$ and $\lambda_1=\lambda_2=1/10$. Note that in this case from the permutation symmetry,
$\tilde{\sigma}_{Z}$ has exactly the same behavior. From Fig.~\ref{fig:temperature}, we note that at zero temperature, both $\tilde{\sigma}_{A}$ and $\tilde{\sigma}_{Z}$ are nontrivial in the regime $M_1/b=M_2/c<0.908$ which indicates the existence of a topological phase transition from the phase with two pairs of Weyl/Z$_2$ nodes at the region $M_1/b=M_2/c<0.908$ to the fully gapped phase at the region $M_1/b=M_2/c>0.908$.
At finite temperature case, when we decrease the temperature, $\sigma_{\text{AHE}}$ approximates closer to zero in the trivial phase.

\vspace{0cm}

\begin{figure}[h!]
  \centering
\includegraphics[width=0.55\textwidth]{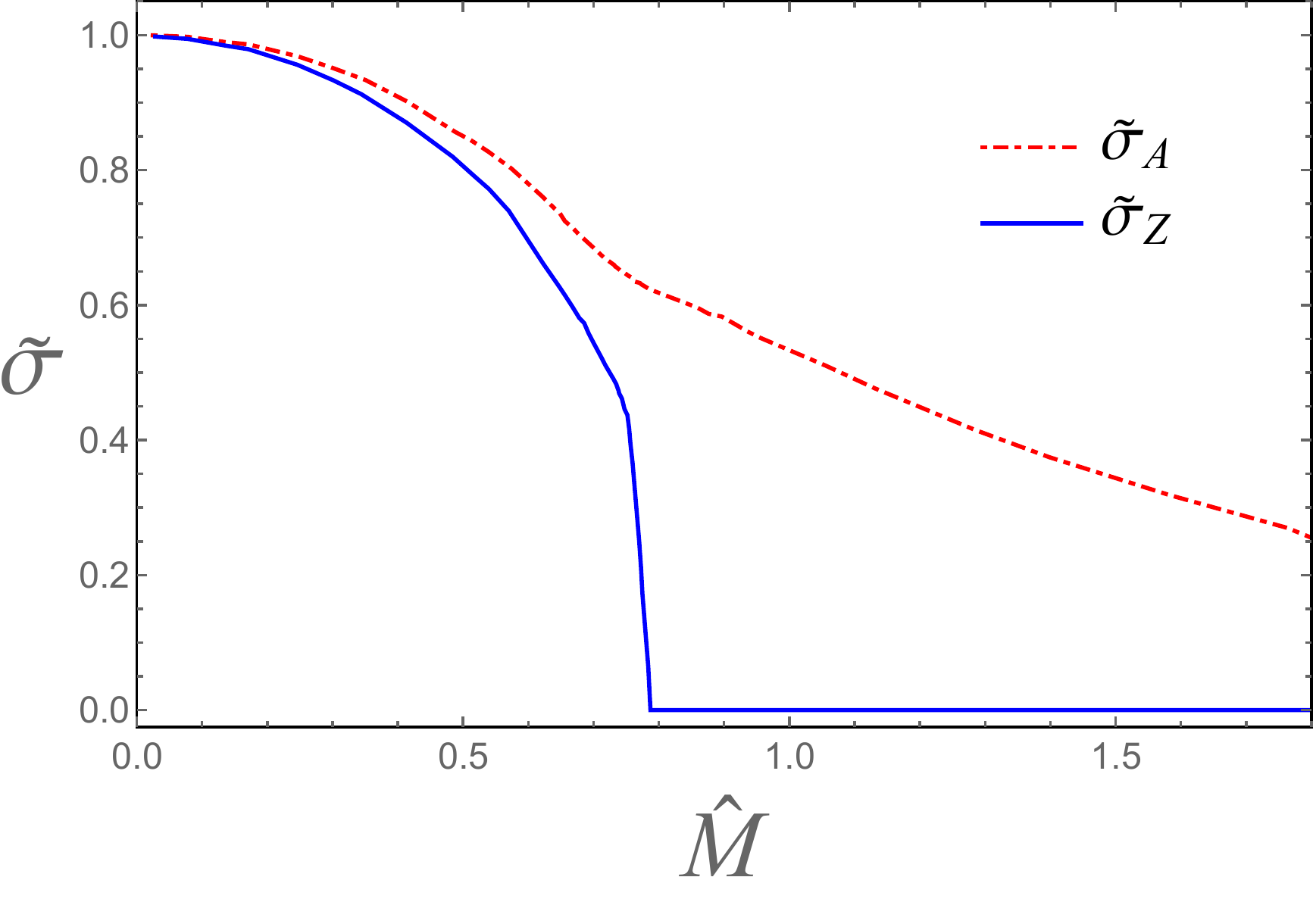}
\vspace{-0.2cm}
  \caption{\small  The anomalous transport as a function of $\hat{M}$ $\equiv \hat{M}_1=\hat{M}_2$, i.e. along the diagonal line in the phase diagram from left bottom to right top, at zero temperature with $c/b=2$. $q_1=q_2=1, \lambda_1=\lambda_2=1/10.$ }
  \label{fig:different}
\end{figure}

Next, we will comment on the effect of $c/b$ on
$\sigma_{\text{AHE}}$ and $\sigma_{\boldmath{Z}_2 \text{AHE}}$.  In the weakly coupled case we have $\tilde{\sigma}_{A}
\propto\sqrt{1-
\frac{M_1^{2}}{b^2}}$ and $\tilde{\sigma}_{Z}
\propto\sqrt{1-\frac{M_{2}^{2}}{c^2}}$ in the Weyl-Z$_2$ semimetal phase. If we fix $\hat{M}_1=\hat{M}_2$ but with $c/b\neq1$, we will have different behaviors of the anomalous transport, i.e., if $c/b>1$ at $\hat{M}_1=\hat{M}_2$, we have $\tilde{\sigma}_{A}>\tilde{\sigma}_{Z}$. We will study the effect of $c/b$ on $\sigma_\text{AHE}$ and $\sigma_{\boldmath{Z}_2 \text{AHE}}$ in holography.
In Fig.~\ref{fig:different} we plotted $\sigma_{\text{AHE}}$ and $\sigma_{\boldmath{Z}_2 \text{AHE}}$ at zero temperature but with fixed $c/b=2$. Here we fix $q_1=q_2=1$ and $\lambda_1=\lambda_2=1/10$
and in this case the critical values for the critical phase are $(M_1/b)_c=1.0044$ and $(M_2/c)_c=0.8414$, respectively. From the plot in Fig.~\ref{fig:different}, we note that in both phases, we have $\tilde{\sigma}_{A}>\tilde{\sigma}_{Z}$, which is qualitatively the same as the behavior in the weak coupling regime.
Compared to Fig.~\ref{fig:temperature},
these transport behaviors indicate that the system is in the phase where one pair of Weyl/Z$_2$ nodes becomes a gap while the other pair of nodes still exists, i.e. the left-up part in the right plot of  Fig.~\ref{fig:phase23}. We also note that at the interval $\hat{M}_1=\hat{M}_2>0.7872$, the nonzero anomalous Hall conductivity slowly decreases while $\sigma_{\boldmath{Z}_2 \text{AHE}}$ already becomes zero at this interval. This behavior is similar to the weakly coupled result. Note that different from the weakly couple case, the critical value of $\hat{M}_2$ here is smaller than that of the critical-critical point. This could naturally be seen from the phase diagram Fig.~\ref{fig:phase23} where the blue dotted phase transition line is curved instead of being straight as in weakly coupled case.

\vspace{0cm}

\begin{figure}[h!]
  \centering
\includegraphics[width=0.48\textwidth]{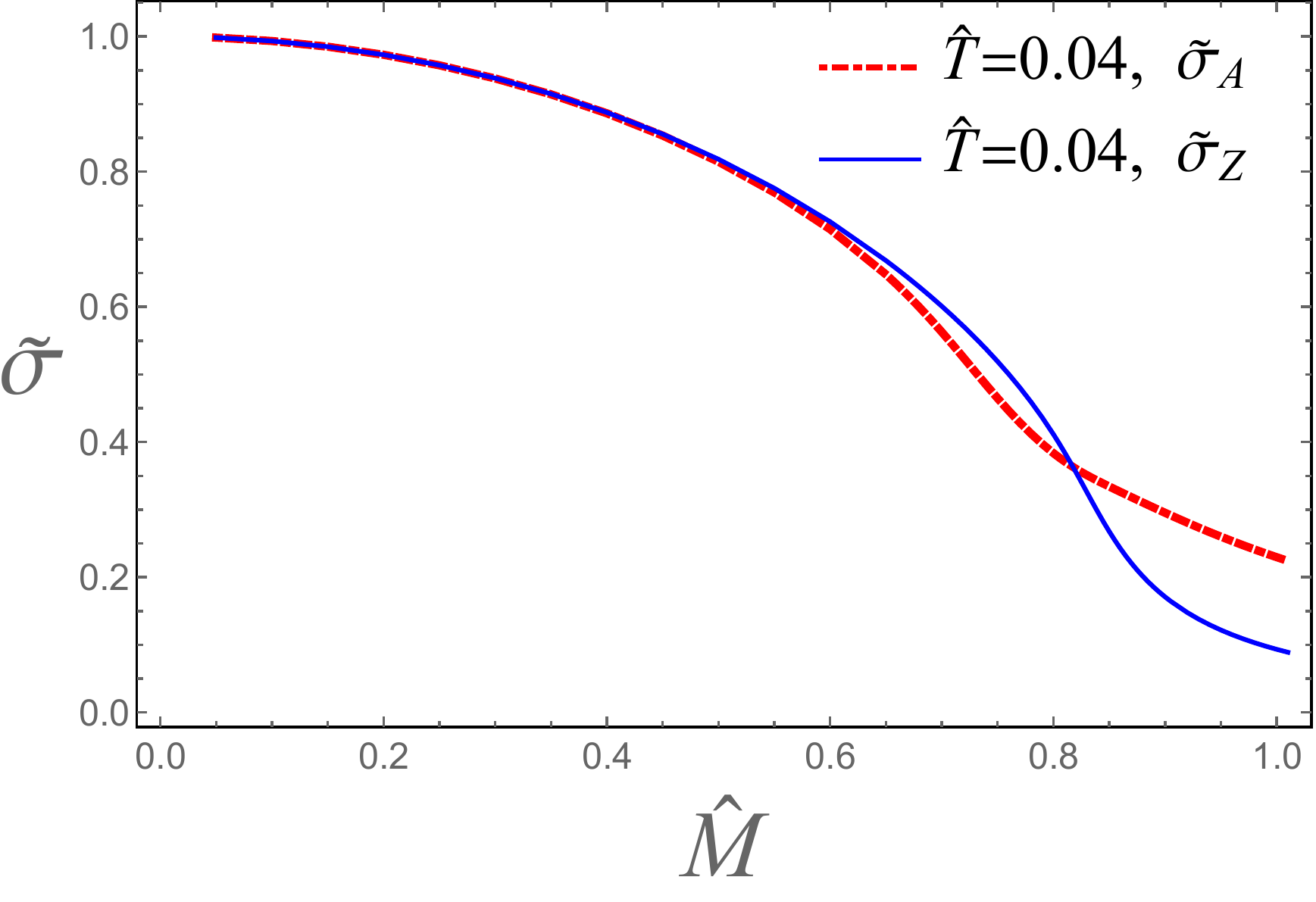}\quad
\vspace{-0.2cm}
   \includegraphics[width=0.48\textwidth]{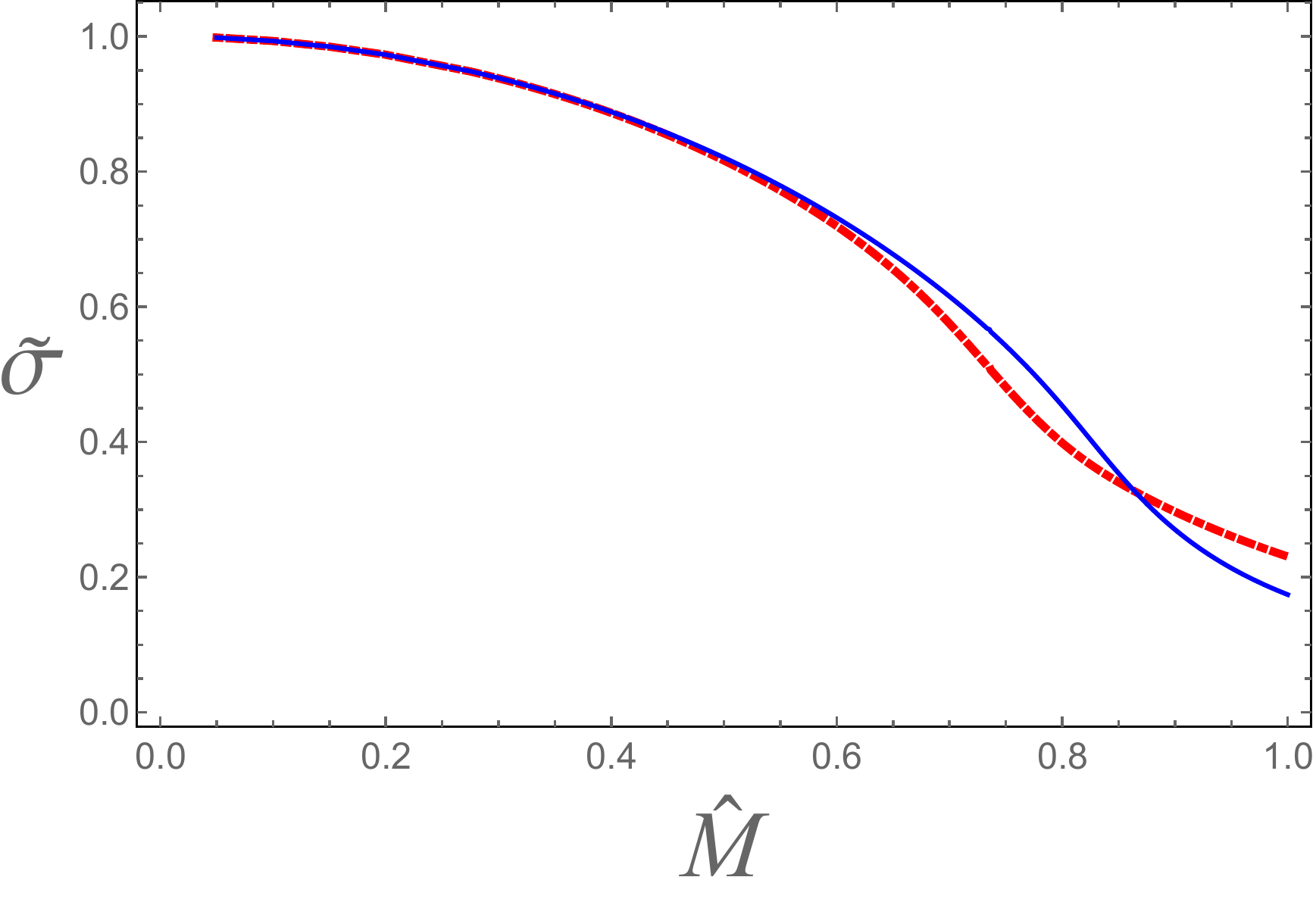}

   \caption{The transport parameters as a function of $\hat{M}$ ( $\equiv \hat{M}_1=\hat{M}_2$, i.e. the diagonal line in the phase diagram.) at $c/b=1, T/b=0.04$, with $q_1=1, \lambda_1=1/10, q_2=3/2, \lambda_2=1/2$ ({\em left}),
   and with $q_1=1, \lambda_1=1/10, q_2=13/10, \lambda_2=45/100$ ({\em right}). }
  \label{fig:different1}
\end{figure}

We will shortly comment on the effect of the couplings $q_1, q_2, \lambda_1, \lambda_2$ on
$\sigma_{\text{AHE}}$ and $\sigma_{\boldmath{Z}_2 \text{AHE}}$.  As is mentioned in \cite{Landsteiner:2016stv,Ji:2019pxx}, at very low temperature, the temperature scaling behavior of anomaly transports in the quantum critical region can be determined from the anisotropic scaling exponents $\alpha_1$ and $\alpha_2$. We also note that the scaling exponents are fully determined by the parameters $q_1,q_2,\lambda_1,\lambda_2$. Hence, if we change these parameters, we will have different locations of the critical values and different transport behaviors. Different from the weakly coupled case, these parameters are taken to be free parameters in holography. Previous studies on the effect of $q, \lambda$ to the critical value of phase transition have been done in \cite{Landsteiner:2015pdh, Copetti:2016ewq}, and here we focus on their effect on Hall conductivities. For simplicity, we choose two groups of parameters where $q_1=1,\lambda_1=1/10$, $q_2=3/2,\lambda_2=1/2$ and $q_1=1$, $\lambda_1=1/10$, $q_2=13/10, \lambda_2=45/100$ separately. With these two group of parameters we calculate the two anomalous transport coefficients at $T=0.04$.
The corresponding results are shown in Fig.~\ref{fig:different1}. Numerically it is not difficult to get a finite temperature solutions for this case, while the zero temperature solution is extremely difficult with $q_1\neq q_2$ and $\lambda_1\neq \lambda_2$.
Nevertheless we could estimate the location of the critical point $\hat{M}_{1,2c}$ in the following way.

Firstly, we calculate the critical values of $\hat{M}_c$ at $q_1=q_2=3/2$ and $\lambda_1=\lambda_2=1/2$ from the zero temperature solution in  \eqref{socritical}, which are obtained to be $(\hat{M}_1)_c=(\hat{M}_2)_c=0.837$.
For the case where $q_1=q_2=13/10$ and $\lambda_1=\lambda_2=45/100$, we find that the critical values are $(\hat{M}_1)_c=(\hat{M}_2)_c=0.881$. Remember that for $q_1=q_2=1$ and $\lambda_1=\lambda_2=1/10$ we have $(\hat{M}_1)_c=(\hat{M}_2)_c=0.908$.
As shown in \cite{Landsteiner:2015pdh, Copetti:2016ewq}, with a relatively larger $q$, the critical value $\hat{M}_c$ becomes smaller and the effect of $q$ is more important than the effect of $\lambda$.  Thus it is naturally expected 
that $0.837<(\hat{M}_2)_c<(\hat{M}_1)_c<0.908$ at $q_1=1,\lambda_1=1/10$, $q_2=3/2,\lambda_2=1/2$ and $0.881<(\hat{M}_2)_c<(\hat{M}_1)_c<0.908$ $q_1=1,\lambda_1=1/10$, $q_2=13/10,\lambda_2=45/100$.
Qualitatively, we can arrive at the conclusion by calculating the solution in \eqref{fwp} which should approach the critical solution in certain limit.
From Fig.~\ref{fig:different1}, we note that at very low temperature the two transports $\tilde{\sigma}_{A}, \tilde{\sigma}_{Z}$ cross and
for large $\hat{M}$ we have $\tilde{\sigma}_{A}>\tilde{\sigma}_{Z}$.
We will show that this is another numerical evidence for the conjecture about the temperature dependence of the anomalous transports found in \cite{Landsteiner:2015pdh, Ji:2019pxx}.
From the conjecture, the anomalous Hall conductivities are temperature dependent, which are given by $\tilde{\sigma}_{A}\propto T^{\alpha_1}$, and $\tilde{\sigma}_{Z}\propto T^{\alpha_2}$.
If
$\alpha_1<\alpha_2$, then we may deduce that $\tilde{\sigma}_{A}>\tilde{\sigma}_{Z}$ near the critical region as Fig.~\ref{fig:different1} shows.
Actually, the value of $\alpha_1$ and the anisotropic scaling exponent $\alpha_2$ for all of the parameters mentioned above can be calculated and we have checked that for all these two groups of parameters $\alpha_2$ is indeed larger than $\alpha_1$.
Strictly speaking, a more general zero temperature solution \eqref{eq:nhcri} of the critical region is needed where we can realize the different anisotropic scaling exponents (also the location of the critical values) caused by different $q_1,q_2$ and $\lambda_1,\lambda_2$. A comparison is also needed with the zero temperature weak coupling field theory, which will be left for future work.
These results show that the difference between the $\tilde{\sigma}_{A}$ and $\tilde{\sigma}_{Z}$ suffers form the contributions both of the mass parameters $M_{1,2}$ and the relative distance $c/b$ of the nodes.
It is interesting to see whether these conclusions from holographic calculations can be verified by further experiments.

\vspace{0cm}

\section{Conclusion and discussion}
\label{sec:cd}

In summary, we have studied two effective field theories for Weyl-Z$_2$ semimetal which carry different types of topological charges. They share similar phase diagrams but different features.
We have also built
a holographic model for a Weyl-Z$_2$ semimetal with two pairs of Weyl/Z$_2$ nodes, carrying Weyl and Z$_2$ topological charges. The dual system has both chiral anomaly and $\boldmath{Z}_2$ anomaly. The novel $\boldmath{Z}_2$ anomaly in the system indicates the non-conservation of the $\boldmath{Z}_2$ charge in the presence of an external electric field, which is quite similar to the chiral anomaly. The holographic model shares both features of the two weakly coupled field theory models.

We found that the system can also have topological phase transitions between topologically nontrivial phases and trivial phases with the order parameters being the anomalous Hall conductivity and the $\boldmath{Z}_2$  anomalous Hall conductivity. In the complete phase diagram, there are nine phases/phase transition lines/critical points and we have obtained nine solutions at zero temperature in the holographic model to correspond to these nine cases. We calculated the anomalous Hall conductivity and the $\boldmath{Z}_2$ anomalous Hall conductivity in the holographic model and found that these two transport parameters are non-vanishing in the topological phases and both become zero at the trivial phase at zero temperature, which indicates the existence of the topological phase transitions.

We also show the different behavior of  $\sigma_{\text{AHE}}$ and $\sigma_{\boldmath{Z}_2 \text{AHE}}$ and find that both of the mass parameters and the relative distance of the nodes will contribute to the difference. This can lead to observable effects in the measure of the magnetoconductivity and may be checked in experiments in the future.

There are still some open questions that we would like to leave for future work. First, the topological invariants in the holographic construction could in principle be calculated using the method of \cite{Liu:2018djq} with appropriate bulk action for the probe fermions. With this calculation one could identify the topological charges of the nodes in the dual theory and make precise connection to the weakly coupled theory. Second, we would like to mention that in the process of our calculation of the anomalous transport in subsection \ref{tp}, $\hat{M}_1=\hat{M}_2$ is chosen for convenience. From the phase diagram we know that $\hat{M}_1$ can take different values from $\hat{M}_2$, so this may lead to a more interesting picture and we leave this in a future work. Finally, we could also try to work out the topological structure and the complete phase diagram for the more general case with an extra SU(2) non-Abelian gauge field with different types of components turned on.

\subsection*{Acknowledgments}
We thank K. Landsteiner and F. Pena-Benitez for discussions.
This work was supported by the National Key R\&D Program of China (Grant No. 2018FYA0305800),
National Natural Science Foundation of China (Grant Nos. 11875083, 12005255, 12035016),
the Strategic Priority Research Program of Chinese Academy of Sciences (Grant No. XDB2800000),
the Key Research Program of Chinese Academy of Sciences (Grant Nos. XDPB08-1, XDPB15).
The work of Y.W.S. has also been partly supported by starting grants from University of Chinese Academy of Sciences and Chinese Academy of Sciences.


\begin{appendix}

\section{The Ward identity from the field theory}
\label{app:a}
In this appendix, we compute the ward identity of the theory (\ref{eq:Lagrangian}). Here we take the mass term as zero for simplicity since the contribution of the mass term can be added directly after we obtain the desired expression.
The massless action is
\begin{eqnarray}
\label{eq:massless}
  S_\text{eff}=\int d^{4}x\Psi^{\dagger}\left(x\right)\Gamma^{0}\Gamma^{\mu}\left[\left(i\partial_{\mu}-e\left(A_{\mu}+S_{z}\hat{A_{\mu}}\right)-\Gamma^{5}b_{\mu}{\mI}_{1}-S_z \hat{\Gamma}^{5}c_{\mu}{\mI}_{2}\right)\right]\Psi\left(x\right)
\end{eqnarray}
where `Gamma' matrices are defined in (\ref{eq:88marix}).
Here, $S_z={\Gamma^{\mu}}^{-1}{\Gamma^{0}}^{-1}\hat{\Gamma}^{0}\hat{\Gamma}^{\mu}$  which is just $\mathbb{I}_{8\times8}$, we drop this in the following.
Under the chiral and $Z_2$ transformation
\begin{eqnarray}
\begin{split}
 \Psi\rightarrow e^{-i\alpha_{1}\theta_{1}\left(x\right)\Gamma^{5}{\mI}_{1}/2-i\alpha_{2}\theta_{2}\left(x\right)\hat{\Gamma}^{5}{\mI}_{2}/2}\Psi\,,\\ \bar{\Psi}\rightarrow\bar{\Psi}e^{-i\alpha_{1}\theta_{1}\left(x\right)\Gamma^{5}{\mI}_{1}/2-i\alpha_{2}\theta_{2}\left(x\right)\hat{\Gamma}^{5}{\mI}_{2}/2}\,,
 \end{split}
\end{eqnarray}
where we follow the notations in Fujikawa's \cite{Fujikawa,Jang} that $\alpha_1\equiv ds_1$,~$\alpha_2\equiv ds_2$, where $s_1(s_2)\in \left[0,1\right]$ to parameterize the infinite sequence of the chiral($\boldmath{Z}_2$) gauge transformations. The theta functions are defined as
$\theta_{1}\left(x\right)=2b^{\mu}x_{\mu}$ and
$\theta_{2}\left(x\right)=2c^{\mu}x_{\mu}$\,.

Then the action becomes
\begin{eqnarray}
S_\text{eff}=\int d^{4}x\,\Psi^{\dagger}\left(x\right)\Gamma^{0} \Gamma^{\mu}\left[i\partial_{\mu}-e\left(A_{\mu}+\hat{A_{\mu}}\right)-\Gamma^{5}{\mI}_{1}\left(1-s_{1}\right)b_{\mu}-\hat{\Gamma}^{5}{\mI}_{2}\left(1-s_{2}\right)c_{\mu}\right] \Psi\left(x\right)\,.\nonumber
\end{eqnarray}
We can define $D\!\!\!\!/=\Gamma^{\mu}\left[\partial_{\mu}+ie\left(A_{\mu}+\hat{A_{\mu}}\right)+i\left(1-s_{1}\right)b_{\mu}\Gamma^{5}{\mI}_{1}+i\left(1-s_{2}\right)c_{\mu}\hat{\Gamma}^{5}{\mI}_{2}\right]$, and assume $\phi_{n}\left(x\right)$ satisfies $D\!\!\!\!/\phi_{n}\left(x\right)=\epsilon_{n}\phi_{n}\left(x\right)$.
We expand the Grassmann variables as $\Psi\left(x\right)=\underset{n}{\sum}c_{n}\phi_{n}\left(x\right)$, $\bar{\Psi}\left(x\right)=\underset{n}{\sum}\phi_{n}^{*}\left(x\right)\bar{c}_{n}$ where ${c}_{n}$ and $\bar{c}_{n}$ are new Grassmann variables. Then the infinitesimal transformation operator can be defined as
\begin{eqnarray}
 U_{nm}=\delta_{nm}-ds\text{\ensuremath{\frac{i}{2}}}\int d^{4}x\,\phi_{n}^{*}\left(x\right)\left(\theta_{1}\left(x\right)\Gamma^{5}{\mI}_{1}+\theta_{2}\left(x\right)\hat{\Gamma}^{5}{\mI}_{2}\right)\phi_{n}\left(x\right).
\end{eqnarray}
Note that we have defined $s_1=s_2=s$ without loss of generality.

The path integral Jacobian can be obtained immediately,
\begin{eqnarray}
J=\text{det}\left(U^{-2}\right)=e^{-ids\int d^{4}x\underset{n}{\sum}\phi_{n}^{*}\left(x\right)\left(\theta_{1}\left(x\right)\Gamma^{5}{\mI}_{1}+\theta_{2}\left(x\right)\hat{\Gamma}^{5}{\mI}_{2}\right)\phi_{n}\left(x\right)}\,.
\end{eqnarray}

Consider the quantity appearing in the exponential and we define the following
\begin{eqnarray}
I_{1}\left(x\right)=\underset{n}{\sum}\phi_{n}^{*}\left(x\right)\Gamma^{5}{\mI}_{1}\phi_{n}\left(x\right)\,,~~~
I_{2}\left(x\right)=\underset{n}{\sum}\phi_{n}^{*}\left(x\right)\hat{\Gamma}^{5}{\mI}_{2}\phi_{n}\left(x\right).
\end{eqnarray}
With the standard method of heat kernel regularization, we have
\begin{eqnarray}
\begin{split}
I_{1}\left(x\right)&=\underset{\text{\ensuremath{\Lambda\to\infty}}}{\lim}\underset{n}{\sum}\phi_{n}^{*}\left(x\right)\Gamma^{5}{\mI}_{1}e^{-\frac{\epsilon_{n}^{2}}{\Lambda^{2}}}\phi_{n}\left(x\right)=\underset{\text{\ensuremath{\Lambda\to\infty}}}{\lim}\underset{n}{\sum}\phi_{n}^{*}\left(x\right)\Gamma^{5}{\mI}_{1}e^{-\frac{D\!\!\!\!/^{2}}{\Lambda^{2}}}\phi_{n}\left(x\right),\\
I_{2}\left(x\right)&=\underset{\text{\ensuremath{\Lambda\to\infty}}}{\lim}\underset{n}{\sum}\phi_{n}^{*}\left(x\right)\hat{\Gamma}^{5}{\mI}_{2}e^{-\frac{\epsilon_{n}^{2}}{\Lambda^{2}}}\phi_{n}\left(x\right)=\underset{\text{\ensuremath{\Lambda\to\infty}}}{\lim}\underset{n}{\sum}\phi_{n}^{*}\left(x\right)\hat{\Gamma}^{5}{\mI}_{2}e^{-\frac{D\!\!\!\!/^{2}}{\Lambda^{2}}}\phi_{n}\left(x\right),
\end{split}
\end{eqnarray}
where
\begin{eqnarray}
D\!\!\!\!/^{2}&=&-D^{\mu}D_{\mu}-\left(1-s\right)^{2}\left(b_{\mu}b^{\mu}+c_{\mu}c^{\mu}\right)+\frac{i\left(1-s\right)}{2}\left[\Gamma^{\mu},\Gamma^{\nu}\right]\left(b_{\mu}D_{\nu}\Gamma^{5}{\mI}_{1}+c_{\mu}D_{\nu}\hat{\Gamma}^{5}{\mI}_{2}\right)\nonumber\\
&+&\frac{ie}{4}\left[\Gamma^{\mu},\Gamma^{\nu}\right]\left(F_{\mu\nu}+\hat{F}_{\mu\nu}\right)+\text{\ensuremath{\frac{i\left(1-s\right)}{4}}}\left[\Gamma^{\mu},\Gamma^{\nu}\right]\left(F_{\mu\nu}^{5}\Gamma^{5}{\mI}_{1}+\hat{F}_{\mu\nu}^{5}\hat{\Gamma}^{5}{\mI}_{2}\right).
\end{eqnarray}

Then we have
\begin{eqnarray}
\begin{split}
&I_1\left(x\right)=\underset{\text{\ensuremath{\Lambda\to\infty}}}{\lim}\int\frac{d^{4}k}{\left(2\pi\right)^{4}}e^{-\frac{k_{\mu}^{2}}{\Lambda^{2}}}\text{tr}\Gamma^{5}{\mI}_{1}\exp\left\{ \frac{\left(ik_{\mu}+D_{\mu}\right)^{2}}{\Lambda^{2}}+\frac{\left(1-s\right)^{2}\left(b_{\mu}b^{\mu}+c_{\mu}c^{\mu}\right)}{\text{\ensuremath{\Lambda^{2}}}}\right.\\
&~~~~~~~\left.-\frac{ie}{4\Lambda^{2}}\left[\Gamma^{\mu},\Gamma^{\nu}\right]\left(F_{\mu\nu}+\hat{F}_{\mu\nu}\right)-\text{\ensuremath{\frac{i\left(1-s\right)}{4\Lambda^{2}}}}\left[\Gamma^{\mu},\Gamma^{\nu}\right]\left(F_{\mu\nu}^{5}\Gamma^{5}{\mI}_{1}+\hat{F}_{\mu\nu}^{5}\hat{\Gamma}^{5}{\mI}_{2}\right)\right.\\
&~~~~~~~\left.+\frac{i\left(1-s\right)}{2\Lambda^{2}}\left[\Gamma^{\mu},\Gamma^{\nu}\right]\left[b_{\mu}\left(ik_{\nu}+D_{\nu}\right)\Gamma^{5}{\mI}_{1}+c_{\mu}\left(ik_{\nu}+D_{\nu}\right)\hat{\Gamma}^{5}{\mI}_{2}\right]\right\}, \\
&I_2\left(x\right)=\underset{\text{\ensuremath{\Lambda\to\infty}}}{\lim}\int\frac{d^{4}k}{\left(2\pi\right)^{4}}e^{-\frac{k_{\mu}^{2}}{\Lambda^{2}}}\text{tr}\hat{\Gamma}^{5}{\mI}_{2}\exp\left\{ \frac{\left(ik_{\mu}+D_{\mu}\right)^{2}}{\Lambda^{2}}+\frac{\left(1-s\right)^{2}\left(b_{\mu}b^{\mu}+c_{\mu}c^{\mu}\right)}{\text{\ensuremath{\Lambda^{2}}}}\right.\\
&~~~~~~~\left.-\frac{ie}{4\Lambda^{2}}\left[\Gamma^{\mu},\Gamma^{\nu}\right]\left(F_{\mu\nu}\hat{F}_{\mu\nu}\right)-\text{\ensuremath{\frac{i\left(1-s\right)}{4\Lambda^{2}}}}\left[\Gamma^{\mu},\Gamma^{\nu}\right]\left(F_{\mu\nu}^{5}\Gamma^{5}{\mI}_{1}+\hat{F}_{\mu\nu}^{5}\hat{\Gamma}^{5}{\mI}_{2}\right)\right.\\
&~~~~~~~\left.+\frac{i\left(1-s\right)}{2\Lambda^{2}}\left[\Gamma^{\mu},\Gamma^{\nu}\right]\left[b_{\mu}\left(ik_{\nu}+D_{\nu}\right)\Gamma^{5}{\mI}_{1}+c_{\mu}\left(ik_{\nu}+D_{\nu}\right)\hat{\Gamma}^{5}{\mI}_{2}\right]\right\}\,.
\end{split}
\end{eqnarray}
When $\Lambda\to\infty$, only $O\left(\frac{1}{\Lambda^4}\right)$ or higher order can contribute to $I_1\left(x\right)$ and $I_2\left(x\right)$, therefore only the term $\left\{ \left[\Gamma^{\mu},\Gamma^{\nu}\right]\left(F_{\mu\nu}+\hat{F}_{\mu\nu}\right)\right\} ^{2}$ makes sense.

Based on $\text{tr}\left(\Gamma^{5}{\mI}_{1}\Gamma^{\mu}\Gamma^{\nu}\Gamma^{\alpha}\Gamma^{\beta}\right)=16i\epsilon^{\mu\nu\alpha\beta}$,~$\text{tr}\left(\hat{\Gamma}^{5}{\mI}_{2}\Gamma^{\mu}\Gamma^{\nu}\Gamma^{\alpha}\Gamma^{\beta}\right)=0$ and $\text{tr}\left(\hat{\Gamma}^{5}{\mI}_{2}\hat{\Gamma}^{\mu}\hat{\Gamma}^{\nu}\hat{\Gamma}^{\alpha}\hat{\Gamma}^{\beta}\right)=16i\epsilon^{\mu\nu\alpha\beta}$ , we have
\begin{eqnarray}
&&I\left(x\right)=I_1\left(x\right)+I_2\left(x\right)
=\frac{1}{32\pi^{2}}\epsilon^{\mu\nu\alpha\beta}\left[F_{\mu\nu}F_{\alpha\beta}+\hat{F}_{\mu\nu}\hat{F}_{\alpha\beta}
+F_{\mu\nu}\hat{F}_{\alpha\beta}+\hat{F}_{\mu\nu}F_{\alpha\beta}\right].
\end{eqnarray}
Integrate the fermionic part in the action, we have
\begin{eqnarray}
S_{\theta}&=&i\int_{0}^{1}ds\int d^{4}x \left[\theta_{1}\left(x\right)I_1\left(x\right)+\theta_{2}\left(x\right)I_2\left(x\right)\right]\nonumber\\
&=&\frac{ie}{32\pi^{2}}\int d^{4}x\left[\epsilon^{\mu\nu\alpha\beta}\theta_{1}\left(x\right)\left(F_{\mu\nu}F_{\alpha\beta}+\hat{F}_{\mu\nu}\hat{F}_{\alpha\beta}\right)+\theta_{2}\left(x\right)\left(F_{\mu\nu}\hat{F}_{\alpha\beta}+\hat{F}_{\mu\nu}F_{\alpha\beta}\right)
\right].
\end{eqnarray}
Integrating by parts and eliminating a total derivative term, we have $S_{\theta}$ in the Chern-Simons form
\begin{eqnarray}
S_{\theta}
=-\frac{e}{4\pi^{2}}\int dtd\mathbf{r}b_{\mu}\epsilon^{\mu\nu\alpha\beta}\left(A_{\nu}\partial_{\alpha}A_{\beta}+\hat{A}_{\nu}\partial_{\alpha}\hat{A}_{\beta}\right)-\frac{e}{4\pi^{2}}\int dtd\mathbf{r}c_{\mu}\epsilon^{\mu\nu\alpha\beta}\left(A_{\nu}\partial_{\alpha}\hat{A}_{\beta}+\hat{A}_{\nu}\partial_{\alpha}A_{\beta}\right),\nonumber
\end{eqnarray}
then the functional derivative of $S_\theta$ with respect to the gauge field $A_{\mu}$ would give the charge current $\mathcal{J}^{\mu}$, while the functional derivative with respect to the gauge field $\hat{A}_{\mu}$ gives the spin current $\mathcal{\hat{J}}^{\mu}$. They are
\begin{eqnarray}
\begin{split}
\mathcal{J}^{\mu}&=-\frac{\delta S}{\delta A_{\nu}}=\frac{e^{2}}{2\pi^{2}}b_{\nu}\epsilon^{\mu\nu\alpha\beta}\partial_{\alpha}A_{\beta}+\frac{e^{2}}{2\pi^{2}}c_{\nu}\epsilon^{\mu\nu\alpha\beta}\partial_{\alpha}\hat{A}_{\beta}\,,\\
\mathcal{\hat{J}}^{\mu}&=-\frac{\delta S}{\delta\hat{A}_{\nu}}=\frac{e^{2}}{2\pi^{2}}b_{\nu}\epsilon^{\mu\nu\alpha\beta}\partial_{\alpha}\hat{A}_{\beta}+\frac{e^{2}}{2\pi^{2}}c_{\nu}\epsilon^{\mu\nu\alpha\beta}\partial_{\alpha}A_{\beta}\,,
\end{split}
\end{eqnarray}
then we have
\begin{eqnarray}
\partial_{\mu}\mathcal{J}^{\mu}=\frac{e^{2}}{32\pi^{2}}\epsilon^{\mu\nu\alpha\beta}\left(F_{\mu\nu}F_{\alpha\beta}^{5}+\hat{F}_{\mu\nu}\hat{F}_{\alpha\beta}^{5}\right)\,,~~~
\partial_{\mu}\mathcal{\hat{J}}^{\mu}=\frac{e^{2}}{32\pi^{2}}\epsilon^{\mu\nu\alpha\beta}\left(F_{\mu\nu}\hat{F}_{\alpha\beta}^{5}+\hat{F}_{\mu\nu}F_{\alpha\beta}^{5}\right).
\label{eq:current}
\end{eqnarray}

To obtain the Ward identity, we define vector currents via
\begin{eqnarray}
\mathcal{J}^{\mu}=\mathcal{J}_{L}^{\mu}+\mathcal{J}_{R}^{\mu}\,,~~~~
\mathcal{\hat{J}}^{\mu}=\mathcal{\hat{J}}_{L}^{\mu}+\mathcal{\hat{J}}_{R}^{\mu},
\end{eqnarray}
then we can obtain the derivative of the currents as
\begin{eqnarray}
\begin{split}
\partial_{\mu}\mathcal{J}_{L,R}^{\mu}&=\pm\frac{e^{2}}{32\pi^{2}}\epsilon^{\mu\nu\alpha\beta}\left(F_{\mu\nu}^{L,R}F_{\alpha\beta}^{L,R}+\hat{F}_{\mu\nu}^{L,R}\hat{F}_{\alpha\beta}^{L,R}\right), \\
\partial_{\mu}\mathcal{\hat{J}}_{L,R}^{\mu}&=\pm\frac{e^{2}}{16\pi^{2}}\epsilon^{\mu\nu\alpha\beta}F_{\mu\nu}^{L,R}\hat{F}_{\alpha\beta}^{L,R}\,.
\end{split}
\end{eqnarray}
Now we define axial currents via
\begin{eqnarray}
\mathcal{J}_{5}^{\mu}=\mathcal{J}_{L}^{\mu}-\mathcal{J}_{R}^{\mu}\,,~~~
\mathcal{\hat{J}}_{5}^{\mu} =\mathcal{\hat{J}}_{L}^{\mu}-\mathcal{\hat{J}}_{R}^{\mu},
\end{eqnarray}
and a basis of vector-like and axial gauge fields
\begin{eqnarray}
\begin{split}
A_{\mu}&=\frac{1}{2}\left(A_{\mu}^{L}+A_{\mu}^{R}\right)\,,~~~
A_{\mu}^{5}=\frac{1}{2}\left(A_{\mu}^{L}-A_{\mu}^{R}\right),\\
\hat{A}_{\mu}&=\frac{1}{2}\left(\hat{A}_{\mu}^{L}+\hat{A}_{\mu}^{R}\right)\,,~~~
\hat{A}_{\mu}^{5}=\frac{1}{2}\left(\hat{A}_{\mu}^{L}-\hat{A}_{\mu}^{R}\right).
\end{split}
\end{eqnarray}
Thus, we have the Ward identities as
\begin{eqnarray}
\begin{split}
\partial_{\mu}\mathcal{J}_{5}^{\mu}&=&\frac{e^{2}}{16\pi^{2}}\epsilon^{\mu\nu\alpha\beta}\left(F_{\mu\nu}F_{\alpha\beta}+F_{\mu\nu}^{5}F_{\alpha\beta}^{5}+\hat{F}_{\mu\nu}\hat{F}_{\alpha\beta}+\hat{F}_{\mu\nu}^{5}\hat{F}_{\alpha\beta}^{5}\right),\\
\partial_{\mu}\mathcal{\hat{J}}_{5}^{\mu}&=&\frac{e^{2}}{16\pi^{2}}\epsilon^{\mu\nu\alpha\beta}\left(F_{\mu\nu}\hat{F}_{\alpha\beta}+F_{\mu\nu}^{5}\hat{F}_{\alpha\beta}^{5}+\hat{F}_{\mu\nu}F_{\alpha\beta}+\hat{F}_{\mu\nu}^{5}F_{\alpha\beta}^{5}\right).
\end{split}
\end{eqnarray}

We note that in (\ref{eq:current}), the vector currents are not conserved. This is because we should write the currents in the form of the consistent current\cite{Landsteiner:2016led} by adding the local counterterms which is the so-called called Bardeen counterterms to our action\cite{Bardeen}. The form of the Bardeen counterterm is
\begin{eqnarray}
\begin{split}
\int d^{4}x\epsilon^{\mu\nu\alpha\beta}\left[A_{\mu}b_{\nu}\left(c_{11}F_{\alpha\beta}+c_{12}F_{\alpha\beta}^{5}+c_{13}\hat{F}_{\alpha\beta}+c_{14}\hat{F}_{\alpha\beta}^{5}\right)
\right. \nonumber \\
\left.  \qquad\qquad
+\hat{A}_{\mu}c_{\nu}\left(c_{21}F_{\alpha\beta}+c_{22}F_{\alpha\beta}^{5}+c_{23}\hat{F}_{\alpha\beta}+c_{24}\hat{F}_{\alpha\beta}^{5}\right)\right].
\end{split}
\end{eqnarray}
If we take $c_{11}=c_{13}=c_{21}=c_{23}=\frac{1}{4}$ and $c_{12}=c_{14}=c_{22}=c_{24}=0$, then we have
\begin{eqnarray}
\begin{split}
\partial_{\mu}\mathcal{J}^{\mu}&=0,\\
\partial_{\mu}\mathcal{\hat{J}}^{\mu}&=0,\\
\partial_{\mu}\mathcal{J}_{5}^{\mu}&=\frac{e^{2}}{32\pi^{2}}\epsilon^{\mu\nu\alpha\beta}\left(3F_{\mu\nu}F_{\alpha\beta}+F_{\mu\nu}^{5}F_{\alpha\beta}^{5}+3\hat{F}_{\mu\nu}\hat{F}_{\alpha\beta}+\hat{F}_{\mu\nu}^{5}\hat{F}_{\alpha\beta}^{5}\right),\\
\partial_{\mu}\mathcal{\hat{J}}_{5}^{\mu}&=\frac{e^{2}}{32\pi^{2}}\epsilon^{\mu\nu\alpha\beta}\left(6F_{\mu\nu}\hat{F}_{\alpha\beta}+2F_{\mu\nu}^{5}\hat{F}_{\alpha\beta}^{5}\right).
\end{split}
\end{eqnarray}
With the mass terms included,
{
we can obtain the formula \eqref{eq:qftwi} in the main text. These Ward identities are very important, which are used in the main text and can also be obtained in our holography model.
}

\section{The equations of motion at zero temperature}
\label{app:b}
The equations of motion of the corresponding field are given as follows:
\begin{eqnarray}
\begin{split}
&u^{\prime\prime}+\frac{u^{\prime}}{2}\left(\frac{f^{\prime}}{f}+\frac{h^{\prime}}{h}+\frac{u^{\prime}}{u}\right)-\frac{u}{3}\left(\frac{A_{z}^{\prime2}}{h}+\frac{C_{y}^{\prime2}}{f}\right)-8+\frac{2}{3}m^{2}\Phi_{1}^{2}+\frac{\lambda_{1}}{3}\Phi_{1}^{4}+\frac{2}{3}m^{2}\Phi_{2}^{2}+\frac{\lambda_{2}}{3}\Phi_{2}^{4}=0\,,\\
&f^{\prime\prime}-\frac{f^{\prime}}{2}\left(\frac{f^{\prime}}{f}-\frac{h^{\prime}}{h}-\frac{3u^{\prime}}{u}\right)-\frac{2hC_{y}^{\prime2}+fA_{z}^{\prime2}}{3h}-\frac{f}{u}\left[8+\frac{2}{3}m^{2}\left(\Phi_{1}^{2}+\Phi_{2}^{2}\right)+\frac{\lambda_{1}}{3}\Phi_{1}^{4}+\frac{\lambda_{2}}{3}\Phi_{2}^{4}\right]=0\,,\\
&-\frac{A_{z}^{\prime2}}{4h}-\frac{C_{y}^{\prime2}}{4f}+\frac{f^{\prime}h^{\prime}}{4fh}+\frac{u^{\prime}}{2u}\left(\frac{f^{\prime}}{f}+\frac{h^{\prime}}{h}+\frac{u^{\prime}}{2u}\right)-\frac{1}{2}\left(\Phi_{1}^{\prime2}+\Phi_{2}^{\prime2}\right)-\frac{6}{u}+\frac{\lambda_{1}\Phi_{1}^{4}+\lambda_{2}\Phi_{2}^{4}}{4u}\\
&~~~~~~~~+\text{\ensuremath{\frac{1}{2u}}}\left[m^{2}\left(\Phi_{1}^{2}+\Phi_{2}^{2}\right)+\frac{q_1^{2}A_{z}^{2}\Phi_{1}^{2}}{h}+\frac{q_2^{2}C_{y}^{2}\Phi_{2}^{2}}{f}\right]=0\,,\\
&A_{z}^{\prime\prime}+A_{z}^{\prime}\left(\frac{3u^{\prime}}{2u}-\frac{h^{\prime}}{2h}+\frac{f^{\prime}}{2f}\right)-\frac{2q_1^{2}A_{z}\Phi_{1}^{2}}{u}=0\,,\\
&C_{y}^{\prime\prime}+C_{y}^{\prime}\left(\frac{3u^{\prime}}{2u}-\frac{f^{\prime}}{2f}+\frac{h^{\prime}}{2h}\right)-\frac{2q_2^{2}C_{y}\Phi_{2}^{2}}{u}=0\,, \\
&\Phi_{1}^{\prime\prime}+\Phi_{1}^{\prime}\left(\frac{3u^{\prime}}{2u}+\frac{h^{\prime}}{2h}+\frac{f^{\prime}}{2f}\right)-\frac{\Phi_{1}}{u}\left(m^{2}+\lambda_{1}\Phi_{1}^{2}+\frac{q_1^{2}A_{z}^{2}}{h}\right)=0\,,\\
&\Phi_{2}^{\prime\prime}+\Phi_{2}^{\prime}\left(\frac{3u^{\prime}}{2u}+\frac{h^{\prime}}{2h}+\frac{f^{\prime}}{2f}\right)-\frac{\Phi_{2}}{u}\left(m^{2}+\lambda_{2}\Phi_{2}^{2}+\frac{q_2^{2}C_{y}^{2}}{h}\right)=0\,,
\end{split}
\end{eqnarray}
where the prime is the derivative with respect to the radial coordinate $r$. We have seven independent ordinary differential equations for seven unknown fields.

\section{The free energy}
\label{app:c}

With the formula of the free energy of the system (\ref{free}), we need to expand the field at the UV. Close to the UV boundary, we can obtain the following behavior of the fields
\begin{eqnarray}
&&u=r^{2}-\frac{M_{1}^{2}+M_{2}^{2}}{3}+\frac{\log r}{18r^{2}}\left(2M_{1}^{4}+4M_{1}^{2}M_{2}^{2}+2M_{2}^{4}+3\lambda_{1}M_{1}^{4}+3\lambda_{2}M_{2}^{4}\right)+\frac{u_{1}}{r^{2}}+\ldots\nonumber\\
&&f=r^{2}-\frac{M_{1}^{2}+M_{2}^{2}}{3}+\frac{\log r}{18r^{2}}\left(2M_{1}^{4}+4M_{1}^{2}M_{2}^{2}+2M_{2}^{4}+3\lambda_{1}M_{1}^{4}+3\lambda_{2}M_{2}^{4}+9c^{2}q_2^{2}M_{2}^{2}\right)+\frac{f_{1}}{r^{2}}+\ldots\nonumber\\
&&h=r^{2}-\frac{M_{1}^{2}+M_{2}^{2}}{3}+\frac{\log r}{18r^{2}}\left(2M_{1}^{4}+4M_{1}^{2}M_{2}^{2}+2M_{2}^{4}+3\lambda_{1}M_{1}^{4}+3\lambda_{2}M_{2}^{4}+9b^{2}q_1^{2}M_{1}^{2}\right)+\frac{h_{1}}{r^{2}}+\ldots\nonumber\\
&&A_{z}=b-\frac{bM_{1}^{2}q_1^{2}\log r}{r^{2}}+\frac{\eta_{1}}{r^{2}}+\ldots\nonumber\\
&&C_{y}=c-\frac{cM_{2}^{2}q_2^{2}\log r}{r^{2}}+\frac{\eta_{2}}{r^{2}}+\ldots\nonumber\\
&&\Phi_{1}=\frac{M_{1}}{r}-\frac{\log r}{6r^{3}}\left(2M_{1}^{3}+2M_{1}M_{2}^{2}+3b^{2}q_1^{2}M_{1}^{2}+3\lambda_{1}M_{1}^{3}\right)+\frac{O_{1}}{r^{3}}+\ldots\nonumber\\
&&\Phi_{2}=\frac{M_{2}}{r}-\frac{\log r}{6r^{3}}\left(2M_{2}^{3}+2M_{2}M_{1}^{2}+3c^{2}q_2^{2}M_{2}^{2}+3\lambda_{2}M_{2}^{3}\right)+\frac{O_{2}}{r^{3}}+\ldots.
\end{eqnarray}
with $h_1=-2f_1+\frac{7M_{1}^{4}}{36}-M_{1}O_{1}+\frac{1}{8}b^{2}q_1^{2}M_{1}^{2}+\frac{\lambda_{1}M_{1}^{4}}{8}$ from the equation of motion.
We have a radially conserved quantity $\frac{uA_{z}A_{z}^{\prime}}{h}+\frac{uC_{y}C_{y}^{\prime}}{f}+\frac{uf^{\prime}}{f}+\frac{uh^{\prime}}{h}-2u^{\prime}=0$, which gives
$f_{1}=\frac{1}{72}(14M_{1}^{4}-72M_{1}O_{1}+18b^{2}q_1^{2}M_{1}^{2}+9c^{2}q_2^{2}M_{2}^{2}+36b\eta_{1}+36c\eta_{2}-144u_{1}+9\lambda_{1}M_{1}^{4})$. We also have another radially conserved relation  which gives
$f_1=\frac{1}{72}(-28M_1^2M_2^2-14M_2^4+72M_2O_2-9c^2M_2^2q_2^2+72u_1-9\lambda_2M_2^4)$.
 Note that one can determine the above expansions only
up to a shift $r\to r+a$. The two equations will lead to one identical relation, which will be used later.

Taking into account the boundary terms and performing a Wick rotation, the free energy
density can be obtained as
{\begin{eqnarray}\label{eC2}
\frac{\Omega}{V}=-\frac{1}{V}S_{ren}&=&-\frac{1}{36}\Big[3M_{2}\left(6c^{2}q_2^{2}M_{2}+7\text{\ensuremath{M_{2}^{3}}}-24O_{2}\right) \nonumber\\
&&+144f_{1}-7M_{1}^{4}+42M_{1}^{2}M_{2}^{2}+72M_{1}O_{1}+18\lambda_{2}M_{2}^{4}-36u_{1}\Big].
\end{eqnarray}
With the two relations of $f_1$ in the above paragraph, we can replace the cross term $M_1^2M_2^2$, and arrive at }
\begin{eqnarray}\label{eC3}
\frac{\Omega}{V}=\frac{1}{8}(8M_1O_1+8M_2O_2+2b^2M_1^2q_1^2+2c^2M_2^2q_2^2+4b\eta_1+4c\eta_2+\lambda_1M_1^4+\lambda_2M_2^4).
\end{eqnarray}

We can also check the relationship between the free energy and total energy density, this can be done as follows. The stress tensor for the dual field theory can be calculated as
\begin{eqnarray}
T_{\mu\nu}=2(K_{\mu\nu}-\gamma_{\mu\nu}K)+\frac{2}{\sqrt{-\gamma}}\frac{\delta S_\text{c.t.}}{\delta \gamma^{\mu\nu}}\,.
\end{eqnarray}
The total energy density is
\begin{eqnarray}\label{eC4}
\begin{split}
\epsilon=\lim_{r\to\infty}\sqrt{-\gamma} \langle T^0_0\rangle&=-\frac{1}{36}\Big[3M_{2}\left(6c^{2}q_2^{2}M_{2}+7\text{\ensuremath{M_{2}^{3}}}-24O_{2}\right) \\
&+144f_{1}-7M_{1}^{4}+42M_{1}^{2}M_{2}^{2}+72M_{1}O_{1}+18\lambda_{2}M_{2}^{4}-36u_{1}\Big].
\end{split}
\end{eqnarray}
Hence, from \eqref{eC2} and \eqref{eC4} we have $\frac{\Omega}{V}=\epsilon$.

\section{The near horizon expansion of the background geometry}
\label{app:d}
At finite temperature, the background geometry has the following near horizon form
\begin{eqnarray}
\begin{split}
&u=4\pi T\left(r-r_{0}\right)+\ldots\\
&v=v_{1}-v_{1}\frac{2\left(\phi_{11}^{2}+\phi_{21}^{2}\right)m^{2}r_{0}^{2}+\lambda_{1}\phi_{1}^{4}+\lambda_{2}\phi_{2}^{4}-24r_{0}^{4}}{12\pi Tr_{0}^{4}}\left(r-r_{0}\right)+\ldots\\
&f=f_{1}+f_{2}\left(r-r_{0}\right)+\ldots\\
&h=h_{1}+h_{2}\left(r-r_{0}\right)+\ldots\\
&A_{z}=A_{z1}+\frac{A_{z1}q_1^{2}\phi_{11}^{2}}{2\pi Tr_{0}^{2}}\left(r-r_{0}\right)+\ldots\\
&C_{y}=C_{y1}+\frac{C_{y1}q_2^{2}\phi_{21}^{2}}{2\pi Tr_{0}^{2}}\left(r-r_{0}\right)+\ldots\\
&r\phi_{1}=\phi_{11}+\frac{h_{1}\phi_{11}m^{2}r_{0}^{2}+\phi_{11}A_{z1}^{2}q_1^{2}r_{0}^{2}+4h_{1}\phi_{11}\pi Tr_{0}+h_{1}\lambda_{1}\phi_{11}^{3}}{4h_{1}\pi Tr_{0}^{2}}\left(r-r_{0}\right)+\ldots\\
&r\phi_{2}=\phi_{21}+\frac{f_{1}\phi_{21}m^{2}r_{0}^{2}+\phi_{21}C_{y1}^{2}q_2^{2}r_{0}^{2}+4f_{1}\phi_{21}\pi Tr_{0}+f_{1}\lambda_{2}\phi_{21}^{3}}{4f_{1}\pi Tr_{0}^{2}}\left(r-r_{0}\right)+\ldots.
\end{split}
\end{eqnarray}

\section{Calculations of conductivities in holography}
\label{app:e}
The corresponding equations of motion for the fluctuations in section \ref{tp} are
\begin{eqnarray}
\label{eq:flu1}
\begin{split}
&v_{x}^{\prime\prime}+\frac{1}{2}\left(\frac{f^{\prime}}{f}-\frac{v^{\prime}}{v}+\frac{h^{\prime}}{h}+\frac{2u^{\prime}}{u}\right)v_{x}^{\prime}+\frac{\omega^{2}}{u^{2}}v_{x}+\frac{8i\alpha\omega\sqrt{fhv}A_{z}^{\prime}}{ufh}\,v_{y}-\frac{8i\text{\ensuremath{\beta}}\omega\sqrt{fhv}C_{y}^{\prime}}{ufh}\hat{v}_{z}=0\,,\\
&v_{y}^{\prime\prime}+\frac{1}{2}\left(\frac{v^{\prime}}{v}-\frac{f^{\prime}}{f}+\frac{h^{\prime}}{h}+\frac{2u^{\prime}}{u}\right)v_{y}^{\prime}+\frac{\omega^{2}}{u^{2}}v_{y}-\frac{8i\alpha\omega\sqrt{fhv}A_{z}^{\prime}}{vhu}\,v_{x}=0\,,\\
&\hat{v}_{z}^{\prime\prime}+\frac{1}{2}\left(\frac{v^{\prime}}{v}-\frac{h^{\prime}}{h}+\frac{f^{\prime}}{f}+\frac{2u^{\prime}}{u}\right)\hat{v}_{z}^{\prime}+\frac{\omega^{2}}{u^{2}}\hat{v}_{z}+\frac{8i\beta\omega\sqrt{fvh}C_{y}^{\prime}}{fuv}\,v_{x}=0\,,
\end{split}
\end{eqnarray}
and
\begin{eqnarray}\label{z2}
\begin{split}
&\hat{v}_{x}^{\prime\prime}+\frac{1}{2}\left(\frac{f^{\prime}}{f}-\frac{v^{\prime}}{v}+\frac{h^{\prime}}{h}+\frac{2u^{\prime}}{u}\right)\hat{v}_{x}^{\prime}+\frac{\omega^{2}}{u^{2}}\hat{v}_{x}+\frac{8i\alpha\omega\sqrt{fvh}A_{z}^{\prime}}{fuh}\,\hat{v}_{y}-\frac{8i\text{\ensuremath{\beta}}\omega\sqrt{fhv}C_{y}^{\prime}}{ufh}v_{z}=0\,,\\
&\hat{v}_{y}^{\prime\prime}+\frac{1}{2}\left(\frac{v^{\prime}}{v}-\frac{f^{\prime}}{f}+\frac{h^{\prime}}{h}+\frac{2u^{\prime}}{u}\right)\hat{v}_{y}^{\prime}+\frac{\omega^{2}}{u^{2}}\hat{v}_{y}-\frac{8i\alpha\omega\sqrt{fvh}A_{z}^{\prime}}{vuh}\,\hat{v}_{x}=0\,,\\
&v_{z}^{\prime\prime}+\frac{1}{2}\left(\frac{v^{\prime}}{v}-\frac{h^{\prime}}{h}+\frac{f^{\prime}}{f}+\frac{2u^{\prime}}{u}\right)v_{z}^{\prime}+\frac{\omega^{2}}{u^{2}}v_{z}+\frac{8i\beta\omega\sqrt{fvh}C_{y}^{\prime}}{vfu}\,\hat{v}_{x}=0\,.
\end{split}
\end{eqnarray}

These six equations are divided into two groups. The first group gives both the electric Hall conductivity and 
the spin Hall conductivity. Therefore, we shall focus on the fluctuations $v_x, v_y, \hat{v}_z$.
We show here
the calculations of the Hall conductivities.

Let us start from the finite temperature case. We work in the small $\omega$ limit $\omega\ll r_0$.
The fluctuations $v_x, v_y, \hat{v}_{z}$ which satisfies the infalling boundary condition can be expanded in the whole radial direction
as \cite{Landsteiner:2015pdh, Landsteiner:2015lsa}
\be
v_i=u^{-\frac{i\omega}{4\pi T}}\left(v_i^{\left(0\right)}+\omega v_i^{\left(1\right)}+\dots\right)\,,~~~~~~~\hat{v}_z=u^{-\frac{i\omega}{4\pi T}}\left(\hat{v}_z^{\left(0\right)}+\omega \hat{v}_z^{\left(1\right)}+\dots\right)\,,\ee where $i=x,y$.  We solve the equations (\ref{eq:flu1}) order by order in $\omega$.

At zeroth order of $\omega$, we have
\begin{eqnarray}
\begin{split}
&v_{x}^{\left(0\right)\prime\prime}+\frac{1}{2}\left(\frac{f^{\prime}}{f}-\frac{v^{\prime}}{v}+\frac{h^{\prime}}{h}+\frac{2u^{\prime}}{u}\right)v_{x}^{\left(0\right)\prime}=0\,,\\
&v_{y}^{\left(0\right)\prime\prime}+\frac{1}{2}\left(\frac{v^{\prime}}{v}-\frac{f^{\prime}}{f}+\frac{h^{\prime}}{h}+\frac{2u^{\prime}}{u}\right)v_{y}^{\left(0\right)\prime}=0\,,\\
&\hat{v}_{z}^{\left(0\right)\prime\prime}+\frac{1}{2}\left(\frac{v^{\prime}}{v}-\frac{h^{\prime}}{h}+\frac{f^{\prime}}{f}+\frac{2u^{\prime}}{u}\right)\hat{v}_{z}^{\left(0\right)\prime}=0\,.
\end{split}
\end{eqnarray}
These three equations can be further simplified as \be \left(\frac{u\sqrt{fh}}{\sqrt{v}}v_{x}^{\left(0\right)\prime}\right)^{\prime}=0\,,~~~
\left(\frac{u\sqrt{vh}}{\sqrt{f}}v_{y}^{\left(0\right)\prime}\right)^{\prime}=0\,,~~~
\left(\frac{u\sqrt{vf}}{\sqrt{h}}\hat{v}_{z}^{\left(0\right)\prime}\right)^{\prime}=0\,,\ee
thus we have $v_{x}^{\left(0\right)}=c_1$,
$v_{y}^{\left(0\right)}=c_2$,
$\hat{v}_{z}^{\left(0\right)}=c_3$, where $c_i$ with $i=1,2,3$ are integration constants and we have used the regularity condition.

At first order of $\omega$, we have
\begin{eqnarray}
\begin{split}
&v_{x}^{\left(1\right)\prime\prime}+\frac{1}{2}\left(\frac{f^{\prime}}{f}-\frac{v^{\prime}}{v}+\frac{h^{\prime}}{h}+\frac{2u^{\prime}}{u}\right)v_{x}^{\left(1\right)\prime}-\frac{i}{4\pi Tu}\left(\frac{u^{\prime\prime}}{u^{\prime}}-\frac{v^{\prime}}{2v}+\frac{h^{\prime}}{2h}+\frac{f^{\prime}}{2f}\right)v_{x}^{\left(0\right)}\\
&~~~~~~-\frac{iu^{\prime}}{2\pi Tu}v_{x}^{\left(0\right)\prime}+\frac{4i\sqrt{fhv}}{fhu}\left(2\alpha v_{y}^{\left(0\right)}A_{z}^{\prime}-2\beta\hat{v}_{z}^{\left(0\right)}C_{y}^{\prime}\right)=0\,,\\
&v_{y}^{\left(1\right)\prime\prime}+\frac{1}{2}\left(\frac{v^{\prime}}{v}-\frac{f^{\prime}}{f}+\frac{h^{\prime}}{h}+\frac{2u^{\prime}}{u}\right)v_{y}^{\left(1\right)\prime}-\frac{i}{4\pi Tu}\left(\frac{u^{\prime\prime}}{u^{\prime}}-\frac{f^{\prime}}{2f}+\frac{h^{\prime}}{2h}+\frac{v^{\prime}}{2v}\right)v_{y}^{\left(0\right)}\\
&~~~~~~-\frac{iu^{\prime}}{2\pi Tu}v_{y}^{\left(0\right)\prime}-\frac{8i\alpha\sqrt{fhv}A_{z}^{\prime}}{fhu}v_{x}^{\left(0\right)}=0\,,\\
&\hat{v}_{z}^{\left(1\right)\prime\prime}+\frac{1}{2}\left(\frac{v^{\prime}}{v}-\frac{h^{\prime}}{h}+\frac{f^{\prime}}{f}+\frac{2u^{\prime}}{u}\right)\hat{v}_{z}^{\left(1\right)\prime}-\frac{i}{4\pi Tu}\left(\frac{u^{\prime\prime}}{u^{\prime}}-\frac{h^{\prime}}{2h}+\frac{f^{\prime}}{2f}+\frac{v^{\prime}}{2v}\right)\hat{v}_{z}^{\left(0\right)}\\
&~~~~~~-\frac{iu^{\prime}}{2\pi Tu}\hat{v}_{z}^{\left(0\right)\prime}+\frac{8i\beta\sqrt{fhv}C_{y}^{\prime}}{fuv}v_{x}^{\left(0\right)}=0\,.
\end{split}
\end{eqnarray}

We focus on the retarded Green function $G^R_{xy}=\langle J_xJ_y\rangle_R$ and $\hat{G}^R_{zx}=\langle \hat{J}_z J_x\rangle_R$. For $G^R_{xy}$, we can choose the source term $c_1=1,c_2=c_3=0$, and we have the equation for $v_{y}^{\left(1\right)}$ as
\begin{eqnarray}
v_{y}^{\left(1\right)\prime\prime}+\frac{1}{2}\left(\frac{v^{\prime}}{v}-\frac{f^{\prime}}{f}+\frac{h^{\prime}}{h}+\frac{2u^{\prime}}{u}\right)v_{y}^{\left(1\right)\prime}-\frac{8i\alpha\sqrt{fhv}A_{z}^{\prime}}{huv}=0\,.
\end{eqnarray}
The above equation can be simplified as $\left(\text{\ensuremath{\frac{u\sqrt{hv}}{\sqrt{f}}}}v_{y}^{\left(1\right)\prime}\right)^{\prime}=8i\alpha \frac{v}{f}A_{z}^{\prime}$, with the regularity boundary condition we have $v_{y}^{\left(1\right)\prime}=\frac{8i\alpha\left(A_{z}\left(r\right)-A_{z}\left(r_{0}\right)\right)\sqrt{v}}{u\sqrt{hf}}$.
Hence, we have $\Im G^R_{xy}=8\omega\alpha\left(A_{z}\left(r\right)-A_{z}\left(r_{0}\right)\right)$. From the Kubo formula, we have $\sigma_{xy}=8\alpha\left(b-A_z\left(r_0\right)\right)$, and $\sigma_{\text{AHE}}=8\alpha b-\sigma_{xy}=8\alpha A_z\left(r_0\right)$.
Similarly, we can obtain the $\boldmath{Z}_2$ anomalous conductivity $\sigma_{\boldmath{Z}_2 \text{AHE}}$ from $\Im \hat{G}^R_{zx}=8\omega\alpha\left(C_{y}\left(r\right)-C_{y}\left(r_{0}\right)\right)$
and obtain $\sigma_{\boldmath{Z}_2 \text{AHE}}=8\beta C_y\left(r_0\right)$.

Now we calculate the anomalous transports at zero temperature following the near-far matching method in \cite{Landsteiner:2015pdh}.
We take the case of four Weyl nodes as an example and the calculations for the other cases are straightforward to calculate.
Using the leading order solution of the near IR solutions in \eqref{fwp}, \eqref{fgp}, \eqref{inter}, the corresponding equations of motion for $v_{i}$ and $\hat{v}_{i}$ have the form
\begin{eqnarray}
\begin{split}
&v_{i}^{\left(n\right)\prime\prime}+\frac{3}{r}v_{i}^{\left(n\right)\prime}+\frac{\omega^{2}}{r^{4}}v_{i}^{\left(n\right)}=0\,,
~~~~
\hat{v}_{z}^{\left(n\right)\prime\prime}+\frac{3}{r}\hat{v}_{z}^{\left(n\right)\prime}+\frac{\omega^{2}}{r^{4}}\hat{v}_{z}^{\left(n\right)}=0\,,
\end{split}
\end{eqnarray}
where $i=x,y$. This form of equations have been discussed in \cite{Landsteiner:2015pdh} and we can expand the in-falling near horizon solutions in the matching regime $\omega\ll r\ll\min\left\{ M_{1},b\right\}\left(\text{or}\,~ \omega\ll r\ll\min\left\{ M_{2},c\right\}\right)$ as
\begin{eqnarray}
&v_{i}^{\left(n0\right)}=1-\frac{\omega^{2}}{4r^{2}}\left(-1+2\gamma+2\ln\left[\frac{-i\omega}{2r}\right]\right)\,,~~~~~~
\hat{v}_{z}^{\left(n0\right)}=1-\frac{\omega^{2}}{4r^{2}}\left(-1+2\gamma+2\ln\left[\frac{-i\omega}{2r}\right]\right)\,.
\end{eqnarray}
Note that the above three solutions are three independent boundary conditions.
The same as the finite temperature case mentioned above, we focus on $G^R_{xy}$ and $\hat{G}^R_{zx}$. We choose the near horizon condition with $v_y^{(n0)}=\hat{v}_z^{(n0)}=0$ while $v_x^{(n0)}$ as above.
We will calculate the far region solution sourced by this infalling near horizon solution.

At the matching region the near horizon solution above gives $v_{x}^{\left(n\right)}=1$, 
$v_{y}^{\left(n\right)}=\omega v_{y}^{\left(n1\right)}$ and
$
\hat{v}_{z}^{\left(n\right)}=\omega \hat{v}_{z}^{\left(n1\right)}$, where
\begin{eqnarray}
&
v_{y}^{\left(n1\right)\prime}=\frac{8i\alpha\left(A_{z}\left(r\right)-A_{z}\left(0\right)\right)}{u_{0}r^{3}}\,,~~~~~~
v_{z}^{\left(n1\right)\prime}=-\frac{8i\beta\left(C_{y}\left(r\right)-C_{y}\left(0\right)\right)}{u_{0}r^{3}}\,.
\end{eqnarray}
These solutions are boundary conditions for the fluctuations in the far region.

In the far region $r \gg \omega$, we have equations
\begin{eqnarray}
\begin{split}
&v_{x}^{\left(f\right)\prime\prime}+\frac{1}{2}\left(\frac{u^{\prime}}{u}+\frac{f^{\prime}}{f}+\frac{h^{\prime}}{h}\right)v_{x}^{\left(f\right)\prime}+\frac{8i\alpha A_{z}^{\prime}}{\sqrt{fhu}}\,v_{y}^{\left(f\right)}-\frac{8i\text{\ensuremath{\beta}}C_{y}^{\prime}}{\sqrt{fhu}}\hat{v}_{z}^{\left(f\right)}=0\,,\\
&v_{y}^{\left(f\right)\prime\prime}+\frac{1}{2}\left(\frac{3u^{\prime}}{u}-\frac{f^{\prime}}{f}+\frac{h^{\prime}}{h}\right)v_{y}^{\left(f\right)\prime}-\frac{8i\alpha\sqrt{f}A_{z}^{\prime}}{u\sqrt{hu}}v_{x}^{\left(f\right)}=0\,,\\
&\hat{v}_{z}^{\left(f\right)\prime\prime}+\frac{1}{2}\left(\frac{3u^{\prime}}{u}+\frac{f^{\prime}}{f}-\frac{h^{\prime}}{h}\right){\hat{v}}_{z}^{\left(f\right)\prime}+\frac{8i\beta\sqrt{h}C_{y}^{\prime}}{u\sqrt{fu}}v_{x}^{\left(f\right)}=0 \,.
\end{split}
\end{eqnarray}
The far regime solutions with the above boundary condition are $v_x^{\left(f\right)}=1$, $v_{y}^{\left(f\right)}=\omega v_{y}^{\left(f1\right)}$ and $\hat{v}_{z}^{\left(f\right)}=\omega \hat{v}_{z}^{\left(f1\right)}$ with
\begin{eqnarray}
\begin{split}
&v_{y}^{\left(f1\right)\prime}=\frac{8i\alpha\left(A_{z}\left(r\right)-A_{z}\left(0\right)\right)}{\sqrt{uhf}}\,,
~~~~~~~~
\hat{v}_{z}^{\left(f1\right)\prime}=-\frac{8i\beta\left(C_{y}\left(r\right)-C_{y}\left(0\right)\right)}{\sqrt{uhf}}\,.
\end{split}
\end{eqnarray}
From the solutions above we obtain $\sigma_{\text{AHE}}=8\alpha A_z\left(0\right)$ and $\sigma_{\boldmath{Z}_2 \text{AHE}}=8\beta C_y\left(0\right)$.
For other phases one can use the same method above and obtain the same results for the anomalous Hall conductivities.
Thus the formulas for the anomalous Hall conductivity ($\boldmath{Z}_2$ anomalous Hall conductivity)
are the same for both the finite temperature and zero temperature.

\end{appendix}
\appendix

\end{document}